\documentclass[
aps,
prx,
reprint,
superscriptaddress,
longbibliography,
nofootinbib
]{revtex4-2}

\usepackage{tikz}
\usetikzlibrary{arrows.meta,positioning,calc}

\usepackage[T1]{fontenc}
\usepackage[utf8]{inputenc}
\usepackage[english]{babel}

\usepackage{amsmath}
\usepackage{amssymb}
\usepackage{amsfonts}
\usepackage{mathtools}
\usepackage{bm}
\usepackage{dsfont}
\usepackage{mathrsfs}

\usepackage{graphicx}
\usepackage{booktabs}
\usepackage{array}
\usepackage{microtype}

\usepackage[dvipsnames]{xcolor}
\usepackage[colorlinks=true,citecolor=blue,urlcolor=blue,linkcolor=blue]{hyperref}
\usepackage[all]{hypcap}
\usepackage{orcidlink}

\allowdisplaybreaks

\begin{document}

\title{Probing Krylov Dynamics in Driven Many-Body Systems}

\author{Andr\'as Grabarits$^{\orcidlink{0000-0002-0633-7195}}$}
\email{andras.grabarits@uni.lu}

\affiliation{Department of Physics and Materials Science, University of Luxembourg, L-1511 Luxembourg}

\author{Adolfo del Campo$^{\orcidlink{0000-0003-2219-2851}}$}\email{adolfo.delcampo@uni.lu}
\affiliation{Department of Physics and Materials Science, University of Luxembourg, L-1511 Luxembourg}
\affiliation{Donostia International Physics Center, E-20018 San Sebasti\'an, Spain}
\affiliation{Institute of Science and Engineering, Kanazawa University, Kakuma-machi, Kanazawa, Ishikawa,  920-1192, Japan}

\date{\today}


\begin{abstract}
Krylov complexity has become a central probe of operator growth, scrambling, and quantum chaos. Yet, a basic fact hinders its experimental study: laboratory quantum systems are routinely driven in time, whereas the Krylov construction assumes a fixed generator and breaks down once the Hamiltonian varies.
Here, we solve the time-dependent many-body Krylov problem exactly and in closed form for scale- and rotation-invariant systems realized in current quantum-simulation and control platforms. We find exact relations that map the driven dynamics onto mutually orthogonal Krylov subspaces and yield a closed-form spread complexity, determined solely by the spectral properties of the initial Hamiltonian and the driving schedule. The construction extends to Liouville space and solves the time-dependent density-matrix Krylov problem, which makes the Krylov dynamics of initial thermal states experimentally accessible. In scale-invariant quantum gases, such as the driven Lieb-Liniger model, the Tonks-Girardeau gas, and the Calogero-Sutherland model, the complexity is set by the width of the atomic cloud and its rate of change, directly accessible via in situ and time-of-flight imaging. In rotation-invariant spin systems such as the Haldane-Shastry and central-spin models, it is governed by collective spin precession that can be probed and read out in trapped-ion, cavity-QED, and circuit-QED platforms.
\end{abstract}

\maketitle


\section{Introduction}
\label{sec:introduction}

Quantum dynamics explores only a small subspace of the full Hilbert space. The Krylov complexity measures the spread of a state or operator along the sequence of states generated by the evolution~\cite{Viswanath_Muller_1994_RecursionMethod, Parker_PRX_2019_UniversalOperatorGrowth, Nandy_PhysRep_2025_KrylovReview, Baiguera_PhysRep_2026_QuantumComplexityReview, Rabinovici_arXiv_2025_KrylovComplexity}. This quantity has become a central probe of information scrambling, operator growth, and quantum chaos~\cite{Parker_PRX_2019_UniversalOperatorGrowth, Barbon_JHEP_2019_OperatorComplexity, Rabinovici_JHEP_2021_EdgeKrylovSpace, Caputa_PRR_2022_GeometryKrylov, Balasubramanian_PRD_2022_ComplexitySpread, Dymarsky_Gorsky_PRB_2020_Delocalization, Dymarsky_Smolkin_PRD_2021_CFT, Rabinovici_JHEP_2022_IntegrabilityChaos, Avdoshkin_JHEP_2024_QFTBeyond, Hornedal_CommPhys_2022_UltimateSpeedLimits,martinezazcona2026}. However, the theory remains largely confined to autonomous systems. It thus lags behind the state of the art in quantum science and technology, where time-dependent interacting many-body systems are routinely engineered in cold-atom, trapped-ion, and cavity-QED platforms. The gap is even sharper in realistic experiments, where finite temperature, preparation errors, dephasing, and residual coupling to uncontrolled modes make the system best described by a density matrix, rather than a pure quantum state~\cite{Caputa_JHEP_2024_DensityMatrixKrylov,
Tan_PRE_2025_FiniteTemperatureKrylov}. Extending Krylov dynamics to density matrices in driven systems is thus a necessary step toward making Krylov complexity an experimentally measurable probe of driven many-body quantum systems.

Krylov subspace methods have recently been applied to quantum control and to nonequilibrium many-body physics from different perspectives. They can be used to systematically engineer the local expansion of counterdiabatic driving required for experimentally realistic shortcuts to adiabaticity~\cite{Takahashi_PRX_2024_STA_Krylov, Visuri_npjQI_2026_DigitizedCD,Bhattacharjee2023}. At the same time, the Krylov complexity has emerged as a probe of dynamical phase transitions, topology, and universal critical growth~\cite{Bento_PRB_2024_LMGDynamicalTransition, Caputa_JHEP_2023_KitaevTransitions, Grabarits_arXiv_2025_UniversalGrowthQPT, Takahashi_PRB_2026_TodaFlows}. Although these developments demonstrate the power of the Krylov subspace methods, they also expose their current limitation. Existing time-dependent constructions~\cite{
Takahashi_PRL_2025_TimeDependentGenerators,%
Grabarits_arXiv_2025_UniversalGrowthQPT,%
Grabarits_arXiv_2026_TDKrylovLie} and density-matrix
formulations~\cite{Caputa_JHEP_2024_DensityMatrixKrylov} have
not yet been combined to allow for an exact treatment of driven
many-body systems in mixed states, which are the relevant
setting in experiments.

This gap in the literature arises from the following challenges. In the static case, quantum dynamics in Krylov space, even for many-body systems, is mapped to a hopping problem on the Krylov lattice or Krylov chain, in which each site is associated with a Krylov basis state, and the hopping amplitudes are set by the Lanczos coefficients determined by the Krylov algorithm \cite{Nandy_PhysRep_2025_KrylovReview,Rabinovici_arXiv_2025_KrylovComplexity}, also known as the recursion method in condensed matter physics \cite{Viswanath_Muller_1994_RecursionMethod}. However, once the generator depends on time, the Krylov basis is no longer tied to a single static
operator, and the Lanczos coefficients are no longer associated with a time-independent
one-dimensional Krylov lattice. In interacting systems, the difficulty becomes even
more severe. The Hilbert-space dimension grows exponentially with the
system size, and so does the cost of computing the Krylov
basis states and the Lanczos coefficients at each instant of
time.
For density matrices, the construction must be performed in
Liouville space with dimension equal to the square of that of the
Hilbert space.

Due to these obstacles, progress on the driven problem has remained scarce and confined to partial and approximate routes. Floquet formulations~\cite{Nizami_PRE_2023_DrivenKrylov, Nizami_PRE_2024_PeriodicallyDriven,
Suchsland_PRB_2025_TrotterTransitions} are restricted to periodic
protocols, Magnus expansions~\cite{Grabarits_arXiv_2025_UniversalGrowthQPT}
effectively hold only at short times or weak driving, and the few exact time-dependent constructions to date~\cite{Takahashi_PRL_2025_TimeDependentGenerators,
Grabarits_arXiv_2026_TDKrylovLie} effectively involve single-particle
systems. Most pressingly, no exact closed-form solution exists for driven interacting systems, either for quantum states or for density
matrices. The establishment of such a framework is the focus of this work.

The central obstacle is that a driven many-body Hamiltonian is generally
not a linear combination of a few operators that close under commutation, as
required by the existing exact constructions. As one of the key messages of the present work, our approach requires such a
linear form only for the generator of the transformation induced by the
driving, not for the Hamiltonian itself.
We consider the kind of scale- and rotation-invariant driving protocols often realized in the laboratory, such as ultracold gases in time-dependent traps or collective fields that simultaneously rotate an ensemble of interacting spins. In both cases,
 driving results in the scaling or rotation of the many-body state, while
all interactions evolve rigidly with it. Due to the presence of dynamical invariants, the driven system can be described in terms of the undriven one and a single dilation or rotation, known in closed
form and controlled directly in experiments.
Technically, these collective transformations can be expressed via the generators of a rank-one algebra, $\mathfrak{su}(1,1)$ for dilations and $\mathfrak{su}(2)$ for rotations.
The initial many-body Hamiltonian itself need not admit
such an algebraic description. It only fixes the initial amplitudes in the multidimensional Krylov basis through the overlaps of its eigenstates with the Krylov basis states. The construction can therefore be applied beyond exactly solvable models.

Our construction thus decomposes the exponentially large Hilbert space into mutually
orthogonal Krylov chains, along which the driven evolution is exact at all
times and at any driving strength. Scale invariance
realizes this mechanism in trapped gases, where the dilation
governs expansion, transport, and collective width
oscillations~\cite{Chen_PRL_2010_FastCooling,
delCampo_PRA_2011_FrictionlessQuenches,
delCampo_Boshier_SciRep_2012_Box,
delCampo_PRL_2013_CounterdiabaticDriving,
Torrontegui_AdvAMOP_2013_STA,
Deffner_PRX_2014_ScaleInvariantSTA} from weakly interacting
Bose gases to strongly correlated quantum fluids such as
the Tonks-Girardeau gas and the Calogero-Sutherland
model~\cite{CastinDum_PRL_1996_BECScaling,
Kagan_PRA_1996_BECExpansion,
Pitaevskii_Rosch_PRA_1997_Breathing,
LewisRiesenfeld_JMP_1969_Invariants,
Girardeau_OptComm_2004_TonksScaling,
Sutherland_PRL_1998_CalogeroScaling,delcampo2016,
ADC_Entropy_2020,Yang2025}.
Rotation invariance occurs in interacting spin systems with isotropic
exchange~\cite{Grabarits_arXiv_2026_TDKrylovLie}, from long-range Heisenberg
chains of Haldane-Shastry type~\cite{Haldane_PRL_1988_HaldaneShastry,
Shastry_PRL_1988_HaldaneShastry} to central-spin Gaudin
magnets~\cite{Gaudin_JPhys_1976_SpinHamiltonians}, where a global rotation
transports every spin direction while leaving the exchange untouched.

The multidimensional Krylov subspace construction can also be extended to the operator dynamics of
arbitrary initial density matrices, including thermal ensembles~\cite{Takahashi_PRL_2025_TimeDependentGenerators,
Grabarits_arXiv_2026_TDKrylovLie}. Along each chain, the spread complexity of states and the
Krylov variance employed in the case of density matrices follow in closed form, controlled solely by the parameters of the driving protocol and the initial amplitudes. In the scale-invariant case, both are fixed
by the expansion of the trapped gas and its rate of change, which
can be read out directly by in situ absorption
imaging~\cite{Ketterle1999,Gemelke_Nature_2009_InSituImaging,Hung_Nature_2011_ScaleInvariance2D}, while releasing the gas and imaging it after
free flight gives the same information from the momentum
distribution \cite{Ketterle1999,Kinoshita_Nature_2006_NewtonsCradle}.
Scale invariance itself has already been measured in numerous experiments with trapped
gases \cite{Hung_Nature_2011_ScaleInvariance2D,Vogt_PRL_2012_ScaleInvariance2DFermi,Holten_PRL_2018_ScaleAnomaly,SaintJalm_PRX_2019_DynamicalSymmetry}.
Likewise, under rotational invariance, the dynamics is fully characterized by the collective rotation angle set by the driving protocol, and can therefore be engineered and probed in trapped-ion and cavity-QED spin platforms~\cite{Monroe_RMP_2021_IonSpinSim,Garttner_NatPhys_2017_OTOC,Britton_Nature_2012_EngineeredIsing,Bohnet_Science_2016_QuantumSpinDynamics,Hosten_Nature_2016_MeasurementNoise,Young2024BCSCavity}. As a result, the spread complexity and the Krylov variance of driven scale- and
rotation-invariant many-body systems can be determined from routine
laboratory measurements.


\section{Unified construction and main results}
\label{sec:general}

In this section, we consider the general framework for scale- and rotation-invariant transformations.
 In each case, the driving acts on the initial Hamiltonian as a dilation or a collective rotation, parametrized by the scaling factor $\gamma(t)$ of the
internal length scale in the former and by the collective rotation angles
$\theta(t)$ and $\phi(t)$ in the latter case. As a result, the time evolution is fully described
by the initial Hamiltonian $H(0)$ and a time-dependent
transformation $\mathcal T(t)$. The transformation is set by
the driving alone and requires no knowledge of the
instantaneous spectrum of $H(t)$.

The analytical solution relies on the fact that the transformation $\mathcal T(t)$ can be written as a linear combination of the generators of the rank-one algebra $\mathfrak{su}(1,1)$ in the scale-invariant case and of $\mathfrak{su}(2)$ in the rotation-invariant case.

The initial Hamiltonian $H(0)$ need not be a linear combination of the same generators as $\mathcal T(t)$. Its
eigenstates only fix the initial amplitudes in mutually
orthogonal Krylov chains, i.e., the irreducible
representations of the underlying algebra. The dynamics thus
maps onto independent one-dimensional chains, and the Krylov
subspace dynamics can be treated exactly, with the explicit
forms given for each realization in
Secs.~\ref{sec:scale_covariant} and~\ref{sec:rot_invariant}. This construction can be generalized to density matrices. With the initial density matrix represented as a vector in Liouville
space, it applies with the adjoint representation of the generators. The driven density matrix thus spreads along
mutually orthogonal Krylov chains in the same way as quantum states.

\subsection{Reduction to the initial Hamiltonian}

In both cases, the driven Hamiltonian is related to the initial one by the transformation $\mathcal T(t)$ as
\begin{eqnarray}
H(t)=\dot s(t)\,\mathcal T(t)\,H(0)\,\mathcal T^\dagger(t)
+i\dot{\mathcal T}(t)\,\mathcal T^\dagger(t),
\label{eq:Ht_frame_general}
\end{eqnarray}
with $s(t)$ depending on the specific form of the driving in the scale-invariant case and $s(t)=t$ for the rotation-invariant case.
Additional details are provided in Sec.~\ref{sec:scale_covariant} and Sec.~\ref{sec:rot_invariant}  for the scale- and rotation-invariant cases, respectively.
 Introducing the transformed wave function of the evolution,
\begin{eqnarray}\label{eq: Int_picture}
i\partial_t|\Psi(t)\rangle=H(t)|\Psi(t)\rangle,\quad|\Psi(t)\rangle=\mathcal T(t)|\chi(t)\rangle,
\end{eqnarray}
the time-dependent Schr\"odinger equation simplifies to
\begin{eqnarray}
i\partial_t|\chi(t)\rangle=
\dot s(t)\,H(0)|\chi(t)\rangle.
\label{eq:chi_general}
\end{eqnarray}
The exact solution can be expressed via the separate actions of $\mathcal T(t)$ and $H(0)$,
\begin{eqnarray}
|\Psi(t)\rangle=
\mathcal T(t)\,e^{-is(t)H(0)}|\Psi(0)\rangle,
\label{eq:Psi_exact_general}
\end{eqnarray}
where the action of the generally unknown $H(0)$ reduces to a phase for any initial eigenstate,
$H(0)|\nu\rangle=E_\nu|\nu\rangle$,
\begin{eqnarray}
|\Psi_\nu(t)\rangle=
e^{-iE_\nu s(t)}\,\mathcal T(t)|\nu\rangle.
\label{eq:Psi_nu_general}
\end{eqnarray}
Thus, the nontrivial
dynamics is encoded in $\mathcal T(t)$, whose action on the Krylov subspace we consider in what follows.

\subsection{Lie algebras and their irreducible representations}
\label{sec:lie_primer}

Before discussing the Krylov dynamics, we briefly summarize the essential
elements of Lie algebras with emphasis on the rank-one algebras $\mathfrak{su}(1,1)$ and $\mathfrak{su}(2)$. A broader overview is given in Refs.~\cite{Gilmore_1974_LieGroups,Perelomov_1986_GeneralizedCoherentStates}.

We consider an arbitrary Lie algebra $\mathfrak g$ of rank $r$ in the
Cartan-Weyl basis, which is spanned by the Cartan elements $H_i\in\mathfrak g$,
$i=1,\dots,r$, and the root elements $E_\alpha\in\mathfrak g$, referred to as
ladder operators in the following, with the commutation relations
\begin{eqnarray}
[H_i,H_j]=0,\,\,
[H_i,E_\alpha]=\alpha(H_i)E_\alpha,\,\,
[E_\alpha,E_{-\alpha}]=H_\alpha,
\label{eq:cartan_weyl_primer}
\end{eqnarray}
where $H_\alpha$ is a Cartan element associated with the root $\alpha$.
Thus, the root operators raise and lower the Cartan eigenvalues.
Of particular interest are the irreducible representations of $\mathfrak g$. Each of them is spanned by
common eigenstates of the Casimir operator $\mathcal C$,
which commutes with all generators,
$[\mathcal C,E_{\pm\alpha}]=0$ and $[\mathcal C,H_i]=0$.
These states provide the Krylov basis for the rest of the
present work. We consider the lowest-weight representations, defined by an eigenstate of the Cartan elements $H_i|\mu\rangle=\mu_i|\mu\rangle$ with $\mu=(\mu_1,\dots,\mu_r)$
that is annihilated by any of the lowering operators, $E_{-\alpha}|\mu\rangle=0$. The remaining basis vectors are generated by the
successive action of the raising operators,
$E_{\alpha}|\mu\rangle\propto|\mu+\alpha\rangle$,
with element-wise addition of $\mu$ and
$\alpha=(\alpha_1,\dots,\alpha_r)$, and are mutually
orthogonal, $\langle\mu|\mu'\rangle=
\delta_{\mu,\mu'}$. As a result, Hamiltonians containing only a single pair of ladder operators generate dynamics along a one-dimensional and mutually orthogonal chain, which allows for the exact analytical treatment of the corresponding Krylov subspace dynamics.

In the following, the focus will be on rank-one algebras, $r=1$, with a single
Cartan generator, $H_1$, and one pair of ladder operators, $E_{\pm1}$. The algebras $\mathfrak{su}(2)$ and $\mathfrak{su}(1,1)$ considered
here are both real forms of the complex algebra $\mathfrak{sl}(2,\mathbb C)$, whose commutation relations, $[H_1,E_{\pm1}]=\pm E_{\pm1}$ and $[E_{+1}, E_{-1}]=2\sigma H_1$, fix only the complex algebra, while the real form is selected by the $*$-structure inherited from Hermitian conjugation. With $H_1^\dagger=H_1$ and $E_{+1}^\dagger=E_{-1}$, the sign $\sigma=+1$ selects the compact algebra $\mathfrak{su}(2)$ and $\sigma=-1$ the noncompact one $\mathfrak{su}(1,1)\cong\mathfrak{sl}(2,\mathbb R)$, which up to isomorphism exhaust the real forms~\cite{Grabarits_arXiv_2026_TDKrylovLie}. Their unitary irreducible representations are finite (compact) and infinite dimensional (noncompact), respectively. The compact case is described by the $\mathfrak{su}(2)$ algebra which is generated by the
collective spin components $J_0\equiv H_1$, $J_\pm\equiv E_{\pm1}$, and $\mathcal C\equiv \mathbf J^2$,
\begin{eqnarray}
[J_0,J_\pm]=\pm J_\pm,\quad
[J_+,J_-]=2J_0,\nonumber\\
\bm J^2=J_0^2+\tfrac12(J_+J_-+J_-J_+).
\label{eq:su2_primer}
\end{eqnarray}
The Casimir $\bm J^2=j(j+1)$ labels the irreducible representations with $j$ integer or half-integer and with $2j+1$ states $|j,q\rangle,\,q=0,\dots,2j$, where the lowest-weight state is given by $J_-|j,0\rangle=0$. Orthogonality holds between all states in each irreducible tower, $\langle j,q|j',q'\rangle=\delta_{jj'}\delta_{qq'}$. Since the total spin $j$ is conserved under a global rotation, the Hilbert space of $N$ spins
decomposes into such spin towers, which remain invariant under the action of $\mathcal T(t)$.

For the noncompact case with the $\mathfrak{su}(1,1)$ algebra, we use the notation $H_1\equiv K_0$ and $E_{\pm1}\equiv K_\pm$, with the commutation relations and Casimir operator,
\begin{eqnarray}
[K_0,K_\pm]=\pm K_\pm,\quad
[K_+,K_-]=-2K_0,\nonumber\\
\mathcal C=K_0^2-\tfrac12(K_+K_-+K_-K_+).
\label{eq:su11_primer}
\end{eqnarray}
First, we consider the discrete irreducible representations labeled by the Bargmann index $k>0$, and with Casimir eigenvalue $\mathcal C=k(k-1)$, lowest-weight state
$K_-|k,0\rangle=0$, and $K_0|k,n\rangle=(k+n)|k,n\rangle$. In contrast to the $\mathfrak{su}(2)$ case, these towers are
infinite, $n=0,1,2,\dots$, but the same orthogonality holds for all states in each Bargmann tower, i.e., $\langle k,n|k',n'\rangle=\delta_{kk'}\delta_{nn'}$.
The $\mathfrak{su}(1,1)$ algebra also admits the principal-series representations, which are labeled by a continuous parameter $\lambda$ and
have discrete bi-infinite Cartan eigenvalues,
$K_0\lvert\lambda,M\rangle=M\lvert\lambda,M\rangle$ and
$K_\pm|\lambda,M\rangle\propto|\lambda,M\pm1\rangle$, with $M\in\mathbb Z$.
They obey the orthogonality relation $\langle\lambda,M\vert\lambda',M'\rangle=
\delta(\lambda-\lambda')\delta_{MM'}$.
As the chain extends in both directions, no lowest-weight state exists, and the towers are distinguished by the Casimir eigenvalues of $\mathcal C=-(\lambda^2+\tfrac14)$. While the discrete version is used for the dynamics of quantum states, the principal-series representation is used for the density-matrix dynamics in Sec.~\ref{sec:density_general}.

\subsection{Dynamics in Krylov space and orthogonal Krylov chains}
\label{sec:krylov_chains}
Next, we show how the generator of $\mathcal T(t)$ decomposes into the
generators of the rank-one algebra and how it generates the Krylov dynamics. Denoting the Cartan element and the ladder operators by
$\{X_0,X_\pm\}$~\cite{Gilmore_1974_LieGroups},
\begin{eqnarray}
[X_0,X_\pm]&=&\pm X_\pm,\qquad
[X_+,X_-]=2\sigma X_0,
\label{eq:CW_general}
\end{eqnarray}
with $\sigma=-1$ for $\mathfrak{su}(1,1)$ (scale-invariant case) and $\sigma=+1$ for
$\mathfrak{su}(2)$ (rotation-invariant case), the transformation $\mathcal T(t)$ can be expressed by the relation
\begin{eqnarray}\label{eq:T_generator}
i\dot{\mathcal T}(t)\,\mathcal T^\dagger(t)
=A_0(t)X_0+A_+(t)X_++A_-(t)X_-.
\end{eqnarray}
The corresponding coefficients depend solely on the driving schedule and follow from
exact decomposition
formulas~\cite{WeiNorman1963,WeiNorman1964,CharzynskiKus2013WeiNorman,Grabarits_arXiv_2026_TDKrylovLie} specific to the choice $\sigma=\pm1$, as will be discussed in more detail in the upcoming sections. Within each irreducible representation, this generator acts as a tridiagonal
matrix identical to the Krylov Liouvillian of the exact  formulation of the time-dependent Krylov subspace approach~\cite{Takahashi_PRL_2025_TimeDependentGenerators,Grabarits_arXiv_2026_TDKrylovLie,Grabarits_arXiv_2025_UniversalGrowthQPT}. In the exact formulation for an arbitrary time-dependent Hamiltonian, the Lanczos algorithm must be performed with the extended generator $H(t)-i\partial_t$ to account for the time dependence of both the Lanczos coefficients and the Krylov states~\cite{Takahashi_PRL_2025_TimeDependentGenerators},
\begin{eqnarray}
b_{n+1}(t)|K_{n+1}\rangle
\!&=&\!\left[H(t)\!-\!i\partial_t\!-\!a_n(t)\right]|K_n\rangle\!-\!b_n(t)|K_{n-1}\rangle,\nonumber\\
a_n(t)&=&\langle K_n|H(t)-i\partial_t|K_n\rangle ,
\label{eq:lanczos_td}
\end{eqnarray}
with $|K_0\rangle=|\Psi(0)\rangle$ as an initial condition,  $b_0=0$ and $b_{n+1}(t)>0$ fixed by
normalization. The
derivative $-i\partial_t$ compensates the motion of the basis, so the evolution
of $|\Psi(t)\rangle=\sum_n\psi_n(t)|K_n(t)\rangle$ reduces to a
nearest-neighbor hopping problem on a one-dimensional lattice,
\begin{eqnarray}
i\dot\psi_n(t)=a_n(t)\psi_n(t)+b_n(t)\psi_{n-1}(t)+b_{n+1}(t)\psi_{n+1}(t).
\label{eq:chain_krylov}
\end{eqnarray}
Dropping the time dependence
throughout, as the $-i\partial_t$ annihilates the static basis, leads to the
time-independent Krylov construction with static Lanczos coefficients and basis states \cite{Viswanath_Muller_1994_RecursionMethod,Parker_PRX_2019_UniversalOperatorGrowth,Balasubramanian_PRD_2022_ComplexitySpread,Nandy_PhysRep_2025_KrylovReview,Rabinovici_arXiv_2025_KrylovComplexity},
in which Krylov complexity has been developed so
far \cite{Barbon_JHEP_2019_OperatorComplexity,Rabinovici_JHEP_2021_EdgeKrylovSpace,Caputa_PRR_2022_GeometryKrylov,Hornedal_CommPhys_2022_UltimateSpeedLimits}.
We exploit this construction and the related algebraic structure of the time-dependent generator~\cite{Grabarits_arXiv_2026_TDKrylovLie}.

For an initial eigenstate $H(0)|\nu\rangle=E_\nu|\nu\rangle$,
the Krylov Liouvillian generating the dynamics is obtained from
Eq.~\eqref{eq:T_generator},
\begin{eqnarray}\label{eq:Heff_general}
H_{\nu,{\rm eff}}(t)
&=&
i\dot{\mathcal T}(t)\mathcal T^\dagger(t)
+
\dot s(t)E_\nu\\
&=&
A_0(t)X_0
+
A_+(t)X_+
+
A_-(t)X_-
+
\dot s(t)E_\nu ,\nonumber
\end{eqnarray}
with the initial eigenenergy appearing as a diagonal term that is the same
for all tower states.

Building on the general rank-one algebraic formulation in Sec.~\ref{sec:lie_primer}, we consider the corresponding tower states of the irreducible representations $\{|\mu,n\rangle\}$~\cite{Perelomov_CMP_1972_CoherentStates,Perelomov_1986_GeneralizedCoherentStates},
with $\mu$ labeling the irreducible representation and $n$ the Krylov
basis state along the tower,
\begin{eqnarray}
X_0|\mu,n\rangle
&=&
(x_\mu+n)|\mu,n\rangle,
\nonumber\\
X_+|\mu,n\rangle
&=&
\beta_{n+1,\mu}|\mu,n+1\rangle,
\nonumber\\
X_-|\mu,n\rangle
&=&
\beta_{n,\mu}|\mu,n-1\rangle,
\label{eq:tower_general}
\end{eqnarray}
with $\mu=j$, $x_j=-j$, and $n=0,\dots,2j$ for
$\mathfrak{su}(2)$, and $\mu=k$, $x_k=k$, and $n=0,1,2,\dots$
for $\mathfrak{su}(1,1)$.

Expanding the initial eigenstate in this basis,
\begin{eqnarray}
|\nu\rangle
&=&
\sum_{\mu,n}c_{n,\mu}^{(\nu)}|\mu,n\rangle,
\qquad
c_{n,\mu}^{(\nu)}=\langle\mu,n|\nu\rangle ,
\label{eq:nu_expansion_general}
\end{eqnarray}
the evolution decomposes into orthogonal irreducible representations, i.e., into independent Krylov chains. The corresponding amplitudes $\psi_{n,\mu}^{(\nu)}(t)=\langle\mu,n|\Psi_\nu(t)\rangle$ evolve according to the tridiagonal equation
\begin{eqnarray}\label{eq:chain_general}
i\dot\psi_{n,\mu}^{(\nu)}
&=&
\left[A_0(t)(x_\mu+n)+\dot s(t)E_\nu\right]\psi_{n,\mu}^{(\nu)}\\
&&+
A_+(t)\beta_{n,\mu}\,\psi_{n-1,\mu}^{(\nu)}
+
A_-(t)\beta_{n+1,\mu}\,\psi_{n+1,\mu}^{(\nu)},\nonumber
\end{eqnarray}
with initial condition $\psi_{n,\mu}^{(\nu)}(0)=c_{n,\mu}^{(\nu)}$.
Thus, from Eq.~\eqref{eq:chain_krylov} it follows that the tower states provide the Krylov basis with the diagonal Lanczos coefficients identical to
$A_0(t)(x_\mu+n)+\dot s(t)E_\nu$,  and the off-diagonal ones to
$A_\pm(t)\beta_{n,\mu}$. Overall, the algebraic decomposition alone contains all analytical information on the time-dependent Krylov dynamics of quantum states.

\subsection{Density-matrix Krylov dynamics}
\label{sec:density_general}
The construction extends directly to mixed states, characterized by a
Hermitian density matrix $\rho\geq0$ with unit normalization,
$\mathrm{Tr}[\rho]=1$. We consider its vectorized version $|\rho)$ in
Liouville space, with Hilbert-Schmidt product
$(A|B)=\mathrm{Tr}(A^\dagger B)$, by which operators become vectors and the
algebraic construction in Sec.~\ref{sec:krylov_chains} can be applied analogously. The time evolution is governed by the von Neumann equation
$i\partial_t\rho=[H(t),\rho]\equiv\mathcal L_H(t)\rho$, with the Liouvillian
superoperator acting as $\mathcal L_H(t)(\cdot)=[H(t),(\cdot)]$. In the exact time-dependent formulation~\cite{Takahashi_PRL_2025_TimeDependentGenerators}, the Krylov basis $\{\hat{\mathcal O}_k(t)\}$ is obtained by using the Lanczos algorithm, Eq.~\eqref{eq:lanczos_td}, with
$\mathcal L_H(t)-i\partial_t$,
\begin{eqnarray}
b_{k+1}(t)\hat{\mathcal O}_{k+1}(t)
\!=\!\big(\mathcal L_H(t)-i\partial_t\big)\hat{\mathcal O}_k(t)
-b_k(t)\hat{\mathcal O}_{k-1}(t),\,\,\,
\label{eq:density_lanczos_recurrence}
\end{eqnarray}
with $b_k(t)=(\hat{\mathcal O}_{k-1}(t)|(\mathcal L_H(t)-i\partial_t)
\hat{\mathcal O}_k(t))$. As a result, the density-matrix evolution is identical to a one-dimensional, nearest-neighbor hopping problem along the Krylov chain, with the
Krylov complexity given by the mean position along this
chain~\cite{Parker_PRX_2019_UniversalOperatorGrowth}.
For scale- and rotation-invariant many-body systems, the
Liouville-space dynamics is generated by the adjoint
representation of the corresponding generators,
\begin{eqnarray}
\mathcal X_a=X_a\otimes\mathbb I-\mathbb I\otimes X_a^{T},
\qquad a\in\{0,+,-\},
\label{eq:adjoint_generators_general}
\end{eqnarray}
which obey the commutation relations of Eq.~\eqref{eq:CW_general} for
$\sigma=\pm1$ and act on the doubled Hilbert space
$\mathcal H\otimes\mathcal H^\ast$. The full Liouville space then decomposes into the corresponding irreducible representations, labeled by the eigenvalues of the adjoint Casimir operator $\mathcal C=\mathcal X_0^2+\sigma(\mathcal
X_+\mathcal X_-+\mathcal X_-\mathcal X_+)/2$, which commutes with all adjoint generators~\cite{Gilmore_1974_LieGroups},
\begin{eqnarray}
\mathcal X_0|\Lambda,M))&=&M\,|\Lambda,M)),
\nonumber\\
\mathcal X_\pm|\Lambda,M))&=&\beta^{\Lambda}_{\pm,M}\,|\Lambda,M\pm1)),
\label{eq:adjoint_tower_general}
\end{eqnarray}
with $\Lambda$ fixing the Casimir eigenvalue and $M$ the position along
the given tower. For $\sigma=+1$ the towers are finite, $\Lambda=L\le2j$ with
$|M|\le L$, while for $\sigma=-1$ the eigenvalues of $\mathcal X_0$ can also take negative values and thus one cannot use the discrete Bargmann towers as irreducible representations. Instead, $\Lambda=\lambda$ runs over the continuum $\lambda\in[0,\infty)$ with $M\in\mathbb Z$, corresponding to the principal-series representation of $\mathfrak{su}(1,1)$. Although no lowest-weight state exists in the principal-series representation~\cite{Bargmann_AnnMath_1947_IrreducibleUnitary,
Perelomov_1986_GeneralizedCoherentStates}, the Krylov dynamics of the
density matrix can still be characterized exactly, as will be shown below.

As for quantum states, Eq.~\eqref{eq:Psi_exact_general}, the time evolution factorizes,
\begin{eqnarray}
|\rho(t))=\mathcal T_{\mathcal L}(t)\,e^{-is(t)\mathcal L_0}|\rho(0)),
\quad
\mathcal T_{\mathcal L}=\mathcal T\otimes\mathcal T^\ast,
\label{eq:rho_evolution_general}
\end{eqnarray}
 for an initial eigenoperator of the Liouvillian $\mathcal L_0[\rho_\nu(0)]=L_\nu\rho_\nu(0)$ with $\mathcal L_0[\,\cdot\,]=[H(0),\,\cdot\,]$. Here, $\mathcal T_{\mathcal L}(t)$ acts in Liouville space as $\mathcal
T_\mathcal{L}(t)|\rho)=|\mathcal T(t)\rho\,\mathcal T^\dagger(t))$, and its
generator is the same linear combination of the adjoint generators, $\{\mathcal X_0,\mathcal X_\pm\}$,
\begin{eqnarray}\label{eq:TL_generator}
i\dot{\mathcal T}_{\mathcal L}(t)\,\mathcal T_{\mathcal L}^\dagger(t)
=A_0(t)\mathcal X_0+A_+(t)\mathcal X_++A_-(t)\mathcal X_-,
\end{eqnarray}
with the same driving coefficients as for quantum states, Eq.~\eqref{eq:T_generator}.
The time evolution thus follows a tridiagonal structure. The Liouville-space tower states provide the Krylov basis, and the amplitudes $\rho_{M,\Lambda}(t)=((\Lambda,M|\rho(t))$ evolve with the Lanczos coefficients given by the generators of Eq.~\eqref{eq:TL_generator},
\begin{eqnarray}
i\dot\rho_{M,\Lambda}
&=&\big[A_0 M+\dot s L_\nu\big]\rho_{M,\Lambda}
+A_+\beta^{\Lambda}_{+,M-1}\rho_{M-1,\Lambda}\nonumber\\
&&+A_-\beta^{\Lambda}_{-,M+1}\rho_{M+1,\Lambda},
\label{eq:density_chain_general}
\end{eqnarray}
where $\dot s L_\nu$ contributes only as a phase~\cite{Grabarits_arXiv_2026_TDKrylovLie}.
We focus on thermal density matrices
of Hamiltonians diagonal in the $\mathfrak{su}(1,1)$ or $\mathfrak{su}(2)$ tower states, $\{|\mu,n\rangle\}$ as common in experiments at finite temperature. Such a thermal state commutes with $X_0$,  $\mathcal X_0|\rho^{(\mu)}_\beta(0))=0$, and thus initially occupies only the
$M=0$ states of the Liouville-space towers,
$\rho^{(\beta)}_{M,\Lambda}(0)=\delta_{M,0}\,\rho^{(\beta)}_{0,\Lambda}(0)$, both
for the principal series of $\mathfrak{su}(1,1)$ and for the
$\mathfrak{su}(2)$ spin towers (see Apps.~\ref{app:density_CG} and~\ref{app:su2_density_CG} for details),
\begin{eqnarray}
    |\rho_\beta^{(\mu)}(0))&=&Z^{-1}\sum_ne^{-\beta E_{\mu,n}}|\mu,n\rangle\langle\mu,n|\nonumber\\
    &=&\sum_{\Lambda}\rho^{(\beta)}_{0,\Lambda}(0)|\Lambda,0)),\,Z=\sum_{n\geq0} e^{-\beta E_{\mu,n}}.
\end{eqnarray}
As the dynamics starts at the center $M=0$ of each
chain and the ladder coefficients are symmetric under
$M\to-M$, the occupations of the Krylov states remain symmetric as well at all times. The mean position
$\sum_{\Lambda,M}M|\rho^{(\beta)}_{M,\Lambda}|^2$ therefore
vanishes, while the Krylov complexity of the density matrix,
given by the mean distance $|M|$ from the initial site, does
not admit a compact form
(App.~\ref{app:density_principal_complexity}). For this reason, the spreading
is captured instead by the Krylov variance, which coincides
with the second moment of the Krylov distribution,
\begin{eqnarray}
M_2[\rho](t)=\frac{1}{\mathrm{Tr}\,\rho^2}
\sum_{\Lambda,M}M^{2}\,\big|\rho_{M,\Lambda}(t)\big|^{2}.
\label{eq:CK_density_sum}
\end{eqnarray}
In this expression, $\sum_\Lambda$ runs over the finite spin towers for $\sigma=+1$ and is understood as the integral over the continuous principal series, $\int_0^\infty d\lambda$ for $\sigma=-1$.
For a thermal state starting from different towers, $M_2$ is the average of the
single-tower results weighted by $p_\mu^2\,{\rm Tr}[\rho_\beta^{(\mu)}(0)]^2$.
The only input from the underlying many-body system is the set of initial amplitudes $\rho^{(\beta)}_{0,\Lambda}(0)$ (details are shown in App.~\ref{app:su2_density_CG} for the $\mathfrak{su}(2)$ case and in App.~\ref{app:density_CG} for the $\mathfrak{su}(1,1)$ case).

Table~\ref{tab:correspondence} summarizes the corresponding quantities
in the scale- and rotation-invariant realizations, which are
illustrated in Fig.~\ref{fig:schematic} and developed in
Secs.~\ref{sec:scale_covariant} and~\ref{sec:rot_invariant}.
\begin{table}[t]
\begin{ruledtabular}
\begin{tabular}{lcc}
 & Scale & Rotation \\
\colrule
$\mathcal T(t)$ & $U_K S_\gamma$ & $R$ \\
$s(t)$ & $\tau(t)$ & $t$ \\
Algebra & $\mathfrak{su}(1,1)$ & $\mathfrak{su}(2)$ \\
Tower label & $k$ & $j$ \\
Coordinate & $n\ge 0$ & $q=0,\ldots,2j$ \\
Chain length & $\infty$ & $2j+1$ \\
Protocol & $\gamma,\dot\gamma$ & $\theta,\phi$ \\
Coherence & $B$ & $M_+$ \\
$C_K$ (single tower) & Eq.~\eqref{eq:Ermakov_CK_single_main}
                      & Eq.~\eqref{eq:K_su2_single_main} \\
$M_2$ (density matrix) & $\propto\sinh^2[2\xi]$
                        & $\propto\sin^2\theta$ \\
Readout & cloud width & spin angle \\
\end{tabular}
\end{ruledtabular}
\caption{Scale- and rotation-invariant realizations of the construction.}
\label{tab:correspondence}
\end{table}

\definecolor{scol}{RGB}{0,120,170}
\definecolor{rcol}{RGB}{170,70,40}
\definecolor{ncol}{RGB}{70,70,70}
\definecolor{icol}{RGB}{210,150,40}

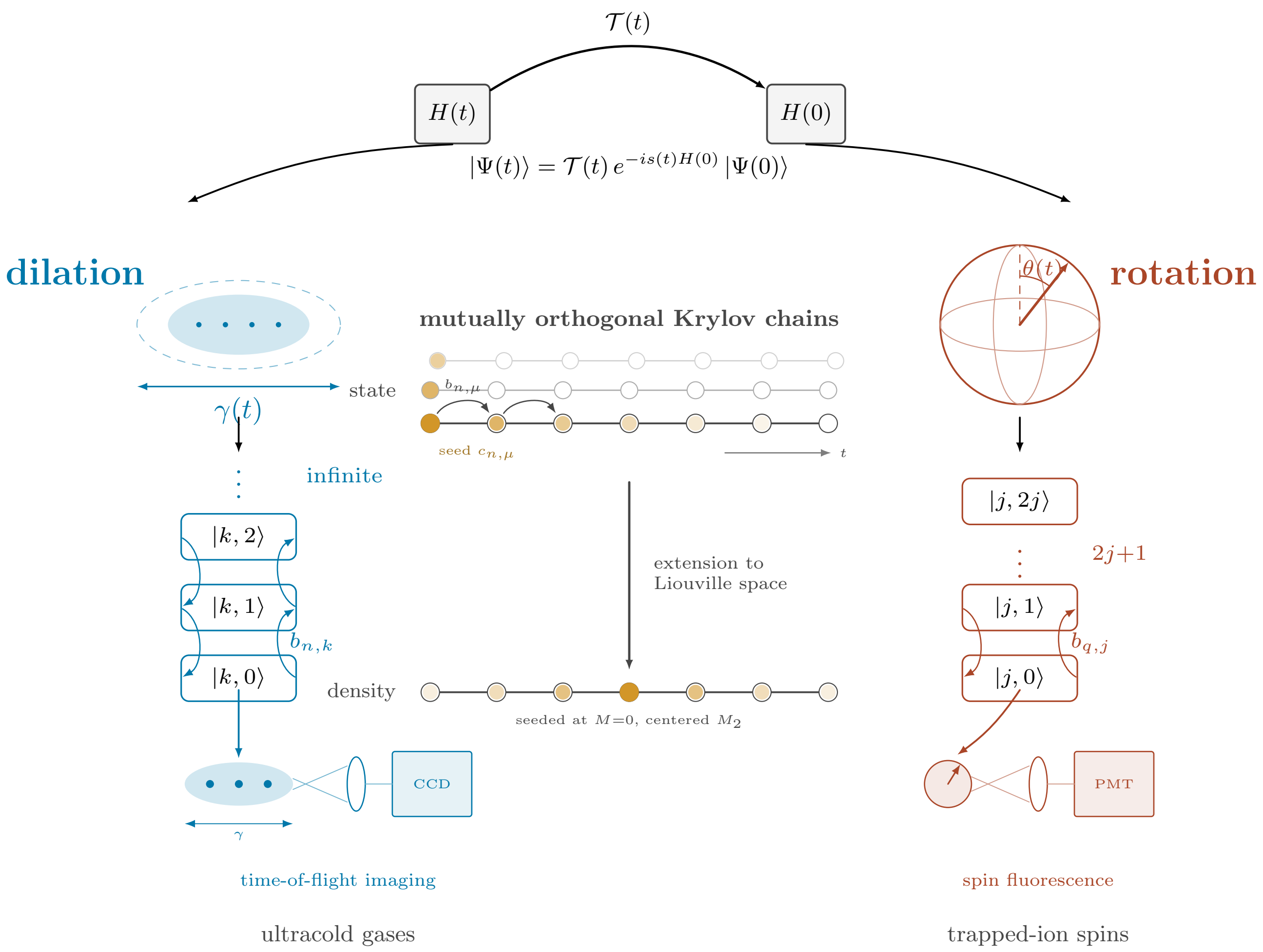
\begin{figure*}[t]
\centering
\begin{tikzpicture}[
  >={Latex[length=1.4mm]},
  font=\small, line cap=round, line join=round,
  every node/.style={align=center},
  hbox/.style={rounded corners=2pt, draw=ncol, line width=0.7pt,
               inner sep=5pt, minimum height=8mm, fill=ncol!6},
  site/.style={rounded corners=3pt, draw, line width=0.6pt, fill=white,
               minimum width=13mm, minimum height=5.2mm, font=\scriptsize},
  hop/.style={->, line width=0.5pt, shorten >=1pt, shorten <=1pt},
]

\node[hbox] (Ht) at (-2.4,7.2) {$H(t)$};
\node[hbox] (H0) at ( 2.4,7.2) {$H(0)$};
\draw[->, line width=0.9pt] (Ht) to[bend left=32]
      node[above,yshift=1pt]{$\mathcal T(t)$} (H0);
\coordinate (mid) at ($(Ht)!0.5!(H0)$);
\node[below=4mm of mid, anchor=north]
      {$|\Psi(t)\rangle=\mathcal T(t)\,e^{-i s(t)H(0)}\,|\Psi(0)\rangle$};
\draw[->, line width=0.7pt] (Ht.south) to[bend right=10] (-6.0,6.0);
\draw[->, line width=0.7pt] (H0.south) to[bend left=10]  ( 6.0,6.0);

\begin{scope}[shift={(-5.3,2.0)},scale=1.20,every node/.style={transform shape,align=center}]
\node[scol] at (-1.85,2.55) {\bfseries {\large dilation}};
\draw[dashed, scol!50] (0.0,1.95) ellipse (1.15 and 0.50);
\fill[scol!18]         (0.0,1.95) ellipse (0.80 and 0.34);
\foreach \x in {-0.45,-0.15,0.15,0.45}{\fill[scol] (\x,1.95) circle (0.03);}
\draw[<->, scol, line width=0.6pt] (-1.15,1.25) -- (1.15,1.25)
      node[midway,below]{$\gamma(t)$};
\draw[->, line width=0.7pt] (0.0,0.90) -- (0.0,0.50);
\node[scol] at (0.0,0.25) {$\vdots$};
\node[scol, anchor=west] at (0.65,0.25) {\scriptsize infinite};
\node[site, draw=scol] (L2) at (0.0,-0.45) {$|k,2\rangle$};
\node[site, draw=scol] (L1) at (0.0,-1.25) {$|k,1\rangle$};
\node[site, draw=scol] (L0) at (0.0,-2.05) {$|k,0\rangle$};
\draw[hop, scol] (L0.east) to[bend left=60]
      node[right]{\scriptsize$b_{n,k}$} (L1.east);
\draw[hop, scol] (L1.east) to[bend left=60] (L2.east);
\draw[hop, scol] (L1.west) to[bend left=60] (L0.west);
\draw[hop, scol] (L2.west) to[bend left=60] (L1.west);
\end{scope}

\begin{scope}[shift={(5.3,2.0)},scale=1.20,every node/.style={transform shape,align=center}]
\node[rcol] at (1.85,2.55) {\bfseries {\large rotation}};
\draw[rcol, line width=0.7pt] (0.0,1.95) circle (0.9);
\draw[rcol!55] (0.0,1.95) ellipse (0.90 and 0.30);
\draw[rcol!55] (0.0,1.95) ellipse (0.30 and 0.90);
\draw[rcol, dashed, line width=0.4pt] (0.0,1.95) -- (0.0,2.85);
\draw[->, rcol, line width=1.0pt] (0.0,1.95) -- ++(52:0.9);
\draw[rcol, line width=0.5pt] (0.0,2.50) arc (90:52:0.55);
\node[rcol] at (0.26,2.57) {\scriptsize$\theta(t)$};
\draw[->, line width=0.7pt] (0.0,0.90) -- (0.0,0.50);
\node[site, draw=rcol] (Rt) at (0.0,-0.05) {$|j,2j\rangle$};
\node[rcol] at (0.0,-0.65) {$\vdots$};
\node[rcol, anchor=west] at (0.70,-0.65) {\scriptsize $2j{+}1$};
\node[site, draw=rcol] (R1) at (0.0,-1.25) {$|j,1\rangle$};
\node[site, draw=rcol] (R0) at (0.0,-2.05) {$|j,0\rangle$};
\draw[hop, rcol] (R0.east) to[bend left=60]
      node[right]{\scriptsize$b_{q,j}$} (R1.east);
\draw[hop, rcol] (R1.west) to[bend left=60] (R0.west);
\end{scope}

\node[ncol, font=\small\bfseries] at (0,4.40) {mutually orthogonal Krylov chains};
\node[ncol, font=\footnotesize, anchor=east] at (-3.05,3.45) {state};
\draw[ncol!25,line width=0.5pt] (-2.60,3.85) -- (2.80,3.85);
\foreach \i in {0,...,6}{\node[circle,draw=ncol!25,fill=white,
   minimum size=2.2mm,inner sep=0pt] at (-2.60+\i*0.90,3.85){};}
\fill[icol!45] (-2.60,3.85) circle (1.0mm);
\draw[ncol!45,line width=0.5pt] (-2.70,3.45) -- (2.70,3.45);
\foreach \i in {0,...,6}{\node[circle,draw=ncol!45,fill=white,
   minimum size=2.3mm,inner sep=0pt] at (-2.70+\i*0.90,3.45){};}
\fill[icol!70] (-2.70,3.45) circle (1.1mm);
\draw[ncol,line width=0.7pt] (-2.70,3.00) -- (2.70,3.00);
\foreach \i in {0,...,6}{\node[circle,draw=ncol,fill=white,
   minimum size=2.5mm,inner sep=0pt] (c\i) at (-2.70+\i*0.90,3.00){};}
\fill[icol] (-2.70,3.00) circle (1.3mm);
\foreach \i/\o in {1/70,2/50,3/34,4/20,5/10}
   {\fill[icol!\o] (-2.70+\i*0.90,3.00) circle (1.0mm);}
\draw[hop,ncol] (c0) to[bend left=55] node[above,font=\tiny]{$b_{n,\mu}$} (c1);
\draw[hop,ncol] (c1) to[bend left=55] (c2);
\node[icol!75!black,font=\tiny,anchor=west] at (-2.70,2.60) {seed $c_{n,\mu}$};
\draw[->,ncol!70,line width=0.5pt] (1.30,2.60) -- (2.74,2.60)
   node[right,font=\tiny,ncol]{$t$};
\draw[->,ncol,line width=0.9pt] (0,2.20) -- (0,-0.35);
\node[ncol,font=\scriptsize,anchor=west,align=left] at (0.22,0.95)
   {extension to\\Liouville space};
\node[ncol, font=\footnotesize, anchor=east] at (-3.05,-0.65) {density};
\draw[ncol,line width=0.7pt] (-2.70,-0.65) -- (2.70,-0.65);
\foreach \i in {0,...,6}{\node[circle,draw=ncol,fill=white,
   minimum size=2.5mm,inner sep=0pt] at (-2.70+\i*0.90,-0.65){};}
\fill[icol] (0,-0.65) circle (1.3mm);
\foreach \k/\o in {1/58,2/32,3/15}
   {\fill[icol!\o] (0.90*\k,-0.65) circle (1.0mm);
    \fill[icol!\o] (-0.90*\k,-0.65) circle (1.0mm);}
\node[ncol,font=\tiny] at (0,-1.05) {seeded at $M{=}0$,\ centered $M_2$};

\draw[->,scol,line width=0.7pt] (-5.3,-0.62) -- (-5.3,-1.55);
\begin{scope}[shift={(-3.95,-2.19)},scale=2.45]
\fill[scol!18] (-0.55,0.12) ellipse (0.30 and 0.12);
\foreach \x in {-0.16,0,0.16}{\fill[scol] (-0.55+\x,0.12) circle (0.022);}
\draw[<->,scol,line width=0.35pt] (-0.85,-0.10) -- (-0.25,-0.10)
   node[midway,below,font=\tiny]{$\gamma$};
\draw[scol!55,line width=0.35pt] (-0.25,0.15) -- (0.05,0.02);
\draw[scol!55,line width=0.35pt] (-0.25,0.09) -- (0.05,0.22);
\draw[scol,line width=0.5pt] (0.10,0.12) ellipse (0.05 and 0.15);
\draw[scol,fill=scol!10,line width=0.5pt,rounded corners=1pt]
   (0.30,-0.06) rectangle (0.74,0.30);
\node[scol,font=\tiny] at (0.52,0.12) {CCD};
\draw[scol!55,line width=0.35pt] (0.15,0.12) -- (0.30,0.12);
\node[scol,font=\scriptsize] at (0.0,-0.42) {time-of-flight imaging};
\node[ncol,font=\small] at (0.0,-0.72) {ultracold gases};
\end{scope}

\draw[->,rcol,line width=0.7pt] (5.3,-0.62) to[bend left=10] (4.45,-1.50);
\begin{scope}[shift={(5.55,-2.19)},scale=2.45]
\draw[rcol,fill=rcol!12,line width=0.5pt] (-0.50,0.12) circle (0.13);
\draw[->,rcol,line width=0.6pt] (-0.50,0.12) -- ++(58:0.13);
\draw[rcol!55,line width=0.35pt] (-0.37,0.16) -- (-0.05,0.02);
\draw[rcol!55,line width=0.35pt] (-0.37,0.08) -- (-0.05,0.22);
\draw[rcol,line width=0.5pt] (0.0,0.12) ellipse (0.05 and 0.15);
\draw[rcol,fill=rcol!10,line width=0.5pt,rounded corners=1pt]
   (0.20,-0.06) rectangle (0.64,0.30);
\node[rcol,font=\tiny] at (0.42,0.12) {PMT};
\draw[rcol!55,line width=0.35pt] (0.05,0.12) -- (0.20,0.12);
\node[rcol,font=\scriptsize] at (0.0,-0.42) {spin fluorescence};
\node[ncol,font=\small] at (0.0,-0.72) {trapped-ion spins};
\end{scope}

\end{tikzpicture}
\caption{Unified construction in its two realizations. A protocol
transformation $\mathcal T(t)$ relates the driven Hamiltonian $H(t)$ to the
initial Hamiltonian $H(0)$, after which the evolution splits into mutually
orthogonal Krylov chains. Left: scale-invariant systems, where $\mathcal T(t)$
is a dilation, the chains are infinite Bargmann towers of $\mathfrak{su}(1,1)$,
and the complexity is set by the scaling factor $\gamma(t)$ of an experimentally realizable cold atomic cloud read from imaging.
Right: rotation-invariant systems, where $\mathcal T(t)$ is a global rotation,
the chains are finite $\mathfrak{su}(2)$ towers of length $2j+1$, and the
complexity is set by the collective-spin rotation angle $\theta(t)$. Middle: Liouville-space construction for initial thermal states, with the dynamics initiated at the middle of the Krylov chains.}
\label{fig:schematic}
\end{figure*}

\section{Krylov dynamics in general scale-invariant many-body systems}
\label{sec:scale_covariant}

In isotropic scale-invariant processes, every measurable quantity,
such as a local correlation function or the density profile,
coincides with its initial value up to a global rescaling of
lengths. All length scales of the system undergo the same
self-similar dilation ~\cite{CastinDum_PRL_1996_BECScaling,
Kagan_PRA_1996_BECExpansion,
Pitaevskii_Rosch_PRA_1997_Breathing,
Chen_PRL_2010_FastCooling,
delCampo_PRL_2013_CounterdiabaticDriving,
Deffner_PRX_2014_ScaleInvariantSTA}.
As a result, the dynamics can be characterized by a single time-dependent
scaling factor $\gamma(t)$, which, together with the form of the Hamiltonian,
determines the coefficients of the generator of $\mathcal T(t)$ in
Eq.~\eqref{eq:T_generator}. The initial Hamiltonian $H(0)$ itself only determines the initial conditions of the dynamics through its eigenstates $H(0)|\nu\rangle=E_\nu|\nu\rangle$. The dynamics then decomposes onto mutually orthogonal Krylov lattices, yielding closed analytical expressions for the dynamics and Krylov complexity in terms of $\gamma(t)$ and the initial conditions, which are directly testable experimentally.

Below, we present the general formulation and demonstrate
it with the Lieb-Liniger gas, whose Hamiltonian is not a
linear combination of the $\mathfrak{su}(1,1)$
generators~\cite{Lieb_Liniger_PR_1963_ExactAnalysis,
Olshanii_PRL_1998_AtomicScattering, Girardeau_OptComm_2004_TonksScaling}. We then
focus on the subclass in which $H(0)$ can itself be expressed as a linear combination of the $\mathfrak{su}(1,1)$ generators and thus its eigenstates coincide with the corresponding tower states. As an illustrative example, we consider the
Moshinsky model with pairwise quadratic potential~\cite{Sutherland_PRL_1998_CalogeroScaling, Brink1992Explicit,
UjinoWadati1994Hidden, GurappaPanigrahi1999Equivalence,
Moshinsky_JMP_1968_HarmonicInteraction}. Throughout these examples, the driving and the
scaling factor $\gamma(t)$ are set
by a time-dependent harmonic trap or a time-dependent pairwise coupling. We also show how the same algebraic construction applies to thermal density matrices in Liouville space, capturing their operator growth as a function of the initial temperature and the scaling factor.

\subsection{General formulation for arbitrary initial states}

Consider $N$ particles ($\hbar=m=1$) in a harmonic trap of
time-dependent frequency $\Omega(t)$, subject to a
scale-invariant on-site potential and pairwise interactions,
\begin{eqnarray}\label{eq: H_gen_t}
H(t)
=
\sum_{i=1}^N
\left[
\frac{\mathbf p_i^2}{2}
+
\frac12\Omega^2(t)\mathbf q_i^2
+
U(\mathbf q_i,t)
\right]
\!+\!
\epsilon(t)V_\kappa,\,\,
\end{eqnarray}
with the scaling properties encoded as $V_\kappa=
\sum_{i<j}
V(\mathbf q_i-\mathbf q_j),\quad V(\gamma\mathbf q)=\gamma^{-\kappa}V(\mathbf q)$ for any $\gamma>0$ and $U(\mathbf q,t)
=
\gamma^{-2}(t)
U\!\left(
\frac{\mathbf q}{\gamma(t)},0\right)$.
Scale invariance is ensured by the consistency equations~\cite{delCampo_PRL_2013_CounterdiabaticDriving,Deffner_PRX_2014_ScaleInvariantSTA}
\begin{eqnarray}\label{eq: consistency_gen}
\epsilon(t)=\gamma^{\kappa-2}(t),\qquad
\Omega^2(t)
=
\frac{\Omega^2_0}{\gamma^4(t)}
-\frac{\ddot\gamma(t)}{\gamma(t)},
\end{eqnarray}
with $\Omega_0\equiv\Omega(0)$, $\gamma(0)=1$, $\dot\gamma(0)=0$, which implies
$\ddot\gamma(0)=0$.

The general construction of Eq.~\eqref{eq:Psi_exact_general} applies with
$\mathcal T(t)=U_K(t)S_\gamma(t)$, where $K=\tfrac12\sum_i\mathbf q_i^2$ is
the quadratic potential, $D=\tfrac12\sum_i\left(\mathbf q_i\!\cdot\!\mathbf
p_i+\mathbf p_i\!\cdot\!\mathbf q_i\right)$ is the generator of dilations, and
\begin{eqnarray}
S_\gamma=e^{-i(\ln\gamma)D},\quad
U_K=e^{i(\dot\gamma/\gamma)K} .
\end{eqnarray}

For illustration, in the coordinate representation with $\mathbf x=(\mathbf q_1,\dots,\mathbf q_N)$ and
$\mathbf q_i\in\mathbb R^d$, the time-evolved wave function of an initial eigenstate $|\nu\rangle$, satisfying $H(0)|\nu\rangle=E_\nu|\nu\rangle$, follows by rescaling the coordinates by $\gamma(t)$ and taking into account a dynamical phase of $\tau(t)E_\nu$~\cite{CastinDum_PRL_1996_BECScaling,
Kagan_PRA_1996_BECExpansion,Deffner_PRX_2014_ScaleInvariantSTA},
\begin{eqnarray}\label{eq:scaling_ansatz}
\Psi_\nu(\mathbf x,t)
&=&
\frac{\exp\!\left[\frac{i\dot\gamma(t)}{2\gamma(t)}\sum_i\mathbf q_i^2-iE_\nu\tau(t)\right]}{\gamma^{Nd/2}(t)}
\Psi_\nu\!\left(\frac{\mathbf x}{\gamma(t)},0\right),\nonumber\\
\tau(t)&=&\int_0^t\frac{dt'}{\gamma^2(t')}.
\end{eqnarray}
In terms of the reduced wave function of Eq.~\eqref{eq: Int_picture}, $\chi_\nu(\mathbf x,\tau)=\langle\mathbf x|\chi_\nu(\tau)\rangle=e^{-iE_\nu\tau}\Psi_\nu(\mathbf x,0)$, this reads $\Psi_\nu(\mathbf x,t)=\gamma^{-Nd/2}(t)\,e^{i\dot\gamma(t)\sum_i\mathbf q_i^2/[2\gamma(t)]}\,\chi_\nu(\mathbf x/\gamma(t),\tau(t))$.
The dilation $S_\gamma$ rescales the configuration and supplies the
normalization factor $\gamma^{-Nd/2}$, and $U_K$ contributes the quadratic phase
set by the expansion rate, while the rescaled time $\tau(t)$ is the
protocol-dependent function $s(t)$ of Eq.~\eqref{eq:chi_general}.

For an initial eigenstate, $H(0)|\nu\rangle=E_\nu|\nu\rangle$, the Krylov Liouvillian in
Eq.~\eqref{eq:Heff_general} becomes~\cite{DodonovManko1989, ADC_Entropy_2020},
\begin{eqnarray}
H_{\nu,\mathrm{eff}}(t)=\frac{\dot\gamma}{\gamma}D-\left(\frac{\ddot\gamma}{\gamma}+\frac{\dot\gamma^2}{\gamma^2}\right)K
+\frac{E_\nu}{\gamma^2}
\label{eq: H_eff_su11}
\end{eqnarray}
 with $T=\sum_i\mathbf p_i^2/2$, closing the $\mathfrak{su}(1,1)$ algebra together with $K$ and $D$,
\begin{eqnarray}\label{eq:TKD_main}
[D,K]=-2iK,\,[D,T]=2iT,\,[T,K]=-iD.
\end{eqnarray}
In the Cartan-Weyl basis of the general construction, Eq.~\eqref{eq:CW_general}, they become
\begin{eqnarray}\label{eq:su11_CW_map}
K_0=\tfrac12\!\left(\Omega_0K+\frac{T}{\Omega_0}\right),
\,
K_\pm=\tfrac12\!\left(\Omega_0K-\frac{T}{\Omega_0}\right)\mp\frac{i}{2}D,\,\,\,
\end{eqnarray}
with the Cartan element acting as $K_0|k,n\rangle=(k+n)|k,n\rangle$ on the Bargmann towers
of Eq.~\eqref{eq:tower_general}. Thus, the general driving coefficients $A_0(t),A_\pm(t)$ follow from
Eq.~\eqref{eq: H_eff_su11} and Eq.~\eqref{eq:su11_CW_map}
(see App.~\ref{app:wei_norman} for more details). Given the initial overlaps
$c_{n,k}=\langle k,n|\nu\rangle$ of Eq.~\eqref{eq:nu_expansion_general}, the time evolution of the
amplitudes $\psi_{n,k}(t)=\langle k,n|\Psi_\nu(t)\rangle$ is governed by the $\mathfrak{su}(1,1)$ version of the general form, Eq.~\eqref{eq:chain_general},
\begin{eqnarray}
i\dot\psi_{n,k}
&=&\left[A_0(t)(k+n)+\frac{E_\nu}{\gamma^2}\right]\psi_{n,k}\nonumber\\
&&
+A_+(t)\sqrt{n(2k+n-1)}\,\psi_{n-1,k}
\nonumber\\
&&+A_-(t)\sqrt{(n+1)(2k+n)}\,\psi_{n+1,k} ,
\label{eq:su11_chain_main}
\end{eqnarray}
with $\psi_{n,k}(0)=c_{n,k}$. As shown in App.~\ref{app:wei_norman} using the Wei-Norman factorization~\cite{WeiNorman1963, WeiNorman1964, CharzynskiKus2013WeiNorman,
Ban_JOSAB_1993_SU11SU2, Grabarits_arXiv_2026_TDKrylovLie}, the time-evolved amplitudes can be expressed as
$\psi_{m,k}(t)
=
e^{-i\tau(t)E_\nu}\sum_{n\ge0}\mathcal D_{mn}^{(k)}\,c_{n,k}$,
with $\mathcal D_{mn}^{(k)}=\langle k,m|U_K S_\gamma|k,n\rangle$, while the spread complexity admits a compact form,
\begin{eqnarray}\label{eq:Ermakov_CK_superposition_main}
&&C_K(t)
=
\frac12
\left[
\gamma^2+\frac{1}{\gamma^2}
+
\left(\frac{\dot\gamma}{\Omega_0}\right)^2
\right]A-\bar k\\
&&\quad+
\frac12
\left[
\gamma^2-\frac{1}{\gamma^2}
+
\left(\frac{\dot\gamma}{\Omega_0}\right)^2
\right]{\rm Re}\,B
-\frac{\dot\gamma}{\Omega_0\gamma}{\rm Im}\,B,\nonumber
\end{eqnarray}
with the auxiliary terms given by
\begin{eqnarray}\label{eq:Ermakov_AB_main}
A&=&\sum_{n,k}|c_{n,k}|^2(k+n),
\qquad
\bar k=\sum_{n,k}|c_{n,k}|^2k,\\
B&=&
\sum_{k,n\ge0}
\sqrt{(n+1)(2k+n)}\,c^*_{n+1,k}c_{n,k},
\nonumber
\end{eqnarray}
which depend only on the initial amplitudes. For an
initial superposition of eigenstates of $H(0)=2\Omega_0K_0$,
the free evolution multiplies $B$ by the phase
$e^{2i\Omega_0\tau(t)}$
(App.~\ref{app:su11_complexity_derivation}).
Note that, due to orthogonality, contributions from different towers are additive, while initial superpositions inside the towers appear as nearest-neighbor coherences via $B$.
For a single tower state $|k,n\rangle$, $B$ vanishes,
leading to the simplified expression
\begin{eqnarray}
C_K^{(k,n)}(t)
=
n+\frac{k+n}{2}
\left[
\left(\gamma-\frac{1}{\gamma}\right)^2
+
\left(\frac{\dot\gamma}{\Omega_0}\right)^2
\right].
\label{eq:Ermakov_CK_single_main}
\end{eqnarray}
As a limiting case, we consider a single particle in a one-dimensional harmonic trap with varying frequency and with $U(\mathbf q,t)=0$, initiated in
the ground state, for which $k=\tfrac14$ and $n=0$. From the general construction, the spread complexity is then
$C_K=\tfrac18[(\gamma-\gamma^{-1})^2+(\dot\gamma/\Omega_0)^2]$. While this result might seem simple, as we will demonstrate below, the key ingredients and the structure of the analytical results remain the same for the introduced broad class of scale-invariant many-body systems.

Finally, we show how the general Liouville-space Krylov construction of Sec.~\ref{sec:density_general} can be applied to an initial thermal density matrix. As different Bargmann indices result in independent Krylov dynamics and contribute independently to the Krylov variance, we consider initial thermal states diagonal in a single tower,
\begin{eqnarray}
\rho_\beta^{(k)}(0)
\!=\!
Z^{-1}\sum_{n} e^{-\beta E_{k,n}}\,|k,n\rangle\langle k,n| ,
\,
Z\!=\!\sum_{n} e^{-\beta E_{k,n}}.\,
\label{eq:thermal_state_general}
\end{eqnarray}
For $\mathfrak{su}(1,1)$, the adjoint generators of
Eq.~\eqref{eq:adjoint_generators_general} are $\mathcal X_a=K_a\otimes\mathbb
I-\mathbb I\otimes K_a^{T}$, $a\in\{0,+,-\}$. They obey the
$\mathfrak{su}(1,1)$ commutation relations and act on the principal series as
\begin{eqnarray}
[\mathcal X_0,\mathcal X_\pm]&\!=\!&\pm\mathcal X_\pm,\quad
[\mathcal X_+,\mathcal X_-]=-2\mathcal X_0,
\nonumber\\
\mathcal X_0|\lambda,M))\!&=&\!M\,|\lambda,M)),\,\,
\mathcal X_\pm|\lambda,M))=\beta^{\lambda}_{\pm,M}\,|\lambda,M\pm1)),
\nonumber\\
\beta^{\lambda}_{\pm,M}\!&=&\!\sqrt{M(M\pm1)+\lambda^2+\tfrac14}\, .
\label{eq:su11_adjoint_principal}
\end{eqnarray}
The thermal state can be resolved in the principal series as
(see App.~\ref{app:density_CG} for details)
$|\rho_\beta^{(k)}(0))=\int_0^\infty d\lambda\,\rho^{(\beta)}_{0,\lambda}(0)\,
|\lambda,0))$ with $\rho^{(\beta)}_{0,\lambda}(0)=(\lambda,0|\rho_\beta^{(k)}(0))$.
In each Krylov chain, the dynamics starts from $M=0$, and the amplitudes
$\rho_{M,\lambda}(t)$ evolve according to the $\mathfrak{su}(1,1)$ form of
Eq.~\eqref{eq:density_chain_general},
\begin{eqnarray}
i\dot\rho_{M,\lambda}
&=&
A_0(t)\,M\,\rho_{M,\lambda}
+A_+(t)\,\beta^{\lambda}_{+,M-1}\,\rho_{M-1,\lambda}
\nonumber\\
&&
+A_-(t)\,\beta^{\lambda}_{-,M+1}\,\rho_{M+1,\lambda} ,
\label{eq:su11_density_chain_main}
\end{eqnarray}
with the same driving coefficients $A_0(t),A_\pm(t)$ as for quantum states, Eq.~\eqref{eq:app_su11_drive_coeffs}, governed by the scaling factor $\gamma(t)$.

As the evolution is symmetric with respect to $\pm M$, Eq.~\eqref{eq:su11_density_chain_main}, the mean displacement vanishes, and the spread is quantified by the Krylov variance of Eq.~\eqref{eq:CK_density_sum}. In particular, as shown in App.~\ref{app:density_principal_complexity}, it admits a compact form,
\begin{eqnarray}\label{eq:density_M2_general}
M_2[\rho_\beta^{(k)}](t)
&=&\frac{1}{\mathrm{Tr}\,\rho^2}\int_0^\infty\mathrm d\lambda\sum_{M\in\mathbb Z}M^{2}\,\big|\rho^{(\beta)}_{M,\lambda}(t)\big|^{2}\\
&=&
\tfrac12\big\langle\lambda^2+\tfrac14\big\rangle_\beta\,\sinh^2[2\xi(t)] ,\nonumber\\
\sinh^2\xi(t)
&=&
\frac14\!\left[
\Big(\gamma-\tfrac1\gamma\Big)^2
+\Big(\tfrac{\dot\gamma}{\Omega_0}\Big)^2
\right],\\
\big\langle\lambda^2+\tfrac14\big\rangle_\beta
&=&
\frac{\displaystyle\int_0^\infty d\lambda\,
\big(\lambda^2+\tfrac14\big)\,
\big|\rho_{0,\lambda}^{(\beta)}(0)\big|^2}
{\displaystyle\int_0^\infty d\lambda\,
\big|\rho_{0,\lambda}^{(\beta)}(0)\big|^2}.
\end{eqnarray}
Here, the squeezing parameter $\xi(t)$ is fixed by the
scaling factor $\gamma(t)$, while the initial temperature
appears only through the thermal average
$\langle\lambda^2+\tfrac14\rangle_\beta$. The time dependence
of the Krylov variance can therefore be measured directly,
for instance, by imaging the width of the time-dependent quantum gas cloud~\cite{Ketterle1999,Schaff2010,Schaff2011,elliott2014conformal,rohringer2015nonequilibrium,deng2016efimovian,Deng2018Efimov,Deng2018,Diao2018,SaintJalm_PRX_2019_DynamicalSymmetry}.

\subsection{Driven trapped Lieb-Liniger gas}
\label{sec:contact_nodal_example}

As a main demonstration, we consider the trapped
Lieb-Liniger gas, whose driving is scale invariant although
$H(0)$ is not a linear combination of the $\mathfrak{su}(1,1)$
generators~\cite{Lieb_Liniger_PR_1963_ExactAnalysis,
Olshanii_PRL_1998_AtomicScattering, Girardeau_OptComm_2004_TonksScaling},
\begin{eqnarray}\label{eq:H_delta_t}
H_\mathrm{LL}(t)
\!=\!
\sum_{i=1}^N
-\frac12\partial_{x_i}^2
\!+\!
\frac12\Omega^2(t)x_i^2
\!+\!
\frac{g}{\gamma(t)}
\sum_{i<j}\delta(x_i-x_j).\,\,
\end{eqnarray}
As $\delta(\gamma x)=\gamma^{-1}\delta(x)$, the contact interaction has scaling
degree $-1$ and thus $H(0)$ does not admit a closed form in terms of the $\mathfrak{su}(1,1)$ generators, $K$, $D$, and $T$, Eq.~\eqref{eq:TKD_main}.
Nonetheless, the driving is scale invariant, as the coupling $g/\gamma(t)$
satisfies the consistency relation of Eq.~\eqref{eq: consistency_gen}, so the general formulation applies.

We consider two classes of exact eigenstates that allow for a compact demonstration of the general framework: a
harmonically confined Tonks-Girardeau
state describing a one-dimensional gas of hard-core bosons in the continuum \cite{Girardeau1960,Girardeau2001}, realized with ultracold atoms in tight one-dimensional traps~\cite{Kinoshita_Science_2004_TonksGas,Paredes_Nature_2004_TonksGas}, and its fermionic counterpart, the ideal spin-polarized Fermi gas~\cite{Girardeau_OptComm_2004_TonksScaling,Minguzzi_PRL_2005_TonksScaling,ADC_Entropy_2020,ADC_UltraCold_2021}
\begin{eqnarray}\label{eq:Phi_F}
\Phi_{\varsigma}(\mathbf x)
\!=\!
\exp\!\left[-\frac{\Omega_0}{2}\sum_i x_i^2\right]
\prod_{i<j}\mathrm{sgn}^{\frac{1-\varsigma}{2}}(x_i-x_j)\,|x_i-x_j|.\nonumber\\
\end{eqnarray}
Here, $\varsigma=+1$ stands for the bosonic Tonks-Girardeau gas with symmetric exchange terms, $\prod_{i<j}|x_i-x_j|$, and $\varsigma=-1$ for the fully antisymmetric fermionic state with $\prod_{i<j}(x_i-x_j)$.
Thanks to the Vandermonde factor and its modulus, both states vanish at contact and are annihilated by the interaction, $\delta(x_i-x_j)\Phi_{\varsigma}=0$, so $\Phi_{\varsigma}$ solves the eigenvalue problem of the harmonic-oscillator part alone, with energy $E=\Omega_0 N^2/2$. For $\varsigma=-1$, it is an exact eigenstate of $H_\mathrm{LL}(0)$ at any coupling $g$, identical to the ground state of the ideal spin-polarized Fermi gas. Likewise, for $\varsigma=+1$, it represents the Tonks-Girardeau ground state in the limit $g\to\infty$. In both cases, $\Phi_{\varsigma}$ becomes the lowest-weight state of a single Bargmann tower, $\Phi_{\varsigma}=|k_0,0\rangle$ with $k_0=N^2/4$, regardless of the exchange statistics of the full many-body wave function.
As $c_{n,k}=\delta_{k,k_0}\delta_{n,0}$ the general formula
Eq.~\eqref{eq:Ermakov_AB_main} applies with $A=\bar k=k_0$ and $B=0$, and the spread complexity is given by
Eq.~\eqref{eq:Ermakov_CK_single_main} with $n=0$ and $k=k_0=N^2/4$.

 For the Liouville-space construction, we consider an initial density matrix diagonal
in the tower states $|k_0,n\rangle$. All of these states
vanish at contact, i.e., are annihilated by the contact potential,
and are thus exact eigenstates of $H_\mathrm{LL}(0)$ with
energies $\Omega_0N^2/2+2\Omega_0n$,
\begin{eqnarray}
\rho_{\beta}^{(k_0)}(0)
\!=\!
(1-Q)\sum_{n\ge0}
Q^n |k_0,n\rangle\langle k_0,n|,
\,\,\,\,
Q=e^{-2\beta\Omega_0}.
\label{eq:LL_tower_thermal}
\end{eqnarray}
The general formula for the Krylov variance in Eq.~\eqref{eq:density_M2_general} applies with the
$\beta$-average given by
$\big\langle\lambda^2+\tfrac14\big\rangle_\beta
=
\frac{2\left[k_0+(1-k_0)Q^2\right]}{(1+Q)^2}$, yielding altogether
\begin{eqnarray}
M_2[\rho_{\beta}^{(k_0)}](t)
=
\frac{k_0+(1-k_0)Q^2}{(1+Q)^2}\,
\sinh^2[2\xi(t)] .
\label{eq:LL_thermal_M2}
\end{eqnarray}

As an analytically exact example, we consider a sudden trap-frequency quench~\cite{Minguzzi_PRL_2005_TonksScaling,delcampo2008,Fang_PRL_2014_1DBreathing},
which has often been used to excite weakly and strongly correlated one-dimensional gases by trap modulation~\cite{Chevy_PRL_2002_TransverseMode,Moritz_PRL_2003_CollectiveOscillations1D,Haller_Science_2009_ExcitedGasPhase}, and has been studied in contexts ranging from Loschmidt chaos \cite{delcampo2016} to quantum control \cite{Deffner_PRX_2014_ScaleInvariantSTA,Dupays21} and quantum thermodynamics \cite{Jaramillo2016,Beau2016}.

 We consider the Tonks-Girardeau case with $\varsigma=+1$,
\begin{eqnarray}
\Omega(t)
=
\begin{cases}
\Omega_0, & t<0,\\
\Omega_f, & t>0 ,
\end{cases}
\qquad
\Omega_f<\Omega_0,
\label{eq:LL_trap_quench}
\end{eqnarray}
where the scaling factor follows in closed form,
\begin{eqnarray}
\gamma^2(t)
=
\cos^2(\Omega_f t)
+
\frac{\Omega_0^2}{\Omega_f^2}
\sin^2(\Omega_f t),
\label{eq:LL_quench_gamma}
\end{eqnarray}
which together with $\dot\gamma(t)$ is directly accessible by imaging the expansion of the gas \cite{Schaff2010,Schaff2011,deng2016efimovian,Deng2018,rohringer2015nonequilibrium,Fang_PRL_2014_1DBreathing}.
Inserting it into the single-tower complexity, Eq.~\eqref{eq:Ermakov_CK_single_main}, gives
\begin{eqnarray}
C_K(t)
=
\frac{k_0}{2}
\left(
\eta-\frac{1}{\eta}
\right)^2
\sin^2(\Omega_f t),
\qquad
k_0=\frac{N^2}{4},
\label{eq:LL_quench_CK_closed}
\end{eqnarray}
with $\eta=\Omega_0/\Omega_f$.

\begin{figure}
    \includegraphics[width=\columnwidth]{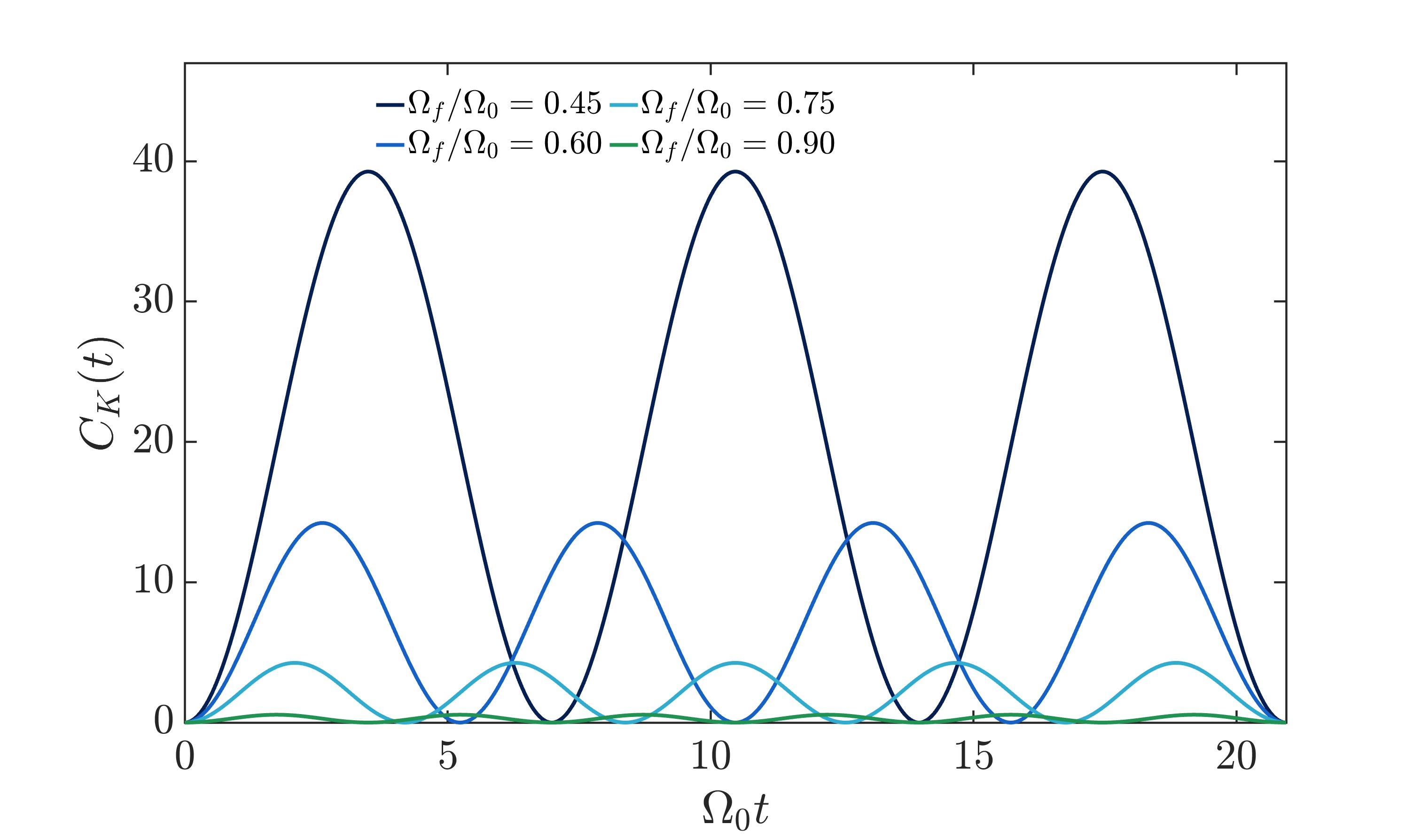}
    \caption{Spread complexity of the Lieb-Liniger gas after a sudden
trap-frequency quench from the Tonks-Girardeau state with $\Omega_0\to\Omega_f$, Eq.~\eqref{eq:LL_quench_CK_closed},
for $\Omega_f/\Omega_0=0.45,\,0.60,\,0.75,\,0.90$ at $N=10$ ($k_0=25$). The
complexity oscillates at twice the post-quench trap frequency, with period
$\pi/\Omega_f$ and an amplitude set by the frequency ratio
$\eta=\Omega_0/\Omega_f$. The amplitude grows with decreasing $\Omega_f/\Omega_0$,
while the period shortens with increasing $\Omega_f$.}
\label{fig:LL_CK}
\end{figure}

Figure~\ref{fig:LL_CK} shows the spread complexity after the quench at
several ratios $\Omega_f/\Omega_0$. It oscillates as $\sin^2(\Omega_f t)$ with the amplitude
$\tfrac{k_0}{2}(\eta-\eta^{-1})^2$ set by the frequency ratio
$\eta=\Omega_0/\Omega_f$. Accordingly, the complexity exhibits larger oscillations with increasing difference between the initial and final frequencies, while their period $\pi/\Omega_f$ shortens with increasing $\Omega_f$.

The Krylov variance of the initial thermal state follows
from the same scaling factor. Inserting
Eq.~\eqref{eq:LL_quench_gamma} into
Eq.~\eqref{eq:LL_thermal_M2} and using
$\sinh^2[2\xi]=4\sinh^2\xi\,(1+\sinh^2\xi)$ gives the closed
form,
\begin{eqnarray}
M_2[\rho_{\beta}^{(k_0)}](t)
&=&
\frac{
k_0+(1-k_0)Q^2
}{
(1+Q)^2
}
\left(
\eta-\frac{1}{\eta}
\right)^2
\sin^2(\Omega_f t)
\nonumber\\
&&\times
\left[
1+
\frac14
\left(
\eta-\frac{1}{\eta}
\right)^2
\sin^2(\Omega_f t)
\right].
\label{eq:LL_quench_M2_closed}
\end{eqnarray}

\begin{figure}
\centering
    \includegraphics[width=.99\columnwidth,trim={0cm 0cm 0cm 0},clip]{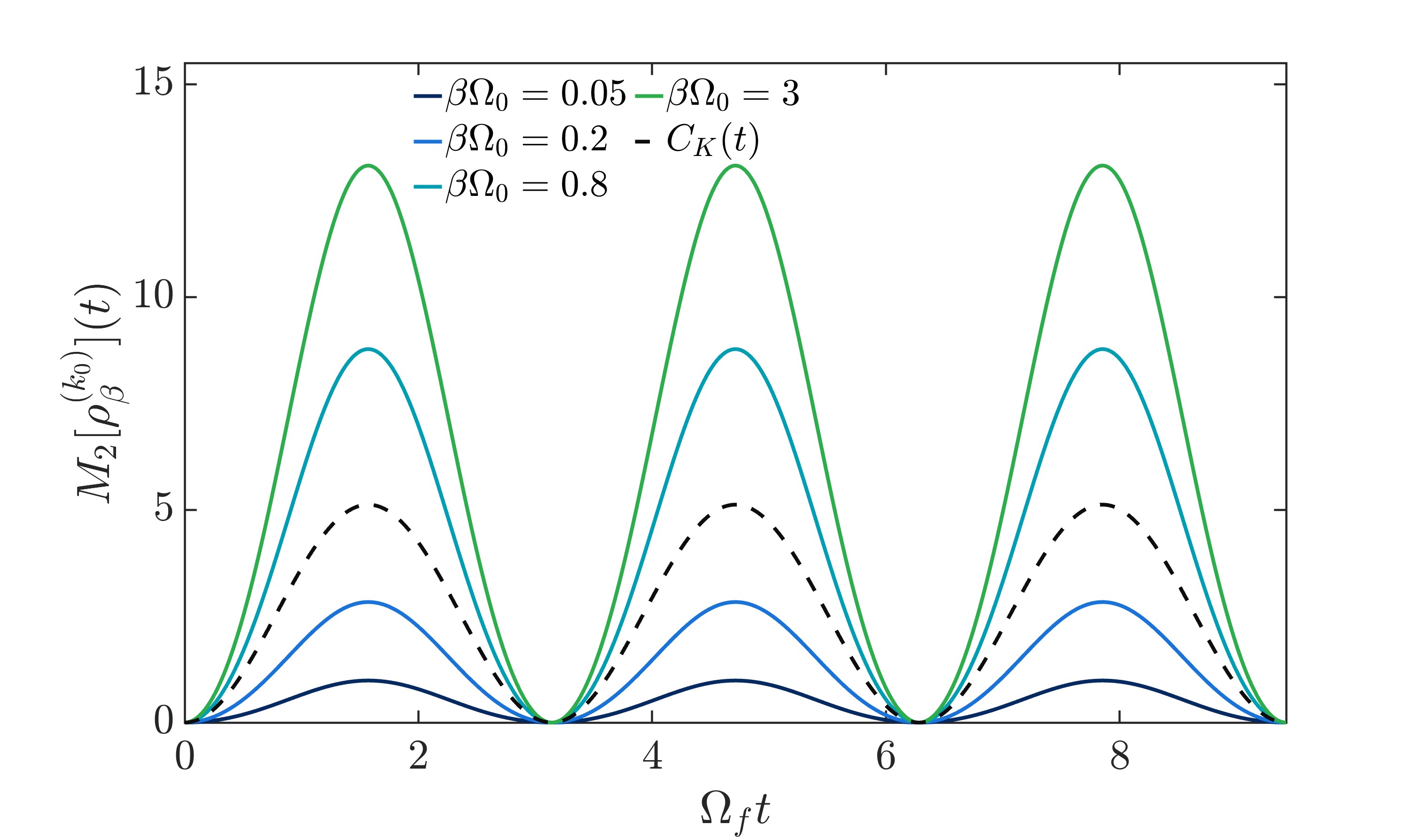}
    \includegraphics[width=.99\columnwidth,trim={0cm 0cm 0 0},clip]{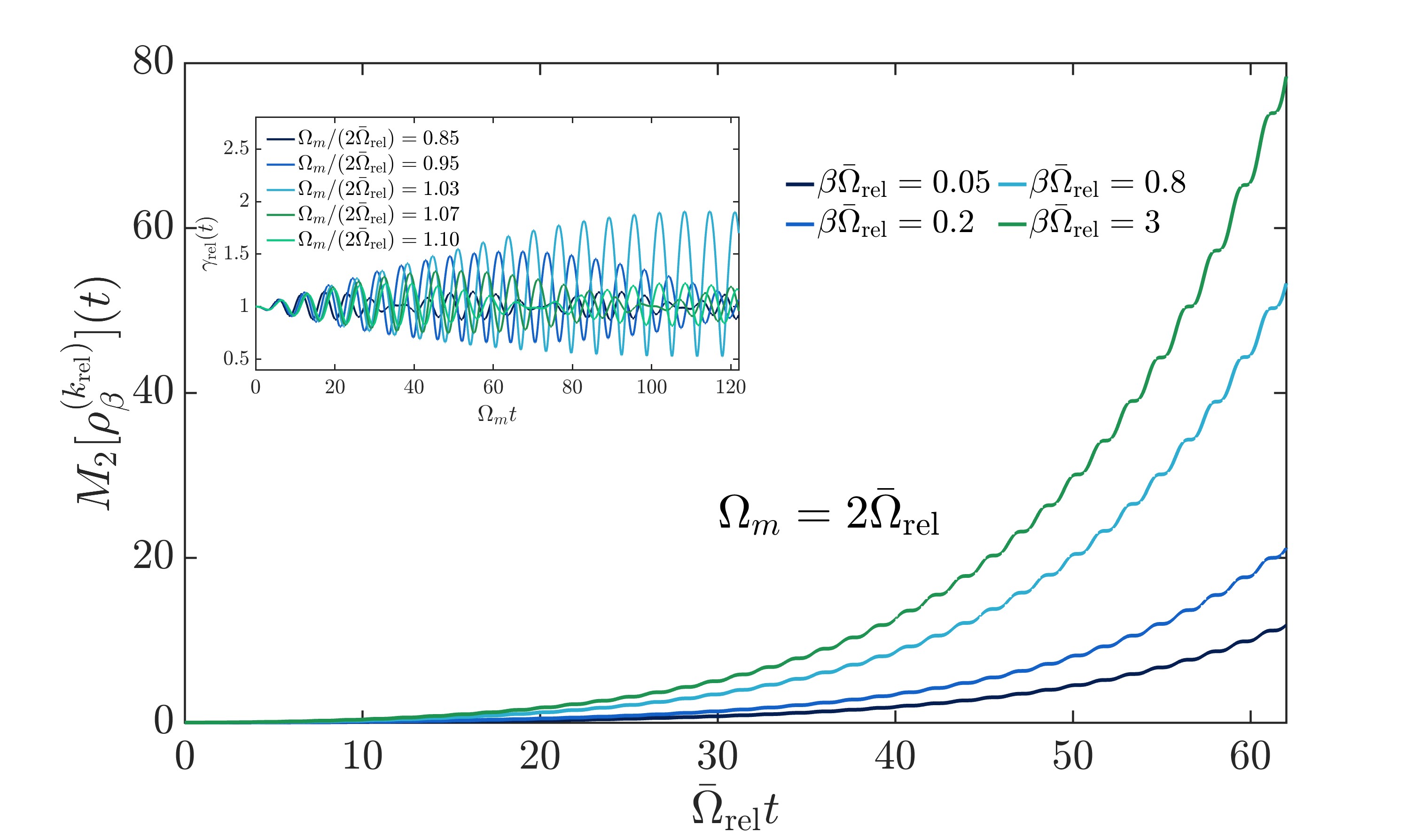}
    \caption{Krylov variance for the initial diagonal density matrix of the Lieb-Liniger gas after a trap quench, Eq.~\eqref{eq:LL_quench_M2_closed}, at $N=6$ ($k_0=9$) and
$\Omega_f/\Omega_0=0.6$. (a) $M_2(t)$ versus $\Omega_f t$ for
$\beta\Omega_0=0.05,\,0.2,\,0.8,\,3$, with the dashed line showing the spread complexity $C_K(t)$ for
reference. The spread shares the period of $C_K$ but peaks more sharply, and
grows with increasing $\beta$. (b) $M_2$ versus $\beta\Omega_0$ at fixed times
$\Omega_f t=0.15\pi,\,0.30\pi,\,0.50\pi,\,0.75\pi$. The spread increases from its $N$-independent infinite-temperature value and
saturates at low temperature at a value proportional to $N^2$.}
\label{fig:LL_M2}
\end{figure}

The density-matrix spread is shown in Fig.~\ref{fig:LL_M2}. The evolution of
$M_2(t)$ at fixed frequency ratio $\Omega_f/\Omega_0$ for several temperatures is shown in panel (a), with the spread complexity $C_K(t)$ 
also highlighted for reference. The Krylov variance keeps the period
of $C_K$ but peaks more sharply through the $\sin^4$ term. Its amplitude grows monotonically with $\beta$.
Panel (b) shows the Krylov variance as a function of
$\beta\Omega_0$ at fixed times. It increases from its
$N$-independent infinite-temperature value and saturates at
low temperature to a value proportional to $k_0=N^2/4$.

\subsection{Algebraic closure and Heisenberg-Weyl extensions}
\label{sec:su11_hw_examples}

Next, we consider the case when $H(0)$ is itself a linear combination of the $\mathfrak{su}(1,1)$ generators, so that its eigenstates coincide with the
tower states and the subsequent dynamics unfolds within the initially occupied tower. This class also admits an exact treatment of an additional quadratic pairwise potential and of a linear driving force, the latter through an extension by the Heisenberg-Weyl algebra. The
corresponding Hamiltonian reads
\begin{eqnarray}
&&H(t)
=\sum_i\frac{\mathbf p_i^2}{2}+\sum_iU_{-2}(\mathbf r_i)
+\varepsilon_{-2}\!\sum_{i<j}V_{-2}(\mathbf q_i-\mathbf q_j)
\nonumber\\
&&+\tfrac12\Omega^2(t)\sum_i\mathbf q_i^2
+\tfrac12\varepsilon_{+2}(t)\!\sum_{i<j}|\mathbf q_i-\mathbf q_j|^2
+\mathbf F(t)\!\cdot\!\mathbf Q,\quad
\label{eq:H_class}
\end{eqnarray}
expressed with the total coordinate $\mathbf Q=\sum_i\mathbf q_i$, the
total momentum $\mathbf P=\sum_i\mathbf p_i$, and the relative coordinates
$\mathbf r_i=\mathbf q_i-\mathbf Q/N$. The on-site potential $U_{-2}$ and the
pair interaction $V_{-2}$ of constant strength $\varepsilon_{-2}$ have scaling degree $-2$, while $\Omega^2(t)$,
$\varepsilon_{+2}(t)$, and $\mathbf F(t)$ set the trap, the quadratic pair coupling, and the center-of-mass force, respectively. Using
$\sum_i\mathbf q_i^2=\sum_i\mathbf r_i^2+|\mathbf Q|^2/N$ and
$\sum_{i<j}|\mathbf q_i-\mathbf q_j|^2=N\sum_i\mathbf r_i^2$, the Hamiltonian
splits into two commuting sectors, $H(t)=H_{\rm rel}(t)+H_{\rm cm}(t)$.
The relative one is described by
\begin{eqnarray}
H_{\rm rel}(t)&=&\sum_i\frac{\bm\pi_i^2}{2}+\sum_iU_{-2}(\mathbf r_i)
+\varepsilon_{-2}\!\sum_{i<j}V_{-2}(\mathbf q_i-\mathbf q_j)\nonumber\\
&&+\Omega_{\rm rel}^2(t)\,K_{\rm rel},
\label{eq:Hrel_admissible}
\end{eqnarray}
where $\bm\pi_i$ is the momentum conjugate to $\mathbf r_i$, the relative frequency is given by $\Omega_{\rm rel}^2(t)=\Omega^2(t)+N\varepsilon_{+2}(t)$, and $K_{\rm rel}=\tfrac12\sum_i\mathbf r_i^2$ is the quadratic potential in the relative coordinates. As a result, $H_{\rm rel}(t)$ is a linear combination of $K_{\rm rel}$, the
dilation generator $D_{\rm rel}$ in the relative coordinates, and $T_{\rm
rel}=\sum_i\bm\pi_i^2/2+\sum_iU_{-2}(\mathbf
r_i)+\varepsilon_{-2}\sum_{i<j}V_{-2}(\mathbf q_i-\mathbf q_j)$, which satisfy
the $\mathfrak{su}(1,1)$ commutation relations of Eq.~\eqref{eq:TKD_main}.

The center-of-mass sector is a driven oscillator with a time-dependent force,
$H_{\rm cm}(t)=\frac{\mathbf P^2}{2N}+\frac{\Omega^2(t)}{2N}|\mathbf Q|^2
+\mathbf F(t)\!\cdot\!\mathbf Q$, whose linear term generates a displacement governed by the Heisenberg-Weyl algebra $\mathfrak h_3$. As shown in App.~\ref{app:HW_su11_dressing} with the variables
$\mathbf X=\mathbf Q/\sqrt N$ and $\bm\Pi=\mathbf P/\sqrt N$, the center-of-mass Hamiltonian becomes
\begin{eqnarray}
H_{\rm cm}(t)
=
\sum_\mu
\left[
\frac{\Pi_\mu^2}{2}
+
\frac{\Omega^2(t)}{2}X_\mu^2
+
\sqrt N\,F_\mu(t)X_\mu
\right].
\label{eq:Hcm_canonical}
\end{eqnarray}
The quadratic part is the usual center-of-mass oscillator, while the force term is linear in the ladder operators, $a_\mu=\sqrt{\Omega_{{\rm cm},0}/2}\,X_\mu+i\Pi_\mu/\sqrt{2\Omega_{{\rm cm},0}}$ with $\Omega_{{\rm cm},0}=\Omega(0)$, satisfying the $\mathfrak h_3$ commutation relations
\begin{eqnarray}
[a_\mu,a_\nu^\dagger]=\delta_{\mu\nu},\qquad
[a_\mu,a_\nu]=[a_\mu^\dagger,a_\nu^\dagger]=0.
\label{eq:Hcm_hw_algebra}
\end{eqnarray}
As shown in App.~\ref{app:HW_su11_dressing},
the multidimensional Krylov dynamics along the orthogonal directions $\mu$ can be obtained by going to the interaction picture with respect to the quadratic parts,
\begin{eqnarray}
X_\mu(t)
=
u_\mu(t)a_\mu+u_\mu^\ast(t)a_\mu^\dagger,
\qquad
\ddot u_\mu+\Omega^2(t)u_\mu=0,
\label{eq:Hcm_mode_function}
\end{eqnarray}
with $u_\mu(0)=1/\sqrt{2\Omega_{{\rm cm},0}}$ and
$\dot u_\mu(0)=-i\sqrt{\Omega_{{\rm cm},0}/2}$. The remaining Hamiltonian is linear
in the $\mathfrak h_3$ generators,
\begin{eqnarray}
H_{{\rm cm},I}(t)
=
\sqrt N\sum_\mu
F_\mu(t)
\left[
u_\mu(t)a_\mu+u_\mu^\ast(t)a_\mu^\dagger
\right],
\label{eq:Hcm_HW_interaction}
\end{eqnarray}
and induces the coherent displacement
\begin{eqnarray}
\alpha_{{\rm cm},\mu}(t)
=
-i\sqrt N
\int_0^t dt'\,
F_\mu(t')u_\mu^\ast(t'),
\label{eq:alpha_cm_admissible}
\end{eqnarray}
where the time-dependent trap frequency appears in the
displacements only through the interaction-picture coefficients, $u_\mu(t)$. The Krylov basis of the center of mass consists of the usual eigenstates of $a^\dagger_\mu a_\mu$, so its spread complexity is
$\sum_\mu\langle a_\mu^\dagger a_\mu\rangle$ in the original
frame. With $\Pi_\mu(t)=\dot X_\mu(t)=\dot u_\mu a_\mu+
\dot u_\mu^*a_\mu^\dagger$, the quadratic part evolves the
ladder operator as $a_\mu(t)=\mathcal U(t)a_\mu+
\mathcal V(t)a_\mu^\dagger$, where
$\mathcal U=\sqrt{\Omega_{{\rm cm},0}/2}\,u_\mu+
i\dot u_\mu/\sqrt{2\Omega_{{\rm cm},0}}$ and
$\mathcal V=\sqrt{\Omega_{{\rm cm},0}/2}\,u_\mu^*+
i\dot u_\mu^*/\sqrt{2\Omega_{{\rm cm},0}}$ do not depend on $\mu$
and satisfy $|\mathcal U|^2-|\mathcal V|^2=1$. As the
displacement shifts the ladder operators as $a_\mu\to a_\mu+\alpha_{{\rm cm},\mu}$, a
center-of-mass Krylov state $|m_\mu\rangle$ contributes
\begin{eqnarray}
C_{{\rm cm},\mu}(t)
&=&
|\mathcal U|^2m_\mu+|\mathcal V|^2(m_\mu+1)\nonumber\\
&&+\big|\mathcal U\alpha_{{\rm cm},\mu}
+\mathcal V\alpha_{{\rm cm},\mu}^*\big|^2.
\label{eq:Ccm_mu}
\end{eqnarray}
For a fixed trap, $\Omega(t)=\Omega_{{\rm cm},0}$, one has
$u_\mu=e^{-i\Omega_{{\rm cm},0}t}/\sqrt{2\Omega_{{\rm cm},0}}$,
$|\mathcal U|=1$, and $\mathcal V=0$, so that
$C_{{\rm cm},\mu}=m_\mu+|\alpha_{{\rm cm},\mu}|^2$
(see App.~\ref{app:HW_su11_dressing} for details).

Overall, the relative sector is the $\mathfrak{su}(1,1)$ problem of Sec.~\ref{sec:scale_covariant}, while the center of mass is a forced oscillator that displaces its $\mathfrak h_3$ ground state. As these two sectors commute, the eigenstates of $H(0)$ factorize, and the total spread complexity decomposes into their sum. In the relative
sector, $H_{\rm rel}(0)|k,n\rangle_{\rm rel}=2\Omega_{{\rm
rel},0}(k+n)|k,n\rangle_{\rm rel}$ with $\Omega_{{\rm rel},0}=\Omega_{\rm
rel}(0)$, while the center of mass has the oscillator levels $|\mathbf
m\rangle=|m_1,\dots,m_d\rangle$, by which the joint eigenstates are
$|k,n;\mathbf m\rangle=|k,n\rangle_{\rm rel}\otimes|\mathbf m\rangle_{\rm
cm}$.
The relative tower contributes to the
single-tower form Eq.~\eqref{eq:Ermakov_CK_single_main} with $\gamma_{\rm rel}(t)$ following from
$\Omega_{\rm rel}(t)$ through the Ermakov equation, Eq.~\eqref{eq: consistency_gen}. The center of mass contributes its own spread
complexity through the squeezing and the displacement of each direction $\mu$,
Eq.~\eqref{eq:Ccm_mu}~\cite{Grabarits_arXiv_2026_TDKrylovLie}. Together,
\begin{eqnarray}
C_K^{(k,n;\mathbf m)}(t)
&=&
n
+
\frac{k+n}{2}
\left[
\left(\gamma_{\rm rel}-\frac{1}{\gamma_{\rm rel}}\right)^2
+
\left(\frac{\dot\gamma_{\rm rel}}{\Omega_{{\rm rel},0}}\right)^2
\right]
\nonumber\\
&&+
\sum_\mu C_{{\rm cm},\mu}(t) ,
\label{eq:admissible_CK_pure}
\end{eqnarray}
where the first line is the contribution of the relative
coordinates and the second line that of the center of
mass, Eq.~\eqref{eq:Ccm_mu}. In the present extended case,
the full spread complexity is fixed only by the scaling
factor $\gamma_{\rm rel}(t)$, the center-of-mass mode
functions $u_\mu(t)$, and the driving force.

Each of the two commuting sectors thus has its own measurable scaling
factor, as we demonstrate below with the Moshinsky model. Further examples, where the force adds a Heisenberg-Weyl displacement of the center of mass, are discussed for the Calogero-Sutherland model in a
translated trap (App.~\ref{app:CS_translated}) and for the Moshinsky model under a uniform force (App.~\ref{app:moshinsky_driven}).

The Moshinsky model is a good example with quadratic interactions~\cite{Moshinsky_JMP_1968_HarmonicInteraction}, describing $N$ particles in a harmonic trap with quadratic pairwise interactions,
\begin{eqnarray}\label{eq:HM_realizations}
H(t)
\!&\!=\!&\!
\sum_{i=1}^N
\left[
-\frac{1}{2}\partial_{x_i}^2
\!+\!
\frac{1}{2}\omega^2(t)x_i^2
\right]
\!+\!
\tfrac12\varepsilon_{+2}(t)
\sum_{i<j}(x_i-x_j)^2.\quad\nonumber\\
\end{eqnarray}
The center-of-mass and relative Hamiltonians can be written in terms of the
$\mathfrak{su}(1,1)$ generators, with frequencies $\Omega_{\rm cm}(t)=\omega(t)$
and $\Omega_{\rm rel}^2(t)=\omega^2(t)+N\varepsilon_{+2}(t)$~\cite{Deffner_PRX_2014_ScaleInvariantSTA,
ADC_Entropy_2020}, with the pairwise coupling renormalizing the relative frequency only. Without the force, the ground state is the lowest-weight state of the relative tower and of the center-of-mass tower $k_{\rm cm}=\tfrac14$, which consists of the even oscillator levels. Both sectors thus contribute the single-tower
result of Eq.~\eqref{eq:Ermakov_CK_single_main}, and the spread complexity is their sum,
\begin{eqnarray}
C_K(t)
&=&
\sum_{s={\rm cm},{\rm rel}}
\frac{k_s}{2}
\left[
\left(\gamma_s-\frac{1}{\gamma_s}\right)^2
+
\left(\frac{\dot\gamma_s}{\Omega_{s,0}}\right)^2
\right],
\nonumber\\
k_{\rm cm}&=&\frac{1}{4},
\qquad
k_{\rm rel}=\frac{N-1}{4},
\label{eq:CK_HM_total}
\end{eqnarray}
with each $\gamma_s(t)$ fixed by the Ermakov relation,
Eq.~\eqref{eq: consistency_gen}, with the corresponding frequencies $\Omega_\mathrm{cm}$ and $\Omega_\mathrm{rel}$, for the center-of-mass and relative sectors, respectively.
This makes the model a direct test for the algebraic construction, as periodic modulations of the interaction strength~\cite{Pollack_PRA_2010_ScatteringLengthModulation} and of the harmonic potential ~\cite{Friebel_PRA_1998_CO2Lattice,Balik_PRA_2009_ImagingParametricResonance} are both standard tools for exciting collective modes in cold-atom experiments.
For the sake of illustration, consider modulating the pair coupling with a fixed trap,
\begin{eqnarray}
\varepsilon_{+2}(t)
&=&
\varepsilon_{+2,0}\bigl[1+\delta\sin(\Omega_m t)\bigr],
\nonumber\\
\Omega_{\rm rel}^2(t)
&=&
\bar\Omega_{\rm rel}^2\bigl[1+h\sin(\Omega_m t)\bigr],
\label{eq:HM_coupling_mod}
\end{eqnarray}
with $h=N\varepsilon_{+2,0}\delta/\bar\Omega_{\rm rel}^2$ and
$\bar\Omega_{\rm rel}^2=\omega^2+N\varepsilon_{+2,0}=\Omega_{{\rm rel},0}^2$, which periodically drives the relative
frequency while leaving the center of mass static. The relative dilation obeys the Ermakov relation
\begin{eqnarray}
\ddot\gamma_{\rm rel}
+
\Omega_{\rm rel}^2(t)\gamma_{\rm rel}
=
\frac{\bar\Omega_{\rm rel}^2}{\gamma_{\rm rel}^3},
\quad
\gamma_{\rm rel}(0)=1,
\quad
\dot\gamma_{\rm rel}(0)=0.\,\,
\label{eq:HM_pinney}
\end{eqnarray}
We consider the solvable limit, where $\gamma_\mathrm{rel}(t)$ stays close to $1$. Linearizing as $\gamma_{\rm rel}=1+\zeta_{\rm rel}$ gives a driven Mathieu equation for the deviation $\zeta_{\rm rel}$ of the scaling factor~\cite{McLachlan_1947_MathieuFunctions},
$\ddot\zeta_{\rm rel}+[\,4\bar\Omega_{\rm rel}^2+h\bar\Omega_{\rm rel}^2
\sin(\Omega_m t)\,]\zeta_{\rm rel}=-h\bar\Omega_{\rm rel}^2\sin(\Omega_m t)$. To leading order in $h$, this is the equation of a harmonic oscillator of
frequency $2\bar\Omega_{\rm rel}$, corresponding to the collective oscillation of the many-body system, subject to a periodic force with a frequency $\Omega_m$. When $\Omega_m\simeq2\bar\Omega_{\rm rel}$, the force is in resonance with the oscillator and the amplitude of $\zeta_{\rm rel}$ grows linearly in time,
$\zeta_{\rm rel}(t)\simeq\frac{h\bar\Omega_{\rm rel}}{4}\,t\cos(2\bar\Omega_{\rm rel}t)$ to leading order in $h$~\cite{LandauLifshitz_Mechanics}. Inserting this into Eq.~\eqref{eq:CK_HM_total}, the oscillating parts cancel, and the spread complexity acquires a quadratic envelope,
\begin{eqnarray}
C_K(t)\simeq\frac{k_{\rm rel}\,h^2\bar\Omega_{\rm rel}^2}{8}\,t^2 .
\label{eq:HM_resonant_CK}
\end{eqnarray}
Away from the resonance, the relative dilation
stays bounded, and the complexity oscillates within a fixed envelope. Close to resonance, the driving frequency $\Omega_m$
differs from the collective frequency $2\bar\Omega_{\rm rel}$
by $\epsilon=\Omega_m-2\bar\Omega_{\rm rel}$ with
$|\epsilon|\ll\bar\Omega_{\rm rel}$. The linear growth of
$\zeta_{\rm rel}$ then holds only for $t\ll1/|\epsilon|$.
Solving the forced oscillator equation with
$\zeta_{\rm rel}(0)=\dot\zeta_{\rm rel}(0)=0$ gives
$\zeta_{\rm rel}(t)\simeq\frac{h\bar\Omega_{\rm rel}}
{2\epsilon}\sin(\epsilon t/2)\cos(2\bar\Omega_{\rm rel}t)$,
which reduces to the linear growth above for $\epsilon\to0$.
Inserting it into Eq.~\eqref{eq:CK_HM_total} yields
$C_K(t)\simeq\frac{k_{\rm rel}h^2\bar\Omega_{\rm rel}^2}
{2\epsilon^2}\sin^2(\epsilon t/2)$. The complexity thus
oscillates with period $2\pi/|\epsilon|$ and amplitude
$k_{\rm rel}h^2\bar\Omega_{\rm rel}^2/(2\epsilon^2)$.

For the initial thermal density matrix diagonal in Bargmann towers in the relative coordinate sector, the Krylov variance of Sec.~\ref{sec:density_general}
follows from the general result Eq.~\eqref{eq:density_M2_general},
\begin{eqnarray}
\label{eqM2}
M_2[\rho_\beta^{(k_{\rm rel})}]
\!=\!\frac{1}{2}
\big\langle\lambda^2+\tfrac14\big\rangle_\beta\sinh^2[2\xi(t)]\,,
\end{eqnarray}
with $\sinh^2\xi(t)=\tfrac14{\cal A}_{\rm rel}(t)$.
Here, we have now used the exact form in terms of $\gamma_\mathrm{rel}$ with
\begin{eqnarray}
{\cal A}_{\rm rel}(t)
=
\left(\gamma_{\rm rel}-\frac{1}{\gamma_{\rm rel}}\right)^2
+
\left(\frac{\dot\gamma_{\rm rel}}{\bar\Omega_{\rm rel}}\right)^2.
\label{eq:HM_Arel_parametric}
\end{eqnarray}
Both the spread complexity and the Krylov variance are
thus fixed by the expansion of the relative sector resonantly enhanced at $\Omega_m\simeq2\bar\Omega_{\rm rel}$,
while the initial temperature appears only through a
multiplicative factor.

\begin{figure}
    \includegraphics[width=.99\columnwidth]{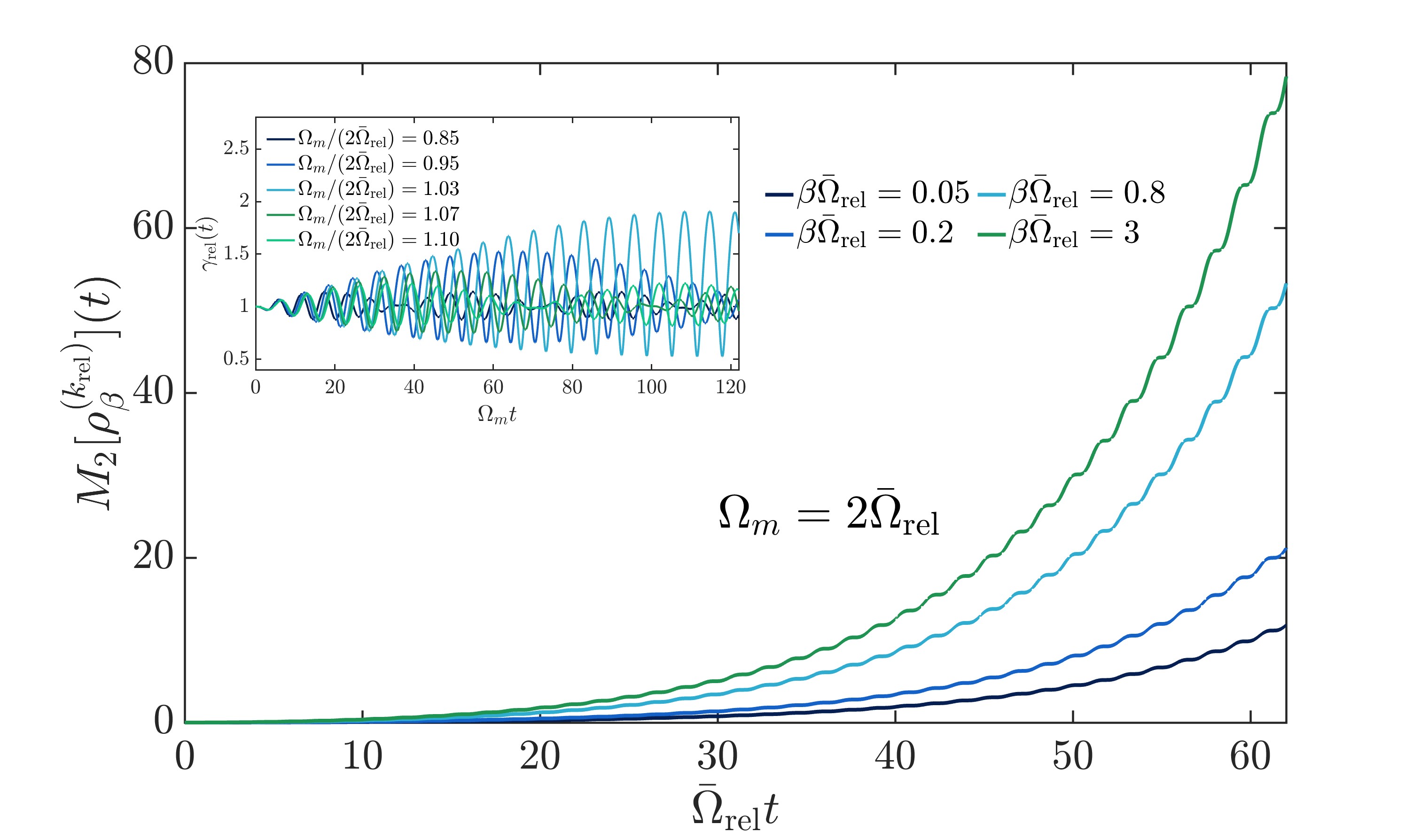}
    \includegraphics[width=.99\columnwidth]{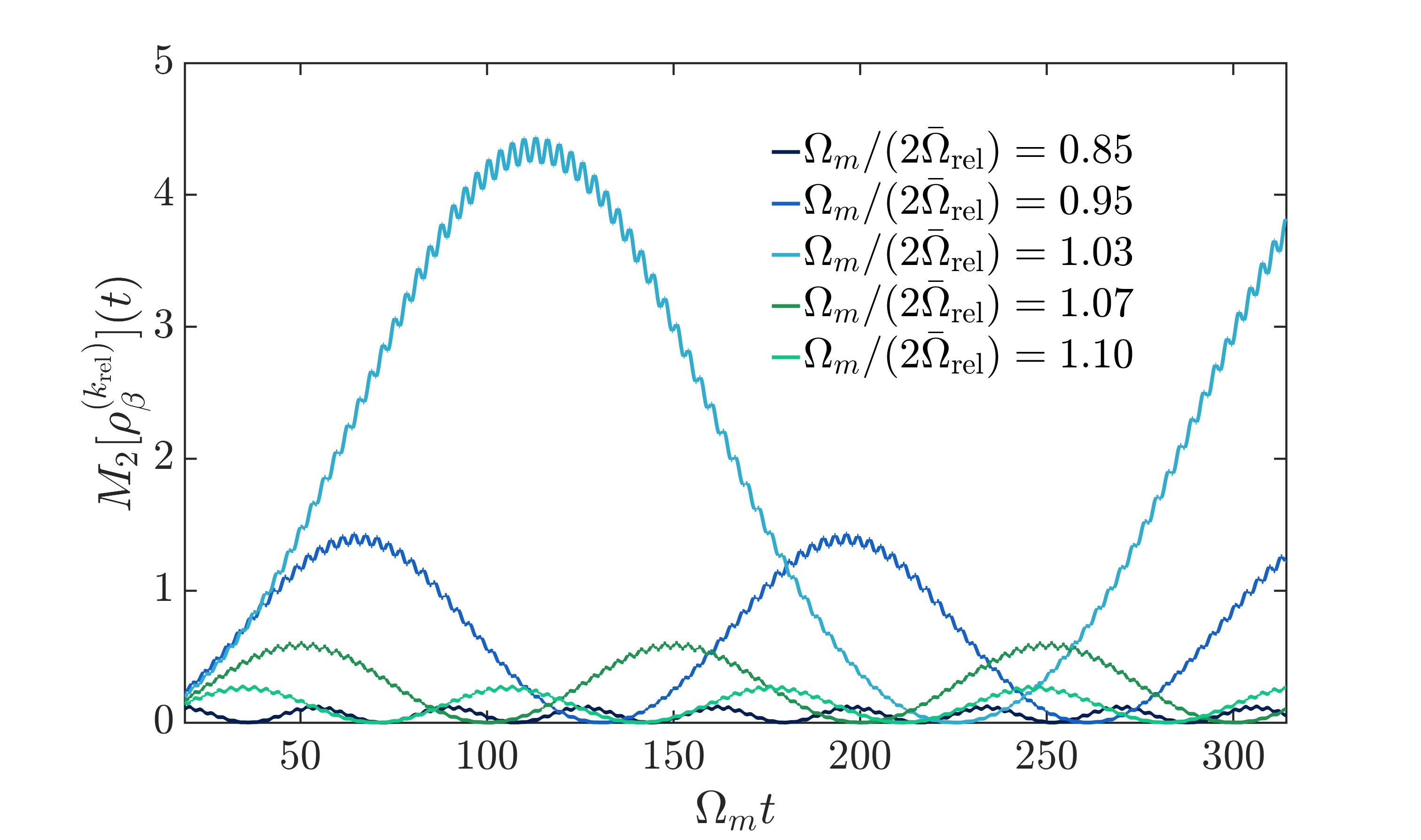}
    \caption{Density-matrix Krylov variance of the parametrically driven
Moshinsky model.  (a) $M_2[\rho_\beta^{(k_{\rm rel})}](t)$ at resonance,
$\Omega_m=2\bar\Omega_{\rm rel}$, for
$\beta\bar\Omega_{\rm rel}=0.05,\,0.2,\,0.8,\,3$.  The spread grows with an envelope and with increasing $\beta$, overall following the predicted quadratic growth.  Inset:
relative dilation $\gamma_{\rm rel}(t)$ for
$\Omega_m/(2\bar\Omega_{\rm rel})=0.85,\,0.95,\,1.00,\,1.10$, showing larger oscillations when closer to resonance.  (b) $M_2[\rho_\beta^{(k_{\rm rel})}](t)$ at fixed temperature for driving frequencies
$\Omega_m/(2\bar\Omega_{\rm rel})=0.85,\,0.95,\,1.03,\,1.07,\,1.10$. The
Krylov variance is bounded off resonance and strongly amplified near
resonance, with the largest response at slight positive detuning.}
\label{fig:moshinsky_parametric}
\end{figure}

Figure~\ref{fig:moshinsky_parametric} shows the density-matrix spread.  At
resonance, the relative dilation is amplified (inset), and $M_2(t)$ grows with a modulated envelope, multiplied by the temperature-dependent prefactor of Eq. \eqref{eqM2}, with the Krylov variance growing monotonically with $\beta$.
Varying the frequency shows that the spread remains bounded away from resonance and is sharply enhanced near $\Omega_m = 2\bar\Omega_{\rm rel}$.

\section{Krylov dynamics in general rotation-invariant many-body systems}
\label{sec:rot_invariant}
Next, we consider rotation-invariant many-body systems as the compact counterpart of the scale-invariant case. While in scale-invariant systems the driving protocol induces a dilation of the internal length scale, rotation-invariant many-body systems undergo global rotations parametrized by a pair of angle variables. We focus on interacting spin systems providing ideal experimental testbeds in current quantum simulators. There, a rotation amounts to the precession of every spin direction, leaving the isotropic exchange couplings unchanged. The time-dependent driving rotates the collective spin, so the transformation $\mathcal T(t)$ of Sec.~\ref{sec:general} is an $\mathrm{SU}(2)$ rotation. The time evolution can thus be expressed in a closed form, without diagonalizing the instantaneous Hamiltonian~\cite{Grabarits_arXiv_2026_TDKrylovLie}. The rotation angles of the driving play the role of the dilation in the scale-invariant case and govern the spread complexity. Unlike the Bargmann towers, each total-spin tower is finite, with $2j+1$ states for integer or half-integer $j$, Eq.~\eqref{eq:su2_primer}, resulting in a bounded complexity. In correspondence to the imaging of trapped gases, collective spin oscillations can be read out directly in trapped-ion, cavity-QED, and circuit-QED
platforms~\cite{Monroe_RMP_2021_IonSpinSim, Garttner_NatPhys_2017_OTOC, Young2024BCSCavity, LewisSwan2021BCSCavity}.

In the following, we develop the general formulation for an arbitrary initial state and apply it to a central spin in an interacting bath. As its initial Hamiltonian is not a linear combination of the $\mathfrak{su}(2)$ generators~\cite{Gaudin_JPhys_1976_SpinHamiltonians}, it can be considered as the counterpart of the Lieb-Liniger gas. We then turn to the case in which $H(0)$ closes over a finite algebra, illustrated by a driven two-mode doublet and by the thermal Krylov variance of the Haldane-Shastry model~\cite{Haldane_PRL_1988_HaldaneShastry, Shastry_PRL_1988_HaldaneShastry}. Throughout these examples, driving involves a collective time-dependent field set by a single rotation angle $\theta(t)$, which governs the spread complexity and the density-matrix spread at any temperature.

\subsection{General formulation for arbitrary initial states}

We consider $N$ spins in time-dependent fields, interacting through isotropic Heisenberg exchange,
\begin{eqnarray}
H(t)
&=&\bm h(t)\!\cdot\!\bm J
+\sum_i\bm u_i(t)\!\cdot\!\bm s_i
+\sum_{i<j}J_{ij}\,\bm s_i\!\cdot\!\bm s_j,
\label{eq:su2_Ht_main}
\end{eqnarray}
with a collective field $\bm h(t)$, local fields $\bm u_i(t)$, Heisenberg couplings $J_{ij}$ with arbitrary connectivity, and total spin $\bm J=\sum_i\bm s_i$.
The general framework, Eq.~\eqref{eq:Psi_exact_general}, applies with $\mathcal T(t)=R(t)=e^{-i\bm\theta(t)\cdot\bm J}$ and $s(t)=t$, where the rotation vector $\bm\theta(t)$ points along the instantaneous rotation axis of $R(t)$ and its magnitude $|\bm\theta(t)|$ sets the rotation angle. The generator of $R(t)$ defines the angular velocity of the driving,
\begin{eqnarray}
i\dot R(t)R^\dagger(t)\equiv\bm\Omega(t)\!\cdot\!\bm J,
\label{eq:su2_angular_velocity_main}
\end{eqnarray}
which is set by $\bm\theta(t)$ through the rotation-vector kinematics (App.~\ref{app:angular_velocity}). As shown in App.~\ref{app:angular_velocity}, given the consistency equations,
\begin{eqnarray}
\dot{\bm u}_i=\bm\Omega(t)\!\times\!\bm u_i,\qquad
\dot{\bm h}=\dot{\bm\Omega}+\bm\Omega\!\times\!(\bm h-\bm\Omega),
\label{eq:su2_consistency}
\end{eqnarray}
by which the magnitudes of the local fields $\bm u_i$ are
preserved and only their directions change with time. Under these conditions, the general construction applies, and the full time evolution can be written as
\begin{eqnarray}
|\Psi(t)\rangle=R(t)\,e^{-itH(0)}|\Psi(0)\rangle .
\label{eq:Psi_exact_su2_main}
\end{eqnarray}
Similarly to the scale-invariant case in Sec.~\ref{sec:scale_covariant}, $H(0)$ only fixes the initial conditions via its eigenstates, $H(0)|\nu\rangle=E_\nu|\nu\rangle$, while $R(t)$ generates the dynamics. In this case, the Cartan-Weyl basis of
Eq.~\eqref{eq:CW_general} is given by the collective spin
components, $J_0=J_z,
\, J_\pm=J_x\pm iJ_y$. The Krylov Liouvillian in  Eq.~\eqref{eq:Heff_general} becomes
\begin{eqnarray}
H_{\nu,\mathrm{eff}}(t)&=&\bm\Omega(t)\!\cdot\!\bm J+E_\nu\\
&=&A_0(t)J_z+A_+(t)J_++A_-(t)J_-+E_\nu,\nonumber
\end{eqnarray}
with $A_0=\Omega_z$ and $A_\pm=\tfrac12(\Omega_x\mp i\Omega_y)$. It acts as a
tridiagonal matrix on the finite tower $\{|j,q\rangle\}$, $q=0,\dots,2j$, of
Eq.~\eqref{eq:tower_general}, with $J_0|j,q\rangle=(-j+q)|j,q\rangle$.

Expanding the initial state over these towers as in
Eq.~\eqref{eq:nu_expansion_general}, the overlaps $c^{(\nu)}_{q,j}=\langle
j,q|\nu\rangle$ set the initial values of the amplitudes
$\psi^{(\nu)}_{q,j}(t)=\langle j,q|\Psi_\nu(t)\rangle$, which obey the
$\mathfrak{su}(2)$ form of Eq.~\eqref{eq:chain_general},
\begin{eqnarray}
&&i\dot\psi^{(\nu)}_{q,j}
=\left[A_0(t)(-j+q)+E_\nu\right]\psi^{(\nu)}_{q,j}
+A_+(t)\,\beta_{q,j}\,\psi^{(\nu)}_{q-1,j}\nonumber\\
&&+A_-(t)\,\beta_{q+1,j}\,\psi^{(\nu)}_{q+1,j},
\quad
\beta_{q,j}=\sqrt{q(2j-q+1)},
\label{eq:su2_chain_main}
\end{eqnarray}
with $\beta_{0,j}=\beta_{2j+1,j}=0$. In terms of the Euler angles of the rotation,
$R(t)=e^{-i\alpha(t)J_z}e^{-i\theta(t)J_y}
e^{-i[\phi(t)-\pi]J_z}$, the amplitudes take a closed form through the
Wigner matrices, while the $\mathfrak{su}(2)$ Wei-Norman
factorization~\cite{WeiNorman1963,WeiNorman1964} gives the lowest-weight
state explicitly (App.~\ref{app:wigner_d}). Here, $\theta(t)$ is the polar angle
of the rotated $z$ axis and $-\phi(t)$ is the azimuth of the $z$ axis in the rotating frame, while the first Euler angle $\alpha(t)$ only
contributes a phase that drops out of the Krylov
probabilities. For rotations around a fixed axis in the $xy$ plane, which
covers all protocols below, $\theta(t)$ coincides with the rotation angle
$|\bm\theta(t)|$ and $\alpha(t)+\phi(t)=\pi$ (App.~\ref{app:wigner_d}). Explicitly, the amplitudes read $\psi^{(\nu)}_{q,j}(t)=e^{-itE_\nu}\sum_{n}\mathcal D^{(j)}_{qn}\,c^{(\nu)}_{n,j}$,
with $\mathcal D^{(j)}_{qn}=\langle j,q|R(t)|j,n\rangle$ (see App.~\ref{app:wigner_d} for the analytical details). The many-body details thus appear only through the coefficients
$c^{(\nu)}_{q,j}$, while the time dependence is set by the Euler angles
$\theta(t)$ and $\phi(t)$. By this, the spread complexity takes the closed form
(App.~\ref{app:su2_complexity_derivation})
\begin{eqnarray}
C_K(t)
=\bar J+\cos\theta(t)\,M_0
+\sin\theta(t)\,{\rm Re}\!\left[e^{i\phi(t)}M_+\right],
\label{eq:K_su2_superposition_main}
\end{eqnarray}
with the auxiliary terms given by
\begin{eqnarray}
\bar J&=&\sum_{j,q}j\,|c^{(\nu)}_{q,j}|^2,
\qquad
M_0=\sum_{j,q}(-j+q)\,|c^{(\nu)}_{q,j}|^2,\nonumber\\
M_+&=&\sum_{j}\sum_{q=0}^{2j-1}
\sqrt{(q+1)(2j-q)}\,(c^{(\nu)}_{q+1,j})^*c^{(\nu)}_{q,j} .
\label{eq:su2_moments_main}
\end{eqnarray}
Similarly to the scale-invariant case, contributions from different towers are additive, while within each tower initial superpositions appear only via the adjacent-level coherence $M_+$. For a single tower state $|j,q\rangle$, $M_+$ vanishes, yielding a simplified expression,
\begin{eqnarray}
C_K^{(j,q)}(t)=q+(j-q)\bigl[1-\cos\theta(t)\bigr],
\label{eq:K_su2_single_main}
\end{eqnarray}
bounded by $2j$. As a simple example, consider the single driven spin-$\tfrac12$ initiated in
its lowest-weight state, $j=\tfrac12$ and $q=0$. The exact formulation gives the spread complexity $C_K=\tfrac12[1-\cos\theta(t)]$. Despite the simplicity, the structure already shows the key steps towards the general framework for rotation-invariant many-body systems.

As in the scale-invariant case, we consider the thermal state within a single total-spin sector $j$, built from eigenstates of $H(0)$ that coincide with its tower states,
\begin{eqnarray}
\rho_\beta^{(j)}(0)
=
Z_j^{-1}\sum_{q=0}^{2j}e^{-\beta E_{j,q}}\,|j,q\rangle\langle j,q| ,
\,\,
Z_j=\sum_{q=0}^{2j}e^{-\beta E_{j,q}} .\quad
\label{eq:su2_thermal_state}
\end{eqnarray}
In Liouville space, the rotation is expressed by the adjoint
generators $\mathcal X_a=J_a\otimes\mathbb I-\mathbb I\otimes
J_a^T$, $a\in\{0,+,-\}$, which obey the $\mathfrak{su}(2)$
commutation relations. Their irreducible representations are
finite towers labeled by $L=0,\dots,2j$ with Casimir
eigenvalues $L(L+1)$. The thermal state decomposes into these
towers as (see App.~\ref{app:su2_density_CG} for details)
\begin{eqnarray}
|\rho_\beta^{(j)}(0))
=
\sum_{L=0}^{2j}
\rho^{(\beta)}_{0,L}(0)\,|L,0)) ,
\quad
\rho^{(\beta)}_{0,L}(0)=((L,0|\rho_\beta^{(j)}(0)),\nonumber\\
\label{eq:su2_rho_decomposition}
\end{eqnarray}
with the dynamics starting from the centers of the chains, $M=0$, with amplitudes $\rho^{(\beta)}_{0,L}(0)$. The Krylov variance, Eq.~\eqref{eq:CK_density_sum}, then takes the form
\begin{eqnarray}
M_2[\rho_\beta^{(j)}](t)
=
\tfrac12\big\langle L(L+1)\big\rangle_\beta\,\sin^2\theta(t).
\label{eq:su2_density_M2_general}
\end{eqnarray}
The thermal average is taken over the initial weights
(App.~\ref{app:su2_density_variance}),
\begin{eqnarray}
\big\langle L(L+1)\big\rangle_\beta
=
\frac{\sum_{L=0}^{2j}L(L+1)\,|\rho^{(\beta)}_{0,L}(0)|^2}
{\sum_{L=0}^{2j}|\rho^{(\beta)}_{0,L}(0)|^2}.
\nonumber
\end{eqnarray}
Thus, the Krylov variance inherits the dependence from the spread complexity in terms of the rotation angle $\theta(t)$, while the finite-temperature effects appear only in a separate multiplicative factor. The Krylov variance can therefore be
measured directly, since the rotation angle of the collective
spin can be read out by several techniques, for instance, by fluorescence detection in trapped-ion crystals~\cite{Monroe_RMP_2021_IonSpinSim,Britton_Nature_2012_EngineeredIsing,Bohnet_Science_2016_QuantumSpinDynamics,Garttner_NatPhys_2017_OTOC} or by dispersive probing of atomic ensembles in cavity QED~\cite{Hosten_Nature_2016_MeasurementNoise,LewisSwan2021BCSCavity,Young2024BCSCavity}.

\subsection{Central spin model with rotating staggered field}
\label{sec:central_spin}

First, we consider a central spin model with a staggered
magnetic field. As the initial Hamiltonian is not a linear
combination of the $\mathfrak{su}(2)$ generators, its
eigenstates must be determined separately to fix the initial
amplitudes on the Krylov chains. The model describes
a single spin coupled to an interacting spin bath, widely used to study spin decoherence in solid-state systems such as NV centers, quantum dots, and donor
qubits~\cite{Gaudin_JPhys_1976_SpinHamiltonians, Breuer_PRB_2004_SpinStar,
Bortz_PRB_2007_InhomogeneousCentralSpin, Childress_Science_2006_CoherentDynamics, Bluhm_NatPhys_2011_DephasingTime, Pla_Nature_2012_SiliconSpinQubit},
\begin{eqnarray}
H(0)=H_B+J\,\mathbf S_B\!\cdot\!\mathbf s_0
+\delta\,\mathbf n_0\!\cdot\!(\mathbf S_B-\mathbf s_0),
\label{eq:central_spin_H0}
\end{eqnarray}
with bath Hamiltonian $H_B=\sum_{i<j\in B}K_{ij}\,\mathbf s_i\!\cdot\!\mathbf s_j$,
bath spin $\mathbf S_B=\sum_{i\in B}\mathbf s_i$, and the axis chosen as
$\mathbf n_0=\hat{\mathbf z}$. The bath couplings $K_{ij}$ are arbitrary and the
central-bath exchange is globally rotation-invariant. The last term provides the staggered coupling to the difference $\mathbf S_B-\mathbf s_0$, acting on
the bath and the central spin with opposite signs along the axis $\mathbf n_0$.
A uniform field along $\mathbf n_0$ would couple to the conserved total
magnetization and simply add a phase, while the staggered field is the minimal
term that mixes the two total-spin sectors and breaks the algebraic closure of the initial Hamiltonian. We employ the global rotation $R(t)=e^{-i\theta(t)J_y}$ around $\hat{\mathbf y}$, which leaves $H_B$ and $\mathbf S_B\cdot\mathbf s_0$ invariant and rotates the staggered axis $\mathbf n_0=\hat{\mathbf z}$. The instantaneous Hamiltonian then takes the form of Eq.~\eqref{eq:su2_Ht_main}, with the staggered axis rotating as
$\mathbf n(t)=(\sin\theta(t),0,\cos\theta(t))$ and the collective field
$\dot\theta(t)\hat{\mathbf y}$ satisfying the consistency conditions, Eq.~\eqref{eq:su2_consistency}.

The time evolution is initiated in an eigenstate of
$H(0)$ that is also an eigenstate of $H_B$, $\mathbf S_B^2$,
and $J_z$, labeled by the bath spin $S$ and the total
magnetization $\mathcal M$. Coupling the central spin-$\tfrac12$ to the bath spin $S$ produces exactly
two total-spin sectors, $j_\pm=S\pm\tfrac12$, on which the central exchange is
diagonal and which get mixed by the staggered field~\cite{Edmonds_1957_AngularMomentum,Varshalovich_1988_QuantumTheoryAngularMomentum}.
At fixed magnetization $\mathcal M$, the staggered field mixes only the tower states $|j_+,q_+\rangle$ and $|j_-,q_-\rangle$ with $q_\pm=j_\pm+\mathcal M$, as $J_0|j,q\rangle=(-j+q)|j,q\rangle$ by Eq.~\eqref{eq:tower_general}. As a result, their superposition becomes an exact eigenstate (see App.~\ref{app:central_spin}),
\begin{eqnarray}
|\Psi_+(0)\rangle
=
\cos\alpha_{\mathcal M}\,|j_+,q_+\rangle
+
\sin\alpha_{\mathcal M}\,|j_-,q_-\rangle,
\label{eq:central_spin_initial}
\end{eqnarray}
with mixing angle~\cite{Schliemann_JPCM_2003_ElectronSpinDynamics,Coish_PRB_2004_HyperfineQuantumDot,Merkulov_PRB_2002_ElectronSpinRelaxation}
\begin{eqnarray}
\tan(2\alpha_{\mathcal M})
&=&
\frac{
\displaystyle
\frac{4\delta}{2S+1}
\sqrt{\left(S+\tfrac12\right)^2-\mathcal M^2}
}{
\displaystyle
\frac{J}{2}(2S+1)-\frac{4\delta\mathcal M}{2S+1}
},
\label{eq:central_spin_alpha}
\end{eqnarray}
providing the overlaps with the initial tower states, i.e., the initial conditions for the multidimensional Krylov dynamics.
The state thus occupies the two orthogonal tower states
$|j_\pm,q_\pm\rangle$ with amplitudes $\cos\alpha_{\mathcal M}$
and $\sin\alpha_{\mathcal M}$. The upper-state branch is fixed by $\alpha_{\mathcal M}\to0$ as $\delta\to0$
(App.~\ref{app:central_spin}). With $M_+=0$ and $M_0=\mathcal M$, as there are no initial coherences, the spread complexity from
Eq.~\eqref{eq:K_su2_superposition_main} gives
\begin{eqnarray}
C_K(t)
=
\cos^2\!\alpha_{\mathcal M}\,j_+
+
\sin^2\!\alpha_{\mathcal M}\,j_-
+
\mathcal M\cos\theta(t).
\label{eq:central_spin_K}
\end{eqnarray}
The interacting bath thus affects the spread complexity
only through the bath spin $S$ and the mixing angle
$\alpha_{\mathcal M}$, while the time dependence is fully controlled by
the rotation angle $\theta(t)$, which makes the prediction
directly testable in experiments.

For illustration, we consider a static transverse coupling with a
periodically modulated longitudinal field,
$\bm\Omega(t)=\Omega_x\hat{\mathbf x}+\Omega_1\cos(\omega_d t)
\hat{\mathbf z}$, which is realized, for instance, by periodically driving the electron spin of an NV center in diamond~\cite{Shu_PRL_2018_FloquetRaman}. The staggered axis then precesses as
$\mathbf n(t)=\mathcal R(t)\hat{\mathbf z}$ with
$\mathbf n(0)=\hat{\mathbf z}$, where $\mathcal R(t)$ is the
$\mathrm{SO}(3)$ rotation matrix associated with $R(t)$ of
Eq.~\eqref{eq:su2_angular_velocity_main}, so that
$\dot{\mathbf n}=\bm\Omega(t)\times\mathbf n$. Going to the rotating frame with $\Phi(t)=\int_0^t\mathrm dt'\,\Omega_1\cos(\omega_dt')=\Omega_1\sin(\omega_dt)/\omega_d$ around $\hat{\mathbf z}$ removes the $J_z$ part and leaves the spread complexity unchanged. The corresponding rotated Hamiltonian reads
\begin{eqnarray}
    H'(t)=\frac{\Omega_x}{2}\left(J_+e^{-i\frac{\Omega_1}{\omega_d}\sin(\omega_d t)}+{\rm h.c.}\right).
\end{eqnarray}
Expanding it according to the Jacobi-Anger formula \cite{AbramowitzStegun_1964_Handbook,Watson_1944_BesselFunctions}, $e^{-ix\sin(\omega_d t)}=\sum_{\ell\in\mathbb Z}\mathrm J_\ell(x)e^{-i\ell\omega_d t}$, where $\mathrm J_\ell$ is the Bessel function of order $\ell$~\cite{NIST_DLMF}, only the nonoscillating $\ell=0$ term survives to leading order in the fast-driving limit of $\omega_d\gg\Omega_x$, which gives $H'\approx \Omega_x\mathrm J_0\left(\Omega_1/\omega_d\right)J_x\equiv \bm\Omega'\cdot\bm J$. This Hamiltonian describes a rotation about the fixed axis $\hat{\mathbf x}$ at the effective frequency $\Omega_x\mathrm J_0(\Omega_1/\omega_d)$. The corresponding rotation angle is thus $\theta(t)\approx\Omega_x \mathrm J_0\left(\Omega_1/\omega_d\right)t$, and the spread complexity follows from the general formula,
\begin{eqnarray}
C_K(t)\!=\!\bar J+\mathcal M\cos\theta(t),
\label{eq:central_spin_floquet_CK}
\end{eqnarray}
with $\bar J=\cos^2\!\alpha_{\mathcal M}\,j_++\sin^2\!\alpha_{\mathcal M}\,j_-$, oscillating between $\bar J\pm\mathcal M$. When the ratio $\Omega_1/\omega_d$ coincides with a zero of $\mathrm J_0$, the effective frequency vanishes, $\mathbf n(t)$ remains along $\hat{\mathbf z}$, and the complexity stays at its initial value, which is the spin analogue of the coherent destruction of tunneling~\cite{Grossmann_PRL_1991_CoherentDestruction, Lignier_PRL_2007_DynamicalControlTunneling, Kierig_PRL_2008_DrivenDoubleWell, Eckardt_RMP_2017_PeriodicallyDriven}.

\begin{figure}
    \includegraphics[width=\columnwidth]{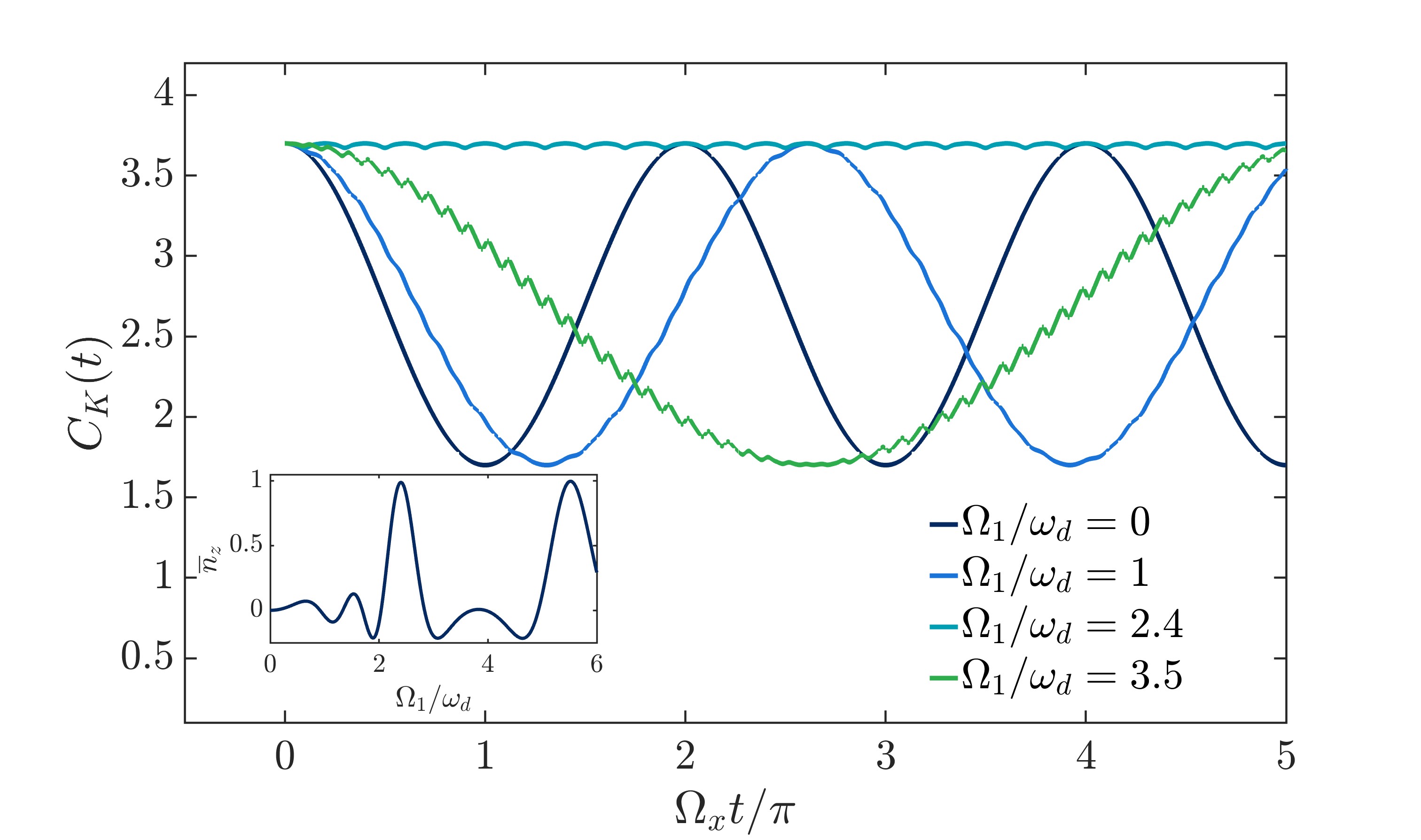}
    \caption{Spread complexity of the central spin model under a periodically modulated longitudinal field with $\Omega_x=0.1\,\omega_d$, for bath spin $S=5/2$ ($j_\pm=3,2$), magnetization $\mathcal M=1$, and mixing $\cos^2\alpha_{\mathcal M}=0.7$ and so $\bar J=2.7$. Main panel: the complexity for driving ratios $\Omega_1/\omega_d=0,\,1,\,2.4,\,3.5$. For $\Omega_1/\omega_d=0$ and $1$, the complexity oscillates over its full
range $\bar J\pm\mathcal M$ at the frequency $\Omega_x\mathrm
J_0(\Omega_1/\omega_d)$, which is largest in the undriven case. At $\Omega_1/\omega_d=2.4$, near the first zero of $\mathrm J_0$, the oscillations are suppressed up to small higher-order corrections. Inset: the time-averaged longitudinal projection $\overline{n}_z$ versus $\Omega_1/\omega_d$, approaching unity at the zeros of $\mathrm J_0$, $\Omega_1/\omega_d\approx2.4,\,5.5$, where the staggered axis remains near $\hat{\mathbf z}$.}
\label{fig:central_spin_floquet}
\end{figure}

Figure~\ref{fig:central_spin_floquet} shows $C_K(t)$ for different values of the ratio $\Omega_1/\omega_d$. For generic values, the complexity oscillates over its full range $\bar J\pm\mathcal M$ with the effective frequency $\Omega_x\mathrm J_0(\Omega_1/\omega_d)$. The
oscillation is the fastest in the undriven case and slows down with decreasing
$|\mathrm J_0(\Omega_1/\omega_d)|$. The higher-order terms of the Jacobi-Anger expansion
add small oscillations at the driving frequency $\omega_d$. Close to the first zero of $\mathrm J_0$, at $\Omega_1/\omega_d\approx2.4$, the effective frequency vanishes, and the oscillation is suppressed. The inset shows the time-averaged $\overline{n}_z$, which is close to unity at the zeros of $\mathrm J_0$, where $\mathbf n(t)$ remains along $\hat{\mathbf z}$, and close to zero elsewhere, where $n_z(t)$ averages out over the precession of $\mathbf n(t)$.

\subsection{Admissible compact Hamiltonians with finite closure}
\label{sec:su2_hw_examples}

Next, we consider the case where $H(0)$ is a linear combination of the
$\mathfrak{su}(2)$ generators, i.e., its eigenstates are spin tower states. Similarly to the scale-invariant case, we extend the
framework by a Heisenberg-Weyl sector formed by $D$
bosonic Schwinger doublets $r=1,\dots,D$, with
a spin-up and a spin-down mode and the corresponding annihilation operators
$b_{r,\uparrow}$ and $b_{r,\downarrow}$. Together
with the collective spin, they generate the semidirect sum
$\mathfrak{su}(2)\ltimes\mathfrak h_3(2D)$, as the collective spin also
rotates the doublets. Each doublet realizes the spin
generators through the Schwinger boson representation, $S_+^{(b)}=b_\uparrow^\dagger
b_\downarrow$, $S_-^{(b)}=b_\downarrow^\dagger b_\uparrow$, and
$S_0^{(b)}=\tfrac12(b_\uparrow^\dagger b_\uparrow-b_\downarrow^\dagger
b_\downarrow)$, mapping two bosonic modes onto one
spin~\cite{Schwinger_1965_AngularMomentum, Auerbach_1994_InteractingElectrons}.  In the absence of the linear terms, the total occupation
$N_b=b_\uparrow^\dagger b_\uparrow+b_\downarrow^\dagger
b_\downarrow$ of each doublet is conserved, which fixes the total spin, $j_b=N_b/2$.
Further allowing for an $\mathrm{SU}(2)$ rotation-invariant term, the general algebraically closed Hamiltonian reads
\begin{eqnarray}
H(t)
&=&A(t)J_0+B(t)J_++B^*(t)J_-
+g_{\rm H}(t)\!\sum_{i<j}\!\mathbf s_i\!\cdot\!\mathbf s_j
\nonumber\\
&&+H_{\rm sp}+\sum_{r=1}^D
\!\left[\mathbf b_{r}^\dagger\!\cdot\!\boldsymbol\Lambda_{r}(t)
+\boldsymbol\Lambda_{r}^\dagger(t)\!\cdot\!\mathbf b_{r}\right],
\label{eq:su2_admissible_H}
\end{eqnarray}
where $A(t),B(t)$ are the collective fields, $g_{\rm H}(t)$ the Heisenberg coupling, $H_{\rm sp}$ a general $\mathrm{SU}(2)$-invariant term, and $\mathbf b_{r}=(b_{r,\uparrow},b_{r,\downarrow})^T$ collects the
annihilation operators of the doublet $r$. The linear
doublet coefficients $\boldsymbol\Lambda_{r}(t)$ are given in the same two-component vector form. The collective generators
$J_a=\sum_i s_a^{(i)}+\sum_{r} S_{a,r}^{(b)}$ act at once on the physical
spins and the Schwinger doublets. The Heisenberg term acts only on the spins,
and $H_{\rm sp}$ commutes with every $J_a$.

The lowest-weight state is the product of the $\lvert j,-j\rangle$ spin state and the $\mathfrak{h}_3$ ground state,
$|\Psi_0\rangle=|j,-j\rangle\otimes|\mathbf 0\rangle_b$. On this state, the Heisenberg coupling and the invariant term $H_{\rm sp}$ contribute
only a phase set by their energy $E_j(t)$ in the total-spin sector $j$, while the remaining evolution is generated by the collective spin $\mathrm{SU}(2)$ rotations according to the general construction in
Sec.~\ref{sec:rot_invariant}. With the same notation for the polar angle $\theta(t)$, the
spin-sector contribution follows directly from Eq.~\eqref{eq:K_su2_single_main}.

The collective rotation also acts on the doublets. In
interaction picture with respect to $R(t)$, the corresponding part of the
Hamiltonian is linear in the Heisenberg-Weyl ladder operators, which generate the Krylov dynamics of the doublet sector. Since each doublet transforms as spin-$\tfrac12$, the rotation acts on the two components of $\mathbf b_{r}$ through its spin-$\tfrac12$ Wigner matrix $R_s(t)$, with $R^\dagger\mathbf b_rR=R_s\mathbf b_r$, which for a rotation axis in the $xy$ plane reads (see details in App.~\ref{app:R_s})
\begin{eqnarray}
R_s(t)=\begin{pmatrix}
\cos\tfrac{\theta}{2} & e^{i\phi}\sin\tfrac{\theta}{2}\\[2pt]
-e^{-i\phi}\sin\tfrac{\theta}{2} & \cos\tfrac{\theta}{2}
\end{pmatrix},
\label{eq:Rs_main}
\end{eqnarray}
by which the interaction-picture part of the Hamiltonian becomes
\begin{eqnarray}
H_{\rm HW}^{I}(t)
=
\sum_{r=1}^{D}
\left[
\mathbf b_{r}^\dagger\!\cdot\!
R_s^\dagger(t)\boldsymbol\Lambda_{r}(t)
+
\boldsymbol\Lambda_{r}^\dagger(t)R_s(t)\!\cdot\!\mathbf b_{r}
\right].
\label{eq:su2_hw_interaction}
\end{eqnarray}
The time-evolved state therefore factorizes into a
rotated spin-tower state and a coherent state of the $2D$
doublet modes, in complete analogy with the relative and center-of-mass split of Sec.~\ref{sec:su11_hw_examples}. Combining Eq.~\eqref{eq:K_su2_single_main} with the Heisenberg-Weyl displacement gives~\cite{Grabarits_arXiv_2026_TDKrylovLie}
\begin{eqnarray}
C_K(t)
=
j\bigl[1-\cos\theta(t)\bigr]
+
\sum_{r=1}^D
|\boldsymbol\alpha_r(t)|^2,
\label{eq:su2_hw_total_K}
\end{eqnarray}
with $\boldsymbol\alpha_r(t)
=
-i\int_0^t dt'\,
R_s^\dagger(t')\boldsymbol\Lambda_{r}(t')$ (for details, see Apps.~\ref{app:HW_su11_dressing} and~\ref{app:R_s}). Note that the doublet contributions depend on the
rotation angles through $R_s^\dagger(t')$ and on the original coefficients $\boldsymbol{\Lambda}_{r}(t')$ inside the time
integral. In the following, we consider two examples without the doublet displacement.
The driven two-mode doublet illustrates the spread complexity, and the
Haldane-Shastry exchange illustrates the Krylov variance of a spin-tower density
matrix.

\subsection{Driven two-mode doublet}
\label{sec:two_mode}

First, we consider a single Schwinger doublet that can be realized in experiments with a two-mode
interferometer, for instance, in a
directional coupler or bosonic Josephson junction~\cite{Harris_NatPhoton_2017_ProgrammableNanophotonic,
Bogaerts_Nature_2020_ProgrammablePhotonicCircuits,
Albiez_PRL_2005_BosonicJosephson, Kierig_PRL_2008_DrivenDoubleWell}.  Writing the doublet as $\mathbf b=(b_\uparrow,b_\downarrow)^T$, where
$b_\uparrow$ and $b_\downarrow$ are the annihilation operators of the two modes with occupations $n_1=b_\uparrow^\dagger b_\uparrow$ and $n_2=b_\downarrow^\dagger b_\downarrow$, the $\mathfrak{su}(2)$ generators read
\begin{eqnarray}
J_0=\tfrac12(b_\uparrow^\dagger b_\uparrow-b_\downarrow^\dagger b_\downarrow),\quad
J_+=b_\uparrow^\dagger b_\downarrow,\quad
J_-=b_\downarrow^\dagger b_\uparrow,
\label{eq:two_mode_schwinger}
\end{eqnarray}
which is the single-doublet case, $D=1$, of Eq.~\eqref{eq:su2_admissible_H}
without the linear terms. We consider a periodically modulated real coupling,
$B(t)=\kappa(t)=\kappa_0+\kappa_1\cos(\Omega_m t)$, with
vanishing longitudinal field, $A(t)=0$, which can be realized
experimentally by a beam-splitter coupling of the two modes,
\begin{eqnarray}
H(t)=\kappa(t)\,b_\uparrow^\dagger b_\downarrow+\kappa^*(t)\,b_\downarrow^\dagger b_\uparrow.
\label{eq:two_mode_H}
\end{eqnarray}

The total occupation $N=n_1+n_2$ commutes with $H(t)$ and fixes the total spin
$j=N/2$. Since $J_0=n_1-j$, the tower state $|j,q\rangle$ is the
Fock state $|n_1=q,\,n_2=N-q\rangle$. The occupation of the first mode thus labels the Krylov states, and the spread
complexity $C_K(t)=\langle n_1(t)\rangle$ follows from the measured population
imbalance, $\langle n_1-n_2\rangle=2(C_K-j)$.

Choosing $\kappa(t)$ real and $J_x=\tfrac12(J_++J_-)$, the beam-splitter Hamiltonian
Eq.~\eqref{eq:two_mode_H} is that of a collective spin in a field along $\hat{\mathbf x}$,
\begin{eqnarray}
H(t)=\kappa(t)\,(J_++J_-)=2\kappa(t)\,J_x,
\quad
H(0)=2\kappa_0J_x .
\label{eq:two_mode_Jx}
\end{eqnarray}
The corresponding spin rotation is given by $R(t)=e^{-i\Theta_R(t)J_x}$. As $R(t)$ commutes with $H(0)$, Eq.~\eqref{eq:Ht_frame_general} reduces to $H(t)=H(0)+i\dot R(t)R^\dagger(t)$ with rotation angle $\Theta_R(t)=(2\kappa_1/\Omega_m)\sin(\Omega_m t)$. The general exact solution in   Eq.~\eqref{eq:Psi_exact_su2_main} combines the two commuting factors into a single rotation by the total angle,
\begin{eqnarray}
|\Psi(t)\rangle&=&R(t)\,e^{-itH(0)}|\Psi(0)\rangle
=e^{-i\Theta(t)J_x}|\Psi(0)\rangle,
\nonumber\\
\Theta(t)
&=&2\kappa_0 t+\frac{2\kappa_1}{\Omega_m}\sin(\Omega_m t) .
\label{eq:two_mode_theta_explicit}
\end{eqnarray}
For a single tower state $|j,q\rangle$, with $M_+=0$, $M_0=q-j$, and
$\theta\to\Theta$ the complexity from Eq.~\eqref{eq:K_su2_single_main} reads
\begin{eqnarray}
C_K(t)=j+(q-j)\cos\Theta(t),
\qquad
C_K\le 2j=N,
\label{eq:two_mode_CK}
\end{eqnarray}
with the oscillations encoded in the angle $\Theta(t)$. To get more insight into the analytical properties of the spread complexity, we again use the Jacobi-Anger expansion,
\begin{eqnarray}
\cos\Theta(t)=\sum_{\ell\in\mathbb Z}
\mathrm J_\ell\!\left(\frac{2\kappa_1}{\Omega_m}\right)
\cos\!\big[(2\kappa_0+\ell\Omega_m)t\big],
\label{eq:two_mode_comb}
\end{eqnarray}
with $\mathrm J_\ell$ denoting the Bessel function of $\ell$-th order \cite{NIST_DLMF}. The complexity is therefore a superposition of
oscillations at the frequencies $2\kappa_0+\ell\Omega_m$, i.e., the static frequency $2\kappa_0$ shifted by integer
multiples of $\Omega_m$, with amplitudes
$\mathrm J_\ell(2\kappa_1/\Omega_m)$. The time average of each term is zero unless its frequency vanishes, which requires $2\kappa_0=n\Omega_m$ with $n=-\ell$ a nonnegative integer. To better capture these features, we consider the time-averaged spread complexity, where only the zero-frequency terms survive,
\begin{eqnarray}
\overline{C}_K-j=(q-j)\,(-1)^n\,\mathrm J_n\!\left(
\frac{2\kappa_1}{\Omega_m}\right).
\label{eq:two_mode_offset}
\end{eqnarray}
In experiments, this time-averaged complexity is directly accessible as the steady offset of the difference between the occupation numbers of the two modes, and it vanishes at the zeros of $\mathrm J_n$. The resonance conditions $2\kappa_0=n\Omega_m$ correspond to photon-assisted resonances
of driven two-mode systems~\cite{Sias_PRL_2008_PhotonAssisted,Eckardt_RMP_2017_PeriodicallyDriven}. The two modes
absorb or emit energy from the modulation in quanta of $\Omega_m$, and a resonance
occurs when an integer number of such quanta matches the
energy splitting $2\kappa_0$.

\begin{figure}
    \includegraphics[width=\columnwidth]{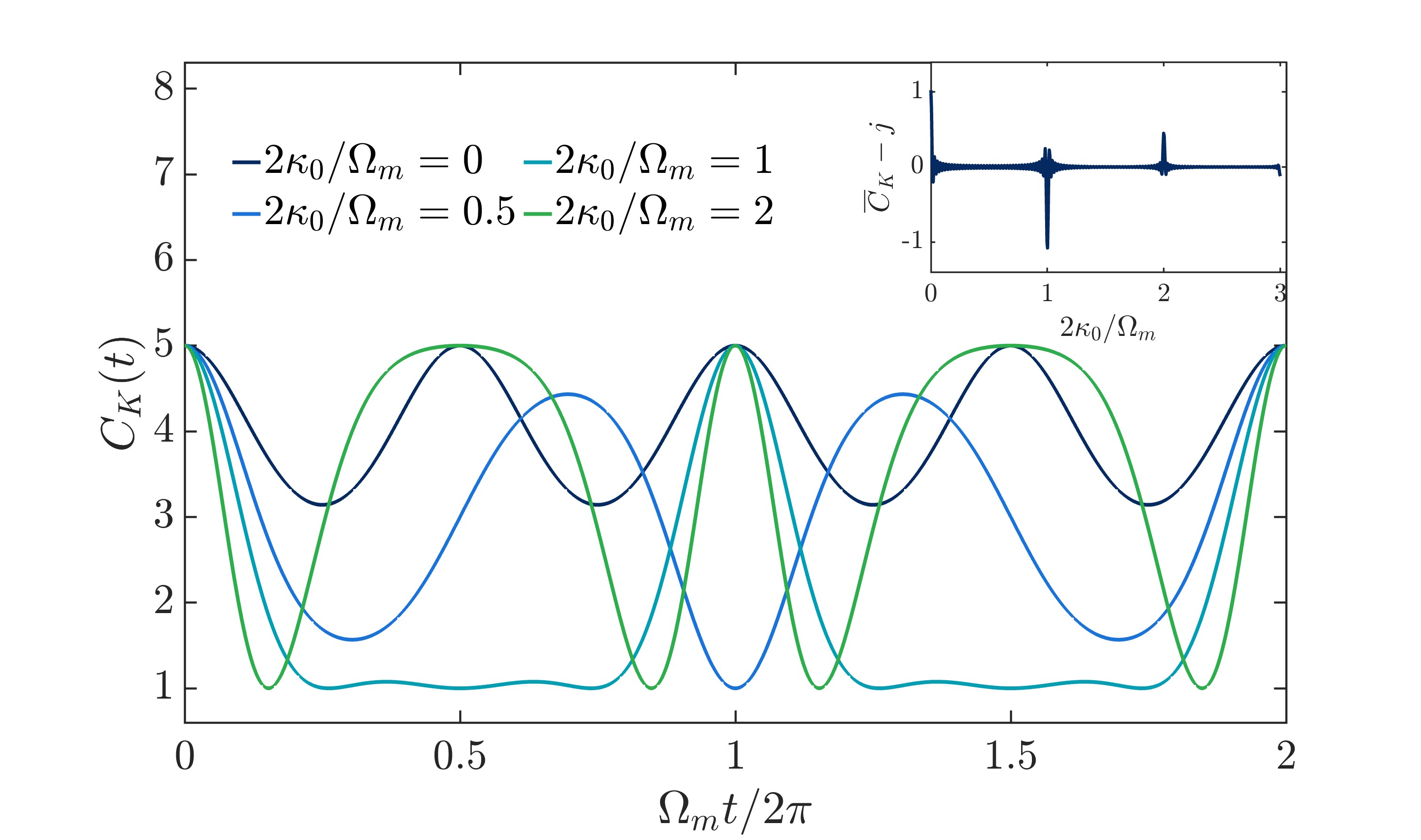}
    \caption{Spread complexity of the driven two-mode doublet for an initial occupation $|n_1,n_2\rangle=|5,1\rangle$ ($N=6$, $j=3$, $q=5$) with $2\kappa_1/\Omega_m=1.5$. Main panel: the complexity at the resonances $2\kappa_0/\Omega_m=0,\,1,\,2$ and off resonance at $2\kappa_0/\Omega_m=0.5$. Away from the resonances, it oscillates around $j$ with vanishing mean. At each resonance, it acquires a nonzero mean set by the corresponding Bessel function. Inset: the time-averaged complexity $\overline{C}_K-j$ as a function of the resonance parameter $2\kappa_0/\Omega_m$, tracing the corresponding
resonances.}
\label{fig:two_mode_doublet}
\end{figure}

Figure~\ref{fig:two_mode_doublet} shows the complexity during the driving. Away from the resonances, $C_K$ oscillates around $j$, so that the time average of $C_K-j$ vanishes. At each resonance $2\kappa_0=n\Omega_m$, it acquires the steady offset $(q-j)(-1)^n\mathrm J_n(2\kappa_1/\Omega_m)$ of Eq.~\eqref{eq:two_mode_offset}, while the inset shows these offsets as the peaks of the time-averaged spread complexity versus the parameter $2\kappa_0/\Omega_m$. This imbalance between the occupations of the modes and its mean can be read out in directional couplers and bosonic Josephson
junctions, which allows for the direct experimental study of spread complexity in driven spin systems.

\subsection{Collective spin tower}
\label{sec:haldane_general}

Finally, we consider a rotation-invariant spin system with an additional $\mathrm{SU}(2)$-invariant term to demonstrate the density-matrix construction of
Sec.~\ref{sec:density_general} on a single collective tower. Spin systems with similar properties can be prepared and read out in cavity-QED
ensembles and trapped-ion crystals~\cite{Monroe_RMP_2021_IonSpinSim,
Garttner_NatPhys_2017_OTOC, Young2024BCSCavity,
Bohnet_Science_2016_QuantumSpinDynamics, Hosten_Nature_2016_MeasurementNoise, Grass_EPJQT_2014_TunableHeisenberg}.
 In particular, for the interactions we consider the Haldane-Shastry exchange~\cite{Haldane_PRL_1988_HaldaneShastry,
Shastry_PRL_1988_HaldaneShastry},
\begin{eqnarray}
H(t)
&=& A(t)J_0+B(t)J_+ +B^*(t)J_-
\nonumber\\
&&+g_{\rm HS}(t)\!\left(\!\tfrac{\pi}{N}\!\right)^{\!2}\!
\sum_{i<j}\frac{\mathbf s_i\!\cdot\!\mathbf s_j}{\sin^2[\pi(i-j)/N]},
\label{eq:HS_collective_H}
\end{eqnarray}
with collective fields $A(t),B(t)$, exchange strength $g_{\rm HS}(t)$, and total
spin $\mathbf J=\sum_i\mathbf s_i$.

We consider the maximal total-spin sector $j=Ns$, where $s$ is the spin of
each site, with the fully polarized state as its lowest-weight state. As the
exchange commutes with every $J_a$, it is constant on all
states $|j,q\rangle$ (with $q=0,\dots,2j$) of this tower and
contributes only to the global phase,
\begin{eqnarray}
\Phi_{\rm HS}(t)=\tfrac{\pi^2 s^2}{6}\!\left(\!N-\tfrac{1}{N}\!\right)\!
\int_0^t\!\!dt'\,g_{\rm HS}(t').
\label{eq:HS_phase}
\end{eqnarray}

For an initial thermal state diagonal in the spin-tower states, we consider
\begin{eqnarray}
\rho_\beta^{(j)}(0)=Z_j^{-1}\sum_{q=0}^{2j}Q^q|j,q\rangle\langle j,q|,
\qquad Q=e^{-\beta A_0},
\label{eq:HS_thermal}
\end{eqnarray}
with $A_0=A(0)$ and $B(0)=0$, so that $H(0)$ is diagonal
in the tower states. The exchange energy is the same for all
states of the tower, so it drops out of the normalized
thermal weights and does not affect the Krylov dynamics.
As in Sec.~\ref{sec:rot_invariant}, the thermal state decomposes into
the adjoint towers $L=0,\dots,2j$ according to
Eq.~\eqref{eq:su2_rho_decomposition}, where initially only the $M=0$ states are occupied in each tower.

For an explicit demonstration, we consider the same driving protocol as in Sec.~\ref{sec:central_spin}, $\bm\Omega(t)=\Omega_x\hat{\mathbf x}+\Omega_1\cos(\omega_d t)\hat{\mathbf z}$ in the fast-driving limit, $\omega_d\gg\Omega_x$. In the rotating frame around $\hat{\mathbf z}$ with $\Phi(t)=\int_0^t\mathrm dt'\Omega_1\cos(\omega_dt')$, the Hamiltonian describes a rotation about the fixed axis $\hat{\mathbf x}$ with the effective frequency $\Omega_x\mathrm J_0(\Omega_1/\omega_d)$ and the corresponding angle $\theta(t)\approx\Omega_x \mathrm J_0\left(\Omega_1/\omega_d\right)t$. From this, the Krylov variance follows from the general formula, Eq.~\eqref{eq:su2_density_M2_general} (for details see Apps.~\ref{app:su2_density_CG} and~\ref{app:su2_density_variance}),
\begin{eqnarray}\label{eq:HS_density_M2}
M_2[\rho_\beta^{(j)}](t)
&=&
P(Q,j)\,\sin^2\theta(t),\\
P(Q,j)&=&
j\,\frac{(1-Q^2)(1-Q)^2}{1-Q^{4j+2}}\,
{}_2F_1\!\left(2,1-2j;-2j;Q^2\right),\nonumber
\end{eqnarray}
with the thermal weight
$P(Q,j)=\tfrac12\langle L(L+1)\rangle_\beta$. Here,
${}_2F_1(a,b;c;x)=\sum_{n\ge0}\frac{(a)_n(b)_n}{(c)_n}\frac{x^n}{n!}$
is the hypergeometric function with the Pochhammer
symbol $(a)_n=a(a+1)\cdots(a+n-1)$, $(a)_0=1$~\cite{NIST_DLMF}. Since $1-2j$ is a nonpositive integer, the
series terminates, so $P(Q,j)$ is a finite polynomial in $Q^2$ of degree $2j-1$.
 Temperature again affects only the prefactor, while the entire time dependence is governed by the single rotation angle $\theta(t)$, with the Krylov variance oscillating between $0$ and $P(Q,j)$. When the ratio $\Omega_1/\omega_d$ coincides with a zero of $\mathrm J_0$, the effective frequency vanishes, the rotation angle $\theta(t)$ stays at
zero, and so does the Krylov variance.

\begin{figure}
    \includegraphics[width=\columnwidth]{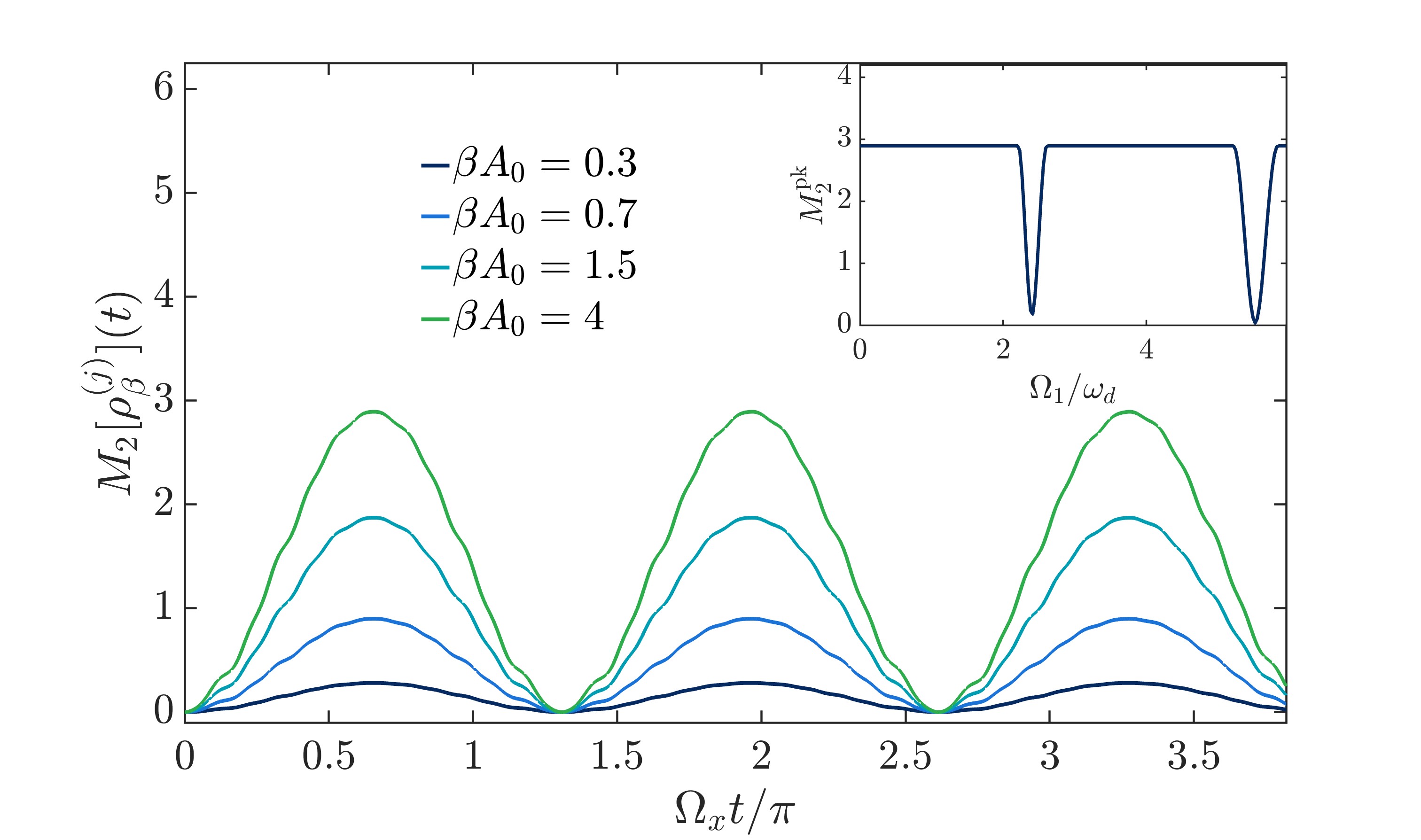}
    \caption{Krylov variance of the thermal state in the maximal total-spin sector of the
driven collective spin,
Eq.~\eqref{eq:HS_density_M2}, for $j=3$ ($N=6$ spin-$\tfrac12$) under the
modulated collective field of Sec.~\ref{sec:central_spin} at
$\Omega_x=0.1\,\omega_d$. Main panel: $M_2(t)$ at fixed drive
$\Omega_1/\omega_d=1$ for temperatures $\beta A_0=0.3,\,0.7,\,1.5,\,4$. All curves follow the same oscillatory character with their peaks increasing with $\beta A_0$. Inset: peak $M_2$ as a function of $\Omega_1/\omega_d$ at the largest $\beta$. Away from the zeros of $\mathrm J_0$, the peak is equal to the thermal weight, while at $\Omega_1/\omega_d\approx2.4,\,5.5$ the effective frequency vanishes.}
\label{fig:collective_thermal_M2}
\end{figure}

The main panel of Fig.~\ref{fig:collective_thermal_M2} shows the Krylov variance $M_2(t)$, oscillating between zero and its maximum $P(Q,j)$ at twice the effective frequency $\Omega_x\mathrm J_0(\Omega_1/\omega_d)$. Each curve follows the same rotation angle $\theta(t)$, with the variance increasing with $\beta$ through the weight $P(Q,j)$. The inset shows the peak of the Krylov variance as a function of the driving ratio $\Omega_1/\omega_d$. Away from the zeros of $\mathrm J_0$, the rotation angle
$\theta(t)$ reaches $\pi/2$ with the effective frequency
$\Omega_x\mathrm J_0(\Omega_1/\omega_d)$ with the peaks given by $P(Q,j)$. At the zeros of $\mathrm J_0$, $\Omega_1/\omega_d\approx2.4,\,5.5$, the effective frequency $\Omega_x\mathrm J_0(\Omega_1/\omega_d)$ vanishes, $\theta(t)$ stays at zero, and the peak of $M_2$ drops to zero.

\section{Conclusions}

In this work, we have established an exact framework for Krylov dynamics in driven interacting quantum many-body systems. The framework relies on driving protocols admitting scale or rotational invariance, which are also feasible for experimental realizations. Using the underlying algebraic structure, we have shown that for systems initiated in an eigenstate of $H(0)$, the time evolution is governed by a single transformation $\mathcal T(t)$ that can be expressed as a linear combination of the generators of the underlying $\mathfrak{su}(1,1)$ and $\mathfrak{su}(2)$ algebras for the scale- and rotation-invariant cases, respectively. The resulting dynamics unfolds in the multidimensional Krylov subspace along mutually orthogonal one-dimensional lattices. The same construction extends to initial mixed states characterized by a density matrix. For the vectorized density matrix, the transformation $\mathcal T(t)$ acts via the adjoint representation of the same algebras in Liouville space, and the density matrix evolves along orthogonal chains with the same driving coefficients as for quantum states. For thermal states, the dynamics starts at the center of each chain, and the spread is captured by the Krylov variance, in which the initial temperature appears only through a multiplicative factor.

In the scale-invariant case, the transformation is a
dilation, and the dynamics decomposes into the irreducible
representations of the $\mathfrak{su}(1,1)$ algebra selected
by the initial state. The complexity follows in closed form from the time-dependent scaling factor of the many-body system and its rate of change. Thus, the spread complexity can be read out directly from the expansion
width of trapped quantum gases using conventional imaging techniques, e.g., time-of-flight imaging. As we demonstrated with the Lieb-Liniger gas, the initial
Hamiltonian, unlike $\mathcal T(t)$, need not be a linear combination of the $\mathfrak{su}(1,1)$ generators. We further presented how the class of algebraically closed initial Hamiltonians with additional linear force and quadratic pairwise potentials can be extended with a Heisenberg-Weyl algebra, while still admitting an exact analytical expression for the spread complexity. This is demonstrated for the Moshinsky model, where the center-of-mass and relative expansions contribute two separately measurable scaling factors. For thermal states, the Krylov variance follows from the same scaling factor as the spread complexity, with the initial temperature appearing only as a multiplicative factor, as illustrated for the same models.

In the rotation-invariant class, the transformation is a global rotation, and the evolution decomposes into finite total-spin towers of the corresponding $\mathfrak{su}(2)$ algebra, spanning the multidimensional Krylov space. The complexity is bounded and oscillatory, determined by the polar and azimuthal angles of the collective spin rotations and the static occupations of each spin sector selected by the eigenstates of $H(0)$. These collective spin rotations can be tested in experiments via trapped-ion and cavity-QED platforms. First, the construction is demonstrated by the central spin model, where $H(0)$ cannot be expressed as a linear combination of the $\mathfrak{su}(2)$ generators. The opposite closed algebraic case is exemplified by the driven doublet and the Haldane-Shastry model. For thermal states of a collective spin tower, as realized in the Haldane-Shastry model, the Krylov variance is governed by the rotation angle alone, up to a prefactor given by half the Casimir average over the initial thermal weights, which interpolates between $j$ for low temperatures and zero in the infinite-temperature limit, where the maximally mixed tower is rotation invariant. The variance remains bounded, as the adjoint towers are finite, and under
periodic driving it oscillates at twice the effective frequency set by the Bessel function $\mathrm J_0$.

These results make the Krylov complexity an experimentally accessible
measure of complexity growth in driven quantum many-body systems for
arbitrary initial quantum states and density matrices, including thermal states, e.g., in cold-atom, trapped-ion, and cavity-QED experiments.

\acknowledgments
We acknowledge financial support from the Luxembourg National Research Fund under Grants No. C22/MS/17132060/BeyondKZM and C22/MS/17132054/AQCQNET. AG acknowledges financial support from the Luxembourg National Research Fund under Grant No. CC25/MS/19559370/FastQOPT.  AdC further acknowledges the Basque Government’s BERC and the Severo Ochoa Centre of Excellence Programmes (CEX2024-001491-S, funded by MICIU/AEI/10.13039/501100011033). 
This work was performed in part at the Aspen Center for Physics, which is supported by the National Science Foundation grant PHY-2210452. Claude (Anthropic) was used to assist with the checking of calculations and references and the editing of the text; the authors assume full responsibility for the content.

\appendix

\widetext

\section{Wei-Norman factorization for $\mathfrak{su}(1,1)$}
\label{app:wei_norman}

In this Appendix, we show the analytical details of the Wei-Norman factorization for the scale-invariant case together with the full form of the Krylov wave functions.

The transformation $\mathcal T=U_KS_\gamma$ with $U_K=e^{i(\dot\gamma/\gamma)K}$ and $S_\gamma=e^{-i(\ln\gamma)D}$ of Sec.~\ref{sec:scale_covariant} can be written in the disentangled form
\begin{eqnarray}
\mathcal T(t)=e^{z(t)K_+}e^{\eta(t)K_0}e^{w(t)K_-} ,
\label{eq:app_WN_general}
\end{eqnarray}
with three complex functions $z(t)$, $\eta(t)$ and $w(t)$ fixed by the
driving. We first derive the driving coefficients and then determine these
functions. The generator of $\mathcal T$ in Eq.~\eqref{eq: H_eff_su11} without the energy shift,  follows from the product rule,
\begin{eqnarray}
i\dot{\mathcal T}\mathcal T^\dagger
=
i\dot U_KU_K^\dagger+U_K\big(i\dot S_\gamma S_\gamma^\dagger\big)U_K^\dagger
=
-\left(\frac{\ddot\gamma}{\gamma}-\frac{\dot\gamma^2}{\gamma^2}\right)K+\frac{\dot\gamma}{\gamma}\,U_KDU_K^\dagger
=
\frac{\dot\gamma}{\gamma}D-\left(\frac{\ddot\gamma}{\gamma}+\frac{\dot\gamma^2}{\gamma^2}\right)K ,
\label{eq:app_su11_generator}
\end{eqnarray}
where we used $U_KDU_K^\dagger=D+i\frac{\dot\gamma}{\gamma}[K,D]=D-2\frac{\dot\gamma}{\gamma}K$ from the $\mathfrak{su}(1,1)$ commutation relations in Eq.~\eqref{eq:TKD_main}. From the Cartan-Weyl basis of Eq.~\eqref{eq:su11_CW_map}, with
$2K_0=\Omega_0K+T/\Omega_0$, $K_++K_-=\Omega_0K-T/\Omega_0$ and $K_+-K_-=-iD$, one
has $\Omega_0K=K_0+\tfrac12(K_++K_-)$ and $D=i(K_+-K_-)$ which from Eq.~\eqref{eq:app_su11_generator} gives
\begin{eqnarray}
A_0(t)=-\frac{1}{\Omega_0}\left(\frac{\ddot\gamma}{\gamma}+\frac{\dot\gamma^2}{\gamma^2}\right),\qquad
A_+(t)=-\frac{1}{2\Omega_0}\left(\frac{\ddot\gamma}{\gamma}+\frac{\dot\gamma^2}{\gamma^2}\right)+i\frac{\dot\gamma}{\gamma},\qquad
A_-(t)=A_+^*(t).
\label{eq:app_su11_drive_coeffs}
\end{eqnarray}
Inserting Eq.~\eqref{eq:app_WN_general} into Eq.~\eqref{eq:T_generator} and using the Baker-Campbell-Hausdorff expansion with the $\mathfrak{su}(1,1)$ commutation relations, $e^{zK_+}K_0e^{-zK_+}=K_0-zK_+$,
$e^{\eta K_0}K_-e^{-\eta K_0}=e^{-\eta}K_-$ and
$e^{zK_+}K_-e^{-zK_+}=K_--2zK_0+z^2K_+$, one recovers the Wei-Norman equations,
\begin{eqnarray}
i\dot w\,e^{-\eta}=A_-,\qquad
i\big(\dot\eta-2z\dot w\,e^{-\eta}\big)=A_0,\qquad
i\big(\dot z-z\dot\eta+z^2\dot w\,e^{-\eta}\big)=A_+ ,
\label{eq:app_WN_odes}
\end{eqnarray}
with $z(0)=\eta(0)=w(0)=0$. For the driving coefficients of
Eq.~\eqref{eq:app_su11_drive_coeffs}, the solution follows from the functions
\begin{eqnarray}
u=\frac12\left(\gamma+\frac1\gamma+i\frac{\dot\gamma}{\Omega_0}\right)
=\cosh\xi\,e^{i\phi_u},\qquad
v=\frac12\left(\gamma-\frac1\gamma+i\frac{\dot\gamma}{\Omega_0}\right)
=\sinh\xi\,e^{i\phi_v},
\label{eq:app_su11_uv}
\end{eqnarray}
which satisfy $|u|^2-|v|^2=1$, $u(0)=1$ and $v(0)=0$. The second equalities
define the phases $\phi_u(t)$ and $\phi_v(t)$ and the squeezing parameter
$\xi(t)$,
\begin{eqnarray}
\sinh^2\xi=|v|^2=\frac14\left[\Big(\gamma-\frac1\gamma\Big)^2
+\Big(\frac{\dot\gamma}{\Omega_0}\Big)^2\right].
\label{eq:app_su11_xi}
\end{eqnarray}
Substituting into Eq.~\eqref{eq:app_WN_odes} gives
\begin{eqnarray}
z=\frac{v}{u^*}=e^{i\varphi_z}\tanh\xi,\qquad
e^{\eta}=\frac{1}{(u^*)^2},\qquad
w=-\frac{v^*}{u^*},
\label{eq:app_su11_WN_params}
\end{eqnarray}
with $\varphi_z=\phi_u+\phi_v$. As $|u|\ge1$, the branch of
$\eta=-2\ln u^*$ is fixed by continuity from $\eta(0)=0$, which determines
$e^{k\eta}=(u^*)^{-2k}$ also for non-half-integer Bargmann indices.

The Krylov wave function follows by applying Eq.~\eqref{eq:app_WN_general} to the tower states. By the ladder coefficients of Eq.~\eqref{eq:su11_chain_main}, the repeated action of the raising and lowering operators gives $K_+^p|k,n\rangle=\sqrt{(n+p)!\,\Gamma(2k+n+p)/[n!\,\Gamma(2k+n)]}\,|k,n+p\rangle$ and $K_-^p|k,n\rangle=\sqrt{n!\,\Gamma(2k+n)/[(n-p)!\,\Gamma(2k+n-p)]}\,|k,n-p\rangle$. From this, the three exponentials of the Wei-Norman form in Eq.~\eqref{eq:app_WN_general} act as
\begin{eqnarray}
\langle k,l|e^{wK_-}|k,n\rangle=\frac{w^{n-l}}{(n-l)!}\sqrt{\frac{n!\,\Gamma(2k+n)}{l!\,\Gamma(2k+l)}},
\,
\langle k,l|e^{\eta K_0}|k,l\rangle=e^{\eta(k+l)},
\,
\langle k,m|e^{zK_+}|k,l\rangle=\frac{z^{m-l}}{(m-l)!}\sqrt{\frac{m!\,\Gamma(2k+m)}{l!\,\Gamma(2k+l)}},\quad
\label{eq:app_su11_factor_elements}
\end{eqnarray}
for $l\le n$ and $l\le m$, respectively. The propagator $\mathcal D^{(k)}_{mn}=\langle k,m|\mathcal T|k,n\rangle$ introduced in Sec.~\ref{sec:scale_covariant} and the Krylov wave function are given by
\begin{eqnarray}
\mathcal D^{(k)}_{mn}(t)
=
\sum_{l=0}^{\min(m,n)}
\frac{z^{m-l}\,e^{\eta(k+l)}\,w^{n-l}}{(m-l)!\,(n-l)!}\,
\frac{\sqrt{m!\,n!\,\Gamma(2k+m)\,\Gamma(2k+n)}}{l!\,\Gamma(2k+l)},\qquad \psi_{m,k}(t)=e^{-i\tau(t)E_\nu}\sum_{n\ge0}\mathcal D^{(k)}_{mn}(t)\,c_{n,k},
\label{eq:app_su11_Dmn_sum}
\end{eqnarray}
for given initial conditions $\psi_{n,k}(0)=c_{n,k}$.
 On the lowest-weight state, $n=0$, only $l=0$ contributes, and with $e^{k\eta}=(u^*)^{-2k}$ one finds
\begin{eqnarray}
\mathcal D^{(k)}_{m0}(t)=
\left[\frac{\Gamma(2k+m)}{m!\,\Gamma(2k)}\right]^{1/2}
\big(u^*\big)^{-2k}z^m
=
\left[\frac{\Gamma(2k+m)}{m!\,\Gamma(2k)}\right]^{1/2}
e^{2ik\phi_u}
(\cosh\xi)^{-2k}\,\big(e^{i\varphi_z}\tanh\xi\big)^{m} ,
\label{eq:app_su11_Mm0}
\end{eqnarray}
normalized by $\sum_{m\ge0}|z|^{2m}\Gamma(2k+m)/[m!\,\Gamma(2k)]=(1-|z|^2)^{-2k}$. For general $n$, the sum in Eq.~\eqref{eq:app_su11_Dmn_sum} is the series representation of the hypergeometric function ${}_2F_1(\dots)$, and in terms of the squeezing parameter and phase it takes the form
\begin{eqnarray}
\mathcal D^{(k)}_{mn}(t)=
\frac{1}{(m-n)!}\left[\frac{m!\,\Gamma(2k+m)}{n!\,\Gamma(2k+n)}\right]^{1/2}
\frac{(e^{i\vartheta}\tanh\xi)^{\,m-n}}{(\cosh\xi)^{2k}}\,
{}_2F_1\!\big(-n,\,2k+m;\,m-n+1;\,\tanh^2\xi\big),\quad m\ge n ,
\label{eq:app_su11_Mmn}
\end{eqnarray}
up to the overall phase $e^{2i\phi_u(k+n)}$.

\section{Derivation of spread complexity in the scale-invariant case}
\label{app:su11_complexity_derivation}

In this Appendix, we show the details of the derivation of Eq.~\eqref{eq:Ermakov_CK_superposition_main}, with the transformation ${\cal T}(t)=U_K(t)S_\gamma(t)$,
$U_K(t)=e^{i[\dot\gamma(t)/\gamma(t)]K}$ and
$S_\gamma(t)=e^{-i(\ln\gamma)D}$, whose driving coefficients are given in App.~\ref{app:wei_norman}.
The spread complexity of a time-evolved initial eigenstate,  
\begin{eqnarray}
|\Psi_\nu(t)\rangle
=
e^{-iE_\nu\tau(t)}{\cal T}(t)|\nu\rangle,\quad H(0)|\nu\rangle=E_\nu|\nu\rangle,
\label{eq:app_eigenstate_solution}
\end{eqnarray}
is defined as
\begin{eqnarray}
C_K^{(\nu)}(t)
=
\langle\nu|{\cal T}^\dagger(t)N_K{\cal T}(t)|\nu\rangle
=
\langle\nu|{\cal T}^\dagger(t)K_0{\cal T}(t)|\nu\rangle
-\langle\nu|\hat k|\nu\rangle,
\label{eq:app_CK_start}
\end{eqnarray}
with the complexity operator
$N_K|k,n\rangle=n|k,n\rangle,\,
N_K=K_0-\hat k$, where $\hat k|k,n\rangle=k|k,n\rangle,\,[\hat k,\mathcal T(t)]=0$ was introduced for convenience.

Using the $\mathfrak{su}(1,1)$ Cartan-Weyl representation,
\begin{eqnarray}
K_0
=
\frac12\left(\Omega_0K+\frac{T}{\Omega_0}\right),
\qquad
K_\pm
=
\frac12\left(\Omega_0K-\frac{T}{\Omega_0}\right)
\mp\frac{i}{2}D ,
\label{eq:app_su11_basis}
\end{eqnarray}
with $[K_0,K_\pm]=\pm K_\pm$ and $[K_+,K_-]=-2K_0$ and
 the conjugation relations with $U_K$ and $S_\gamma$,
\begin{eqnarray}
U_K^\dagger K U_K&=&K,\qquad
U_K^\dagger D U_K=D+2\frac{\dot\gamma}{\gamma}K,\qquad
U_K^\dagger T U_K=T+\frac{\dot\gamma}{\gamma}D+\frac{\dot\gamma^2}{\gamma^2}K,
\nonumber\\
S_\gamma^\dagger K S_\gamma&=&\gamma^2K,\qquad
S_\gamma^\dagger T S_\gamma=\frac{T}{\gamma^2},\qquad
S_\gamma^\dagger D S_\gamma=D,
\label{eq:app_US_actions}
\end{eqnarray}
one finds
\begin{eqnarray}
{\cal T}^\dagger K_0{\cal T}
&=&
\frac12
\left[
\left(\Omega_0\gamma^2+\frac{\dot\gamma^2}{\Omega_0}\right)K
+
\frac{T}{\Omega_0\gamma^2}
+
\frac{\dot\gamma}{\Omega_0\gamma}D
\right].
\label{eq:app_K0_intermediate}
\end{eqnarray}
Furthermore, from
\begin{eqnarray}
\Omega_0K=K_0+\frac12(K_++K_-),\qquad
\frac{T}{\Omega_0}=K_0-\frac12(K_++K_-),\qquad
D=i(K_+-K_-),
\label{eq:app_inverse_basis}
\end{eqnarray}
one obtains
\begin{eqnarray}
{\cal T}^\dagger K_0{\cal T}
=
a(t)K_0+b(t)K_++b^*(t)K_-,
\label{eq:app_K0_adjoint}
\end{eqnarray}
where the auxiliary variables read
\begin{eqnarray}
a(t)
=
\frac12
\left[
\gamma^2+\frac{1}{\gamma^2}
+
\left(\frac{\dot\gamma}{\Omega_0}\right)^2
\right],\quad
b(t)
=
\frac14
\left[
\gamma^2-\frac{1}{\gamma^2}
+
\left(\frac{\dot\gamma}{\Omega_0}\right)^2
\right]
+
\frac{i}{2}\frac{\dot\gamma}{\Omega_0\gamma}.
\label{eq:app_ab_coefficients}
\end{eqnarray}
First, we expand the initial eigenstate in the Bargmann tower states and define the following auxiliary functions of the given expansion coefficients,
\begin{eqnarray}
|\nu\rangle
&=&
\sum_{k,n}c_{n,k}^{(\nu)}|k,n\rangle,
\qquad
c_{n,k}^{(\nu)}=\langle k,n|\nu\rangle,\\
A_\nu
&=&
\sum_{k,n}|c_{n,k}^{(\nu)}|^2(k+n),
\quad
\bar k_\nu=\sum_{k,n}|c_{n,k}^{(\nu)}|^2k,\quad
B_\nu=\sum_{k,n\ge0}
\sqrt{(n+1)(2k+n)}
\left(c_{n+1,k}^{(\nu)}\right)^*
c_{n,k}^{(\nu)} .
\label{eq:app_AB_initial}
\end{eqnarray}
Using the action of the generators on the tower states,
$K_0|k,n\rangle=(k+n)|k,n\rangle,\quad
K_+|k,n\rangle
=
\sqrt{(n+1)(2k+n)}\,|k,n+1\rangle,\quad
K_-|k,n\rangle
=
\sqrt{n(2k+n-1)}\,|k,n-1\rangle$, the auxiliary functions are the expectation values
\begin{eqnarray}
\langle\nu|K_0|\nu\rangle=A_\nu,\qquad
\langle\nu|\hat k|\nu\rangle=\bar k_\nu,\qquad
\langle\nu|K_+|\nu\rangle=B_\nu,\qquad
\langle\nu|K_-|\nu\rangle=B_\nu^* .
\label{eq:app_expectations_initial}
\end{eqnarray}
By Eq.~\eqref{eq:app_K0_adjoint}, $\langle\nu|{\cal T}^\dagger K_0{\cal T}|\nu\rangle=aA_\nu+2\,{\rm Re}(bB_\nu)$, and substitution into Eq.~\eqref{eq:app_CK_start} yields
\begin{eqnarray}
C_K^{(\nu)}(t)
&=&
\frac12
\left[
\gamma^2+\frac{1}{\gamma^2}
+
\left(\frac{\dot\gamma}{\Omega_0}\right)^2
\right]A_\nu-\bar k_\nu+
\frac12
\left[
\gamma^2-\frac{1}{\gamma^2}
+
\left(\frac{\dot\gamma}{\Omega_0}\right)^2
\right]{\rm Re}\,B_\nu
-
\frac{\dot\gamma}{\Omega_0\gamma}{\rm Im}\,B_\nu.
\label{eq:app_CK_su11_final}
\end{eqnarray}
 For a single tower state $|\nu\rangle=|k,n\rangle$, one finds $A_\nu=k+n$,
$\bar k_\nu=k$ and $B_\nu=0$, so that Eq.~\eqref{eq:app_CK_su11_final}
becomes $C_K^{(k,n)}(t)=n+\tfrac{k+n}{2}\big[(\gamma-\tfrac1\gamma)^2
+(\tfrac{\dot\gamma}{\Omega_0})^2\big]$,
Eq.~\eqref{eq:Ermakov_CK_single_main}. For a superposition of tower states
evolving under $H(0)=2\Omega_0K_0$, the free evolution $e^{-i\tau H(0)}$ in Eq.~\eqref{eq:Psi_exact_general} multiplies $c_{n,k}$ by $e^{-2i\Omega_0(k+n)\tau}$, so that adjacent levels acquire the relative phase $B_\nu\to B=e^{2i\Omega_0\tau(t)}B_\nu$, as in Eq.~\eqref{eq:Ermakov_AB_main}.

\section{Expansion of the density matrix in the vectorized
Liouville space}
\label{app:density_CG}

In this section, we show the analytical details for the expansion in the irreducible representations of a thermal density matrix diagonal in a single Bargmann tower $k_0$. In Liouville space, the action on the density matrix, $\mathcal X_a\rho=[K_a,\rho]$, is defined by the adjoint representations of the $\mathfrak{su}(1,1)$ generators,
\begin{eqnarray}
    \mathcal X_a=K_a\otimes\mathbb I-\mathbb I\otimes K_a^{T},\quad a=\pm,0.
\end{eqnarray}

We consider thermal operators diagonal in a single Bargmann tower,
\begin{eqnarray}
\rho_\beta^{(k_0)}(0)=(1-Q)\sum_{n\ge0}Q^{n}\,D_n,\qquad
Q=e^{-2\beta\Omega_0},\quad D_n=|k_0,n\rangle\langle k_0,n|.
\label{eq:APPC_thermalstate}
\end{eqnarray}
As $[K_0,D_n]=0$, the adjoint Casimir operator $\mathcal C=\mathcal X_0^{2}-\tfrac12(\mathcal X_+\mathcal X_-+\mathcal X_-\mathcal X_+)$ acts on them as $\mathcal C D_n=-\tfrac12\big([K_+,[K_-,D_n]]+[K_-,[K_+,D_n]]\big)$, which gives in the $k_0$ tower
\begin{eqnarray}
\mathcal C D_n
&=&
\beta_{n+1,k_0}^2D_{n+1}
-
(\beta_{n+1,k_0}^2+\beta_{n,k_0}^2)D_n
+
\beta_{n,k_0}^2D_{n-1},
\nonumber\\
\beta_{n,k_0}^2&=&n(2k_0+n-1),
\qquad
\beta_{0,k_0}^2=0 .
\label{eq:app_jacobi}
\end{eqnarray}
On the orthonormal operators $|D_n)$, with
$(D_m|D_n)=\mathrm{Tr}(D_mD_n)=\delta_{mn}$, the Casimir thus acts as a
symmetric tridiagonal matrix, and its eigenvectors
$|\lambda)=\sum_n\psi_n(\lambda)|D_n)$ obey the three-term recursion of the continuous dual Hahn
polynomials
$S_n(\lambda^2;a,b,c)=(a+b)_n(a+c)_n\,{}_3F_2(-n,a+i\lambda,a-i\lambda;
a+b,a+c;1)$~\cite{NIST_DLMF}, with the recurrence coefficients
$(n+a+b)(n+a+c)=\beta_{n+1,k_0}^2$ and
$n(n+b+c-1)=\beta_{n,k_0}^2$. This fixes the constants and, through the eigenvalue $-(a^2+\lambda^2)$ of this
recurrence, the Casimir eigenvalue as
\begin{eqnarray}
(a,b,c)=\big(\tfrac12,\ \tfrac12,\ 2k_0-\tfrac12\big),\quad
\mathcal C|\lambda)
=
-\left(\lambda^2+\frac14\right)|\lambda),
\label{eq:app_Casimir_lambda}
\end{eqnarray}
with the continuous dual Hahn weight and norm
\begin{eqnarray}
w_{k_0}(\lambda)
=
\frac{1}{2\pi}
\left|
\frac{
\Gamma(\tfrac12+i\lambda)^2
\Gamma(2k_0-\tfrac12+i\lambda)
}{
\Gamma(2i\lambda)
}
\right|^2 ,\quad
h_n
=
(n!)^2\,\Gamma(n+2k_0)^2,
\end{eqnarray}
by which the components $\psi_n(\lambda)=\sqrt{w_{k_0}(\lambda)/h_n}\,S_n(\lambda^2;a,b,c)$ are orthonormal, $\int_0^\infty\psi_m(\lambda)\psi_n(\lambda)d\lambda=\delta_{mn}$.

Projecting the thermal vector onto $|\lambda)$ gives
\begin{eqnarray}\label{eq:rho_sum}
\rho^{(\beta)}_{0,\lambda}(0)
&=&
(\lambda|\rho_\beta^{(k_0)}(0))
=
(1-Q)\sum_{n\ge0}Q^n\psi_n(\lambda)
=
\frac{(1-Q)\sqrt{w_{k_0}(\lambda)}}{\Gamma(2k_0)}
\sum_{n\ge0}Q^n\,
{}_3F_2\!\left(\!\begin{array}{c}
-n,\,\tfrac12+i\lambda,\,\tfrac12-i\lambda\\
1,\,2k_0
\end{array};1\right),
\end{eqnarray}
where $S_n/\sqrt{h_n}=\Gamma(2k_0)^{-1}\,{}_3F_2(\cdots)$ was used. Writing the ${}_3F_2(\dots)$ hypergeometric function in its finite-sum representation 
over $p=0,\dots,n$, exchanging the order of the sums over $n$ and $p$,
and resumming the geometric series with
$\sum_{n\ge p}Q^n(-n)_p=(-1)^pp!Q^p(1-Q)^{-p-1}$, the double sum becomes
\begin{eqnarray}
\sum_{n\ge0}Q^n\,
{}_3F_2\!\left(\!\begin{array}{c}
-n,\,\tfrac12+i\lambda,\,\tfrac12-i\lambda\\
1,\,2k_0
\end{array};1\right)
=
\frac{1}{1-Q}\,
{}_2F_1\!\left(
\tfrac12+i\lambda,\tfrac12-i\lambda;
2k_0;
\frac{-Q}{1-Q}
\right).
\end{eqnarray}
From this, the principal-series amplitude
$\rho^{(\beta)}_{0,\lambda}(0)$ used in
Sec.~\ref{sec:scale_covariant} reads
\begin{eqnarray}
\rho^{(\beta)}_{0,\lambda}(0)
=
\frac{\sqrt{w_{k_0}(\lambda)}}{\Gamma(2k_0)}
\,{}_2F_1\!\left(
\tfrac12+i\lambda,\tfrac12-i\lambda;
2k_0;
\frac{-Q}{1-Q}
\right).
\label{eq:app_thermal_closed}
\end{eqnarray}

For completeness, we show that the Parseval identity recovers the purity. As the functions $\psi_n(\lambda)$ are orthonormal, $\int_0^\infty d\lambda\,\psi_n(\lambda)\psi_m(\lambda)=\delta_{nm}$, the cross terms in the sum of $|\rho^{(\beta)}_{0,\lambda}(0)|^2$ in Eq.\eqref{eq:rho_sum} integrate to zero, and only the geometric series of the squared thermal weights survives,
\begin{eqnarray}
\int_0^\infty
|\rho^{(\beta)}_{0,\lambda}(0)|^2d\lambda
=(1-Q)^2\sum_{n\ge0}Q^{2n}
=\frac{(1-Q)^2}{1-Q^2}
=
\frac{1-Q}{1+Q}
=
\mathrm{Tr}\,\rho^2 .
\label{eq:app_density_purity}
\end{eqnarray}

\section{Principal-series derivation of the density-matrix Krylov variance}
\label{app:density_principal_complexity}

We derive the density-matrix analogue of
App.~\ref{app:su11_complexity_derivation} for the thermal state of
Eq.~\eqref{eq:APPC_thermalstate}, diagonal in a single Bargmann tower, with
the adjoint generators $\mathcal X_a$, $a\in\{0,\pm\}$, of App.~\ref{app:density_CG}.
The transformation acts on vectorized operators through
${\cal T}_{\cal L}={\cal T}\otimes{\cal T}^*$. As the adjoint generators close
the same algebra as the $K_a$, conjugation by ${\cal T}_{\cal L}$ reproduces
Eq.~\eqref{eq:app_K0_adjoint} with identical coefficients,
\begin{eqnarray}
{\cal T}_{\cal L}^\dagger\mathcal X_0{\cal T}_{\cal L}
=
a(t)\mathcal X_0+b(t)\mathcal X_++b^*(t)\mathcal X_- ,
\label{eq:app_density_xi}
\end{eqnarray}
with $a(t)$ and $b(t)$ of Eq.~\eqref{eq:app_ab_coefficients} and the squeezing parameter $\xi(t)$ of Eq.~\eqref{eq:app_su11_xi}, related by $a=\cosh2\xi$ and $|b|=\tfrac12\sinh2\xi$.

Under the adjoint generators, the diagonal operators of one Bargmann tower
span the principal-series towers $|\lambda,M))$, $M\in\mathbb Z$, built on the
eigenoperators $|\lambda,0))\equiv|\lambda)$ of App.~\ref{app:density_CG}, with
\begin{eqnarray}
\mathcal X_0|\lambda,M))&=&M|\lambda,M)),
\,
\mathcal X_+|\lambda,M))
=
\sqrt{\lambda^2+\left(M+\frac12\right)^2}\,
|\lambda,M+1)),\,
\mathcal X_-|\lambda,M))
=
\sqrt{\lambda^2+\left(M-\frac12\right)^2}\,
|\lambda,M-1)),\qquad
\label{eq:app_density_principal_ladders}
\end{eqnarray}
and Casimir eigenvalue
\begin{eqnarray}
{\cal C}|\lambda,M))
=
-\left(\lambda^2+\frac14\right)|\lambda,M)).
\end{eqnarray}
The initial conditions at $M=0$ are set by Eq.~\eqref{eq:app_thermal_closed},
\begin{eqnarray}
|\rho_\beta^{(k_0)}(0))
=
\int_0^\infty d\lambda\,
\rho_{0,\lambda}^{(\beta)}(0)|\lambda,0)),
\quad
\rho_{0,\lambda}^{(\beta)}(0)
=
\frac{\sqrt{w_{k_0}(\lambda)}}{\Gamma(2k_0)}
\,{}_2F_1
\left(
\tfrac12+i\lambda,\tfrac12-i\lambda;
2k_0;
\frac{-Q}{1-Q}
\right),
\label{eq:app_density_initial_amp}
\end{eqnarray}
while the exact evolved amplitudes are
\begin{eqnarray}
\rho_{M,\lambda}^{(\beta)}(t)
=
{\cal D}_{M0}^{(\lambda)}(t)\rho_{0,\lambda}^{(\beta)}(0),
\qquad
{\cal D}_{M0}^{(\lambda)}(t)
=
((\lambda,M|{\cal T}_{\cal L}(t)|\lambda,0)),
\label{eq:app_density_evolved}
\end{eqnarray}
since ${\cal T}_{\cal L}$ commutes with the Casimir and does not mix different $\lambda$.
For $M\ge0$, the principal-series matrix elements give
\begin{eqnarray}
\left|
{\cal D}_{M0}^{(\lambda)}(t)
\right|^2
&=&
\frac{
\left|
\Gamma\!\left(M+\frac12+i\lambda\right)
\right|^2
}{
\left|
\Gamma\!\left(\frac12+i\lambda\right)
\right|^2
}
\frac{
\tanh^{2M}\xi(t)
}{
(M!)^2
}
\left|
{}_2F_1
\left(
M+\frac12+i\lambda,\,
M+\frac12-i\lambda;\,
M+1;\,
-\sinh^2\xi(t)
\right)
\right|^2 ,
\label{eq:app_density_Dq0_closed}
\end{eqnarray}
which for $M=0$ reduces to $|P_{-1/2+i\lambda}(\cosh2\xi)|^2$ with the Legendre function $P_{-1/2+i\lambda}$. The probabilities are symmetric under $M\to-M$,
\begin{eqnarray}
\left|
{\cal D}_{-M,0}^{(\lambda)}(t)
\right|^2
=
\left|
{\cal D}_{M0}^{(\lambda)}(t)
\right|^2,
\end{eqnarray}
by which the mean position along the chain vanishes. Thus, we consider instead the Krylov variance, defined as
\begin{eqnarray}
M_{2,\lambda}(t)
&=&
\sum_{M\in\mathbb Z}
M^2
\left|
{\cal D}_{M0}^{(\lambda)}(t)
\right|^2=((\lambda,0|
{\cal T}_{\cal L}^\dagger(t)\mathcal X_0^2{\cal T}_{\cal L}(t)
|\lambda,0)).
\label{eq:app_density_M2_def}
\end{eqnarray}
Using Eq.~\eqref{eq:app_density_xi} and the relations
\begin{eqnarray}
\mathcal X_0|\lambda,0))=0,\qquad
((\lambda,0|\mathcal X_+\mathcal X_-|\lambda,0))
=
((\lambda,0|\mathcal X_-\mathcal X_+|\lambda,0))
=
\lambda^2+\frac14 ,
\end{eqnarray}
by which the cross terms vanish and only the
$\mathcal X_+\mathcal X_-+\mathcal X_-\mathcal X_+$ part of the squared
generator contributes, one finds
\begin{eqnarray}
M_{2,\lambda}(t)
&=&
2|b(t)|^2
\left(\lambda^2+\frac14\right)
=
\frac12
\left(\lambda^2+\frac14\right)
\sinh^2\!\left[2\xi(t)\right].
\label{eq:app_density_M2_lambda}
\end{eqnarray}

For the full density-matrix spread, we insert Eq.~\eqref{eq:app_density_evolved} into the definition of the Krylov variance, Eq.~\eqref{eq:CK_density_sum}, $M_2[\rho](t)=\frac{1}{\mathrm{Tr}\,\rho^2}
\sum_{\lambda,M}M^{2}\,\big|\rho_{M,\lambda}(t)\big|^{2}$ with the superscript $(\beta)$ omitted for brevity. As the principal-series towers are mutually orthogonal and ${\cal T}_{\cal L}$ is unitary, the normalization is given by the initial amplitudes, $\mathrm{Tr}\,\rho^2=(\rho|\rho)=\int_0^\infty d\lambda\,|\rho_{0,\lambda}^{(\beta)}(0)|^2$, while each sector contributes $|\rho_{0,\lambda}^{(\beta)}(0)|^2M_{2,\lambda}(t)$. The time dependence thus factors out of the $\lambda$ integral, and the rest is the thermal average of the $-\mathcal C$ Casimir eigenvalue in the principal series. Using
Eq.~\eqref{eq:app_density_M2_lambda} in the continuous decomposition gives
\begin{eqnarray}
M_2[\rho_\beta^{(k_0)}](t)
&=&
\frac12
\sinh^2\!\left[2\xi(t)\right]\,
\big\langle-{\cal C}\big\rangle_\beta,
\nonumber\\
\big\langle-{\cal C}\big\rangle_\beta
&=&
\frac{(\rho_\beta^{(k_0)}(0)|-{\cal C}|\rho_\beta^{(k_0)}(0))}{(\rho_\beta^{(k_0)}(0)|\rho_\beta^{(k_0)}(0))}=
\frac{
\int_0^\infty d\lambda\,
\left(\lambda^2+\frac14\right)
|\rho_{0,\lambda}^{(\beta)}(0)|^2
}{
\int_0^\infty d\lambda\,
|\rho_{0,\lambda}^{(\beta)}(0)|^2
}.
\label{eq:app_density_M2_Cavg}
\end{eqnarray}
The Casimir average follows without the amplitudes of
Eq.~\eqref{eq:app_thermal_closed}, directly in the diagonal basis, $|\rho_\beta^{(k_0)}(0))=\sum_{n\ge0}f_n|D_n)$ with $f_n=(1-Q)Q^n$ and $D_n=|k_0,n\rangle\langle k_0,n|$ of Eq.~\eqref{eq:APPC_thermalstate}. From
Eq.~\eqref{eq:app_jacobi}, $-\mathcal C$ has the diagonal elements
$\beta_{n,k_0}^2+\beta_{n+1,k_0}^2$ and the neighboring elements
$-\beta_{n+1,k_0}^2$. Relabeling $n\to n+1$ in the sum over
$\beta_{n,k_0}^2f_n^2$ leaves it unchanged as $\beta_{0,k_0}=0$, so that the
diagonal and neighboring terms combine into the finite-difference norm
\begin{eqnarray}
(\rho_\beta^{(k_0)}|-{\cal C}|\rho_\beta^{(k_0)})
=
\sum_{n=0}^{\infty}
\beta_{n+1,k_0}^2
\left|f_{n+1}-f_n\right|^2 ,\qquad \beta_{n+1,k_0}^2=(n+1)(2k_0+n),
\label{eq:app_density_C_quad}
\end{eqnarray}
with
\begin{eqnarray}
f_{n+1}-f_n=-(1-Q)^2Q^n,
\end{eqnarray}
and therefore, using $\sum_{n\ge0}(n+1)Q^{2n}=(1-Q^2)^{-2}$ and $\sum_{n\ge0}(n+1)^2Q^{2n}=(1+Q^2)(1-Q^2)^{-3}$,
\begin{eqnarray}
(\rho_\beta^{(k_0)}|-{\cal C}|\rho_\beta^{(k_0)})
&=&
(1-Q)^4
\sum_{n=0}^{\infty}
(n+1)(2k_0+n)Q^{2n}
=
(1-Q)^4
\frac{
2\left[k_0+(1-k_0)Q^2\right]
}{
(1-Q^2)^3
}.
\label{eq:app_density_C_num}
\end{eqnarray}
Dividing by the purity from Eq.~\eqref{eq:app_density_purity} gives
\begin{eqnarray}
\big\langle-{\cal C}\big\rangle_\beta
=
\frac{
2\left[k_0+(1-k_0)Q^2\right]
}{
(1+Q)^2
},
\label{eq:app_density_C_avg_closed}
\end{eqnarray}
interpolating between $2k_0$ and $\tfrac12$ in the $Q\to0$ and $Q\to1$ limits, respectively.
The Krylov variance consequently reduces to
\begin{eqnarray}
M_2[\rho_\beta^{(k_0)}](t)
=
\frac{
k_0+(1-k_0)Q^2
}{
(1+Q)^2
}
\sinh^2\!\left[2\xi(t)\right]=4\frac{
k_0+(1-k_0)Q^2
}{
(1+Q)^2
}
\left(\sinh^2\!\left[\xi(t)\right]+1\right)\sinh^2\!\left[\xi(t)\right],
\label{eq:app_density_M2_closed}
\end{eqnarray}
with $\sinh^2\xi(t)
=
\frac14
\left[
\left(\gamma-\frac{1}{\gamma}\right)^2
+
\left(\frac{\dot\gamma}{\Omega_0}\right)^2
\right]$ encoding the scaling factor, in agreement with Eq.~\eqref{eq:LL_thermal_M2}.

\section{Calogero-Sutherland model with expansion}
\label{app:CS_pure}

In this section, we provide a further example by the (rational) Calogero-Sutherland model in a time-dependent harmonic
trap ($\hbar=m=1$) \cite{Sutherland_PRL_1998_CalogeroScaling,delcampo2016},
\begin{eqnarray}
H_{\rm CS}(t)
&=&\sum_{i=1}^N\!\Big[\!-\tfrac{1}{2}\partial_{x_i}^2
+\tfrac{1}{2}\omega^2(t)x_i^2\Big]+\sum_{i<j}\frac{g(g-1)}{(x_i-x_j)^2},
\label{eq:HCS_pure}
\end{eqnarray}
with a degree-$(-2)$ inverse-square interaction. The generators in the Hamiltonian are given by
$H_c=\tfrac{1}{2}\sum_i p_i^2+\sum_{i<j}g(g-1)/(x_i-x_j)^2$, so that
$H_{\rm CS}(t)=H_c+\omega^2(t)K$ with $K=\tfrac{1}{2}\sum_i x_i^2$.
The triple $\{H_c,K,D\}$ inherits the $\mathfrak{su}(1,1)$ commutators
of Eq.~\eqref{eq:TKD_main}, and the Cartan-Weyl basis Eq.~\eqref{eq:su11_CW_map} applies with $T\to H_c$, so that
\begin{eqnarray}
H_{\rm CS}(t)=g_0(t)K_0+f_0(t)(K_++K_-),\quad
g_0(t)=\frac{\omega_0^2+\omega^2(t)}{\omega_0},\quad
f_0(t)=\frac{\omega^2(t)-\omega^2_0}{2\omega_0}.
\label{eq:HCS_KW}
\end{eqnarray}
At $t=0$, $H_{\rm CS}(0)=2\omega_0 K_0$, and the
ground state
$\Phi_{\rm CS}=\prod_{i<j}|x_i-x_j|^g e^{-\omega_0\sum_i x_i^2/2}$
is the lowest-weight state with Bargmann index
$k=\frac{E_0}{2\omega_0}=\frac{N[1+g(N-1)]}{4}$ and $n=0$.
 The scaling factor obeys the Ermakov equation of Sec.~\ref{sec:scale_covariant}, Eq.~\eqref{eq: consistency_gen},
\begin{eqnarray}
\ddot \gamma(t)+\omega^2(t)\,\gamma(t)=\frac{\omega_0^2}{\gamma^3(t)},
\quad \gamma(0)=1,\;\dot \gamma(0)=0,
\label{eq:Ermakov_CS}
\end{eqnarray}
 and the single-tower formula, Eq.~\eqref{eq:Ermakov_CK_single_main}, applies with $n=0$,
\begin{eqnarray}
C_K(t)=\frac{k}{2}\!\left[
\Bigl(\gamma\!-\!\tfrac{1}{\gamma}\Bigr)^{\!2}
\!+\!\Bigl(\tfrac{\dot \gamma}{\omega_0}\Bigr)^{\!2}
\right].
\label{eq:CK_CS}
\end{eqnarray}

\section{Interaction picture, Heisenberg-Weyl displacement, and the dragged Calogero-Sutherland model}
\label{app:HW_su11_dressing}

In this Appendix, we first derive the Heisenberg-Weyl algebraic structure in the interaction picture and its contribution to the spread complexity used in Sec.~\ref{sec:su11_hw_examples}, and then apply it to the dragged Calogero-Sutherland model, while the driven Moshinsky model follows in App.~\ref{app:moshinsky_driven}.

\subsection{Interaction picture for a forced center-of-mass mode}
\label{app:HW_interaction}

We consider a harmonic oscillator with time-dependent
frequency $\Omega(t)$ driven by a linear force $F(t)$,
\begin{eqnarray}
H_{\rm cm}(t)
=
\frac{P^2}{2}
+
\frac12\Omega^2(t)X^2
+
F(t)X,
\qquad
a=\sqrt{\frac{\Omega_0}{2}}X+\frac{i}{\sqrt{2\Omega_0}}P ,
\label{eq:app_Hcm_force}
\end{eqnarray}
where $a$ is the ladder operator
with $\Omega_0=\Omega(0)$. We split the Hamiltonian as
$H_{\rm cm}(t)=H_{\rm osc}(t)+H_{\rm HW}(t)$. The first term,
$H_{\rm osc}(t)=P^2/2+\Omega^2(t)X^2/2$, is the oscillator
without the force and is a linear combination of the
$\mathfrak{su}(1,1)$ generators of Eq.~\eqref{eq:TKD_main}.
The second term is the force, which is linear in the
Heisenberg-Weyl generators, as
$X=(a+a^\dagger)/\sqrt{2\Omega_0}$,
\begin{eqnarray}
H_{\rm HW}(t)
=
\Lambda(t)a^\dagger+\Lambda^*(t)a,
\qquad
\Lambda(t)=\frac{F(t)}{\sqrt{2\Omega_0}} .
\label{eq:app_HW_force}
\end{eqnarray}
The interaction picture is understood with respect to $H_{\rm osc}(t)$  with the corresponding evolution operator given by Eq.~\eqref{eq:Psi_exact_general}
$U_{\rm su}(t)={\cal T}(t)e^{-i\tau(t)H_0}$, ${\cal T}=U_KS_\gamma$, and
$H_0=H_{\rm osc}(0)=\Omega_0(a^\dagger a+\tfrac12)$. Using
$U_K^\dagger PU_K=P+(\dot\gamma/\gamma)X$,
$S_\gamma^\dagger XS_\gamma=\gamma X$, and
$S_\gamma^\dagger PS_\gamma=P/\gamma$, one has
${\cal T}^\dagger X{\cal T}=\gamma X$ and
${\cal T}^\dagger P{\cal T}=P/\gamma+\dot\gamma X$, so that
the ladder operator transforms as
\begin{eqnarray}
{\cal T}^\dagger(t)\,a\,{\cal T}(t)
=
u(t)a+v(t)a^\dagger,\quad
u(t)
&=&
\frac12
\left[
\gamma+\frac{1}{\gamma}
+i\frac{\dot \gamma}{\Omega_0}
\right],
\qquad
v(t)
=
\frac12
\left[
\gamma-\frac{1}{\gamma}
+i\frac{\dot \gamma}{\Omega_0}
\right],
\label{eq:app_uv_b}
\end{eqnarray}
where $\gamma(t)$ obeys the Ermakov equation,
Eq.~\eqref{eq: consistency_gen}, with $\gamma(0)=1$ and
$\dot\gamma(0)=0$, and the coefficients satisfy
$|u|^2-|v|^2=1$ and $u^*+v=\gamma$. Together with the evolution part induced by $H_0$, $e^{i\tau H_0}ae^{-i\tau H_0}=e^{-i\Omega_0\tau}a$, one finds
\begin{eqnarray}
U_{\rm su}^\dagger(t)\,a\,U_{\rm su}(t)
=
e^{-i\Omega_0\tau(t)}u(t)\,a+e^{i\Omega_0\tau(t)}v(t)\,a^\dagger,
\label{eq:app_bogoliubov_full}
\end{eqnarray}
by which the final interaction-picture Hamiltonian
$H_I=U_{\rm su}^\dagger H_{\rm HW}U_{\rm su}$ becomes
\begin{eqnarray}
H_I(t)
=
\widetilde\Lambda(t)a^\dagger+
\widetilde\Lambda^*(t)a,
\qquad
\widetilde\Lambda(t)
=
\gamma(t)e^{i\Omega_0\tau(t)}\Lambda(t)
=
\frac{\gamma(t)e^{i\Omega_0\tau(t)}F(t)}{\sqrt{2\Omega_0}} ,
\label{eq:app_Lambda_tilde_real}
\end{eqnarray}
valid for real-valued force.
As $[H_I(t),H_I(t')]$ is constant, the Magnus expansion of the interaction-picture time-evolution operator terminates at second order, and up to a global phase it is the displacement $U_I(t)=e^{\alpha(t)a^\dagger-\alpha^*(t)a}$, with
\begin{eqnarray}
\alpha(t)
=
-i\int_0^t dt'\,\widetilde\Lambda(t').
\label{eq:app_xi_HW_general}
\end{eqnarray}

The resulting spread complexity thus becomes the expectation value of the number operator, which is the spread complexity operator of the Heisenberg-Weyl chain with the Fock ladder as Krylov basis~\cite{Grabarits_arXiv_2026_TDKrylovLie}. For an initial Fock state $|m\rangle$, as $U_I^\dagger aU_I=a+\alpha$ and $\langle m|a|m\rangle=0$, one finds
\begin{eqnarray}
C_{\rm HW}^{(m)}(t)
=
\langle m|U_I^\dagger(t)\,a^\dagger a\,U_I(t)|m\rangle
=
\langle m|(a^\dagger+\alpha^*)(a+\alpha)|m\rangle
=
m+|\alpha(t)|^2 ,
\label{eq:app_HW_complexity}
\end{eqnarray}
in which the entire time dependence is contained in the displacement, Eq.~\eqref{eq:app_xi_HW_general}, and the initial state contributes only the constant offset $m$. In $d$ dimensions, the same holds componentwise, $C_{\rm HW}^{(\mathbf m)}(t)=\sum_\mu m_\mu+|\bm\alpha(t)|^2$ with $|\bm\alpha|^2=\sum_\mu|\alpha_\mu|^2$.

\subsection{Dragged Calogero-Sutherland model}
\label{app:CS_translated}

As a first example, we consider the Calogero-Sutherland model dragged by a linear force $f(t)$,
\begin{eqnarray}
H(t)
&=&\sum_{i=1}^N\!\Big[\!-\tfrac{1}{2}\partial_{x_i}^2
+\tfrac{1}{2}\omega^2(t)\bigl(x_i-f(t)\bigr)^2\Big]+\sum_{i<j}\frac{g(g-1)}{(x_i-x_j)^2}.
\label{eq:HCS_translated}
\end{eqnarray}
The Hamiltonian can be written as
$H(t)=H_{\rm CS}^{(0)}(t)+\beta(t)\sum_ix_i+\chi(t)$, with
$\beta(t)=-\omega^2(t)f(t)$ and the scalar term
$\chi(t)=\tfrac{N}{2}\omega^2(t)f^2(t)$. In terms of the center-of-mass and relative coordinates, $X=Q/\sqrt N$ with $Q=\sum_ix_i$, $\Pi=P/\sqrt N$ with $P=\sum_ip_i$, and $r_i=x_i-Q/N$ with conjugate momenta $\pi_i$, one has $\sum_ix_i^2=\sum_ir_i^2+X^2$ and $\sum_ip_i^2=\sum_i\pi_i^2+\Pi^2$, while the inverse-square interaction depends on the differences $x_i-x_j=r_i-r_j$ only. The Hamiltonian thus splits into two commuting sectors,
\begin{eqnarray}
H(t)&=&H_{\rm rel}(t)+H_{\rm cm}(t),\\
H_{\rm rel}(t)&=&\sum_i\frac{\pi_i^2}{2}+\frac12\omega^2(t)\sum_ir_i^2+\sum_{i<j}\frac{g(g-1)}{(r_i-r_j)^2},\quad
H_{\rm cm}(t)=\frac{\Pi^2}{2}+\frac12\omega^2(t)X^2+\sqrt N\beta(t)X+\chi(t),
\label{eq:HCS_split}
\end{eqnarray}
of which the center-of-mass one is the forced oscillator of Eq.~\eqref{eq:app_Hcm_force} with $F(t)=\sqrt N\beta(t)=-\sqrt N\omega^2(t)f(t)$. As the frequency is the same for both parts, the scaling factors $\gamma(t)$ also obey the same Ermakov equation, Eq.~\eqref{eq:Ermakov_CS}.

The relative sector is the $\mathfrak{su}(1,1)$ problem
of App.~\ref{app:CS_pure} in the relative coordinates. Its
Bargmann index follows from the full ground-state energy
$E_0=\tfrac{\omega_0}{2}N[1+g(N-1)]$ after subtracting the
center-of-mass ground-state energy $\omega_0/2$, which reduces
the index $k$ of App.~\ref{app:CS_pure} by
$k_{\rm cm}=\tfrac14$,
\begin{eqnarray}
k_{\rm rel}=k-\tfrac{1}{4}=\frac{(N-1)(1+gN)}{4}.
\label{eq:k_rel_CS}
\end{eqnarray}
Its contribution to the spread complexity is the single-tower result, Eq.~\eqref{eq:Ermakov_CK_single_main} at $n=0$ with $k\to k_{\rm rel}$.

The center-of-mass sector is treated in the same way as in App.~\ref{app:HW_interaction}. With $F=\sqrt N\beta$, the translation drives the center-of-mass sector with Heisenberg-Weyl
coupling $\Lambda(t)=-\sqrt{N/(2\omega_0)}\,\omega^2(t)f(t)$, Eq.~\eqref{eq:app_HW_force}, by which the coupling in interaction picture with respect to the quadratic part becomes $\widetilde\Lambda(t)=\gamma(t)e^{i\omega_0\tau(t)}\Lambda(t)$, Eq.~\eqref{eq:app_Lambda_tilde_real}. As a result, the coherent displacement, Eq.~\eqref{eq:app_xi_HW_general}, governing the complexity is given by
\begin{eqnarray}
\alpha_{\rm cm}(t)=i\sqrt{\tfrac{N}{2\omega_0}}\!\int_0^t\!\!dt'\,\gamma(t')e^{i\omega_0\tau(t')}\omega^2(t')f(t').
\label{eq:xi_CS_translated}
\end{eqnarray}
 In the original frame, the
center-of-mass state is $U_{\rm su}(t)U_I(t)|0\rangle$, and
the Krylov basis consists of the eigenstates $|n\rangle$ of
$a^\dagger a$. Using Eq.~\eqref{eq:app_bogoliubov_full} with
$\Omega_0\to\omega_0$ and $U_I^\dagger aU_I=a+\alpha_{\rm cm}$,
the spread complexity of the center-of-mass ground state,
$m=0$, reads
\begin{eqnarray}
C_{\rm cm}(t)
=
\langle0|U_I^\dagger U_{\rm su}^\dagger\,a^\dagger a\,
U_{\rm su}U_I|0\rangle
=
|v|^2+\big|u\,e^{-i\omega_0\tau}\alpha_{\rm cm}
+v\,e^{i\omega_0\tau}\alpha_{\rm cm}^*\big|^2,
\label{eq:Ccm_lab}
\end{eqnarray}
with $u(t)$ and $v(t)$ of Eq.~\eqref{eq:app_uv_b} at
$\Omega_0\to\omega_0$, and
$|v|^2=\tfrac14[(\gamma-\gamma^{-1})^2+
(\dot\gamma/\omega_0)^2]$. Without the force, only the even
states $|2n\rangle$ are occupied. They form the
$k_{\rm cm}=\tfrac14$ tower with Krylov index $n$, and the
squeezing contributes $|v|^2/2$, in agreement with
Eq.~\eqref{eq:Ermakov_CK_single_main} at $k=\tfrac14$. With
the force, the odd states $|2n+1\rangle$ are occupied as
well, with the Krylov states labeled by the eigenvalues of $a^\dagger a$.
For a fixed trap frequency, $\gamma\equiv1$, so that $u=1$,
$v=0$, and $C_{\rm cm}=|\alpha_{\rm cm}|^2$.

As the two sectors commute, the initial ground state is the product $|0\rangle_{\rm cm}\otimes|k_{\rm rel},0\rangle_{\rm rel}$, the two Krylov chains are orthogonal, and the spread complexity is the sum of the two contributions,
\begin{eqnarray}
C_K(t)
=\frac{k_{\rm rel}}{2}\!\left[
\Bigl(\gamma\!-\!\tfrac{1}{\gamma}\Bigr)^{\!2}
\!+\!\Bigl(\tfrac{\dot \gamma}{\omega_0}\Bigr)^{\!2}
\right]+C_{\rm cm}(t).
\label{eq:CK_CS_translated}
\end{eqnarray}

\section{Driven Moshinsky model with linear force}
\label{app:moshinsky_driven}

As a second example also incorporating the quadratic pairwise potential, we consider the Moshinsky model of Sec.~\ref{sec:su11_hw_examples}, Eq.~\eqref{eq:HM_realizations}, with an additional linear force
\begin{eqnarray}
H(t)=\sum_{i=1}^N
\left[
-\frac{1}{2}\partial_{x_i}^2
+
\frac{1}{2}\omega^2(t)x_i^2
\right]
+
\frac12\varepsilon_{+2}(t)
\sum_{i<j}(x_i-x_j)^2
+\beta(t)\sum_{i=1}^N x_i .
\label{eq:HM_driven_H}
\end{eqnarray}
In terms of the center-of-mass and relative coordinates of App.~\ref{app:CS_translated}, $X=Q/\sqrt N$, $\Pi=P/\sqrt N$ and $r_i=x_i-Q/N$, one has $\sum_ix_i^2=\sum_ir_i^2+X^2$, $\sum_ip_i^2=\sum_i\pi_i^2+\Pi^2$, $\sum_ix_i=\sqrt NX$, and $\sum_{i<j}(x_i-x_j)^2=N\sum_ix_i^2-Q^2=N\sum_ir_i^2$. The Hamiltonian splits again into two commuting sectors,
\begin{eqnarray}
H(t)=H_{\rm rel}(t)+H_{\rm cm}(t),\qquad
H_{\rm rel}(t)=\sum_i\frac{\pi_i^2}{2}+\frac12\Omega_{\rm rel}^2(t)\sum_ir_i^2,\qquad
H_{\rm cm}(t)=\frac{\Pi^2}{2}+\frac12\omega^2(t)X^2+\sqrt N\beta(t)X,
\label{eq:HM_driven_split}
\end{eqnarray}
where $\Omega_{\rm rel}^2(t)=\omega^2(t)+N\varepsilon_{+2}(t)$, the force couples again only to the center-of-mass sector, and the pairwise interaction renormalizes the relative frequency only.
The center-of-mass Heisenberg-Weyl ladders are given again by
$a=\sqrt{\omega_0/2}\,X+(i/\sqrt{2\omega_0})\,\Pi$, $[a,a^\dagger]=1$,
which commute with the relative generators.

The relative sector consists of $N-1$ independent oscillators of frequency $\Omega_{\rm rel}(t)$, whose kinetic and potential parts are the  $\mathfrak{su}(1,1)$ generators, Eq.~\eqref{eq:TKD_main}, in the relative coordinates. Its ground state, with energy $(N-1)\Omega_{\rm rel,0}/2$ at $\Omega_{\rm rel,0}=\Omega_{\rm rel}(0)$, is the lowest-weight state of a single Bargmann tower with index $k_{\rm rel}=(N-1)/4$, and its scaling factor $\gamma_{\rm rel}(t)$ obeys the Ermakov equation, Eq.~\eqref{eq: consistency_gen}, with frequency $\Omega_{\rm rel}(t)$. Its contribution to the spread complexity is given by the general result, Eq.~\eqref{eq:Ermakov_CK_single_main} for a single tower with $n=0$ and $k\to k_{\rm rel}$.

The center-of-mass sector is the same forced oscillator as in Eq.~\eqref{eq:app_Hcm_force} with frequency $\omega(t)$ and $F(t)=\sqrt N\beta(t)$, and is treated in interaction picture as in the example above, with $ H_I (t)=\widetilde\Lambda(t)a^\dagger+\widetilde\Lambda^*(t)a$ and the interaction-picture coupling of Eq.~\eqref{eq:app_Lambda_tilde_real},
\begin{eqnarray}
\widetilde\Lambda(t)=\sqrt{\tfrac{N}{2\omega_0}}\,\gamma_{\rm cm}(t)\,e^{i\omega_0\tau_{\rm cm}(t)}\,\beta(t),
\qquad
\tau_{\rm cm}(t)=\int_0^t\frac{dt'}{\gamma_{\rm cm}^2(t')},
\label{eq:Lambda_tilde_HM}
\end{eqnarray}
where $\gamma_{\rm cm}(t)$ obeys the Ermakov equation at the trap
frequency $\omega(t)$, $\ddot\gamma_{\rm cm}+\omega^2(t)
\gamma_{\rm cm}=\omega_0^2/\gamma_{\rm cm}^3$,
$\gamma_{\rm cm}(0)=1$, $\dot\gamma_{\rm cm}(0)=0$. The center-of-mass coherent displacement, Eq.~\eqref{eq:app_xi_HW_general}, is then
\begin{eqnarray}
\alpha_{\rm cm}(t)=-i\sqrt{\tfrac{N}{2\omega_0}}\!\int_0^{t}\!\!dt'\,
\gamma_{\rm cm}(t')\,e^{i\omega_0\tau_{\rm cm}(t')}\,\beta(t'),
\label{eq:xi_HM_driven}
\end{eqnarray}
and for the center-of-mass ground state its contribution to the spread complexity is $C_{\rm cm}(t)$ of
Eq.~\eqref{eq:Ccm_lab}, with $\gamma\to\gamma_{\rm cm}$ and
$\tau\to\tau_{\rm cm}$ in $u$, $v$, and the phases.

Similarly to the dragged Calogero-Sutherland model, the lowest-weight state factorizes as $|0\rangle_{\rm cm}\otimes|k_{\rm rel},0\rangle_{\rm rel}$, and the spread complexity is the sum of the relative and the center-of-mass contributions,
\begin{eqnarray}
C_K(t)
=
\frac{k_{\rm rel}}{2}\!\left[
\Bigl(\gamma_{\rm rel}\!-\!\tfrac{1}{\gamma_{\rm rel}}\Bigr)^{\!2}
\!+\!\Bigl(\tfrac{\dot \gamma_{\rm rel}}{\Omega_{\rm rel,0}}\Bigr)^{\!2}
\right]+C_{\rm cm}(t),\qquad k_{\rm rel}=\frac{N-1}{4}.
\label{eq:CK_HM_driven}
\end{eqnarray}
For the spring-quench protocol
$\varepsilon_{+2,{\rm i}}\to\varepsilon_{+2,{\rm f}}$ at fixed trap frequency
$\omega(t)=\omega_0$, one has $\gamma_{\rm cm}\equiv1$ and
$\tau_{\rm cm}=t$, so that $C_{\rm cm}=|\alpha_{\rm cm}|^2$
and Eq.~\eqref{eq:xi_HM_driven} reduces to
$\alpha_{\rm cm}(t)=-i\sqrt{N/(2\omega_0)}\int_0^tdt'\,
e^{i\omega_0t'}\beta(t')$, while the scaling factor in the relative sector is set by
the Ermakov equation, Eq.~\eqref{eq: consistency_gen}, at the frequency
$\Omega_{\rm rel}(t)$.

\section{Rotating frame, angular velocity, and the admissible drive}
\label{app:angular_velocity}

In this section, we derive the rotation-vector kinematics and the consistency conditions for rotation-invariant systems in Eq.~\eqref{eq:su2_consistency} of Sec.~\ref{sec:rot_invariant}. The transformation is a single $\mathrm{SU}(2)$
rotation acting identically on every spin, whose generator defines the angular velocity $\bm\Omega(t)$,
\begin{eqnarray}
\mathcal T(t)=R(t)=e^{-i\bm\theta(t)\cdot\bm J},\qquad \bm J=\sum_i\bm s_i,\quad
i\dot R(t)R^\dagger(t)\equiv\bm\Omega(t)\cdot\bm J ,
\label{eq:app_angular_velocity_def}
\end{eqnarray}
fixed by $\bm\theta(t)$ as follows. Writing $R=e^{A}$ with $A=-i\bm\theta\cdot\bm J$, the derivative of the exponential of a time-dependent generator reads
\begin{eqnarray}
\dot R\,R^\dagger
=
\int_0^1ds\,e^{sA}\dot A\,e^{-sA}
=
\sum_{n\ge0}\frac{1}{(n+1)!}\,{\rm ad}_A^{\,n}\dot A ,
\qquad
{\rm ad}_A X=[A,X] .
\label{eq:app_dexp}
\end{eqnarray}
By the commutation relations $[\bm a\cdot\bm J,\bm b\cdot\bm J]=i(\bm a\times\bm b)\cdot\bm J$, the adjoint action closes on the linear combinations of the spin components, ${\rm ad}_A(\bm v\cdot\bm J)=(\bm\theta\times\bm v)\cdot\bm J$, so that in Eq.~\eqref{eq:app_dexp} it acts on the coefficient vector as the matrix $\Theta$ defined by $\Theta\bm v\equiv\bm\theta\times\bm v$. With $i\dot A=\dot{\bm\theta}\cdot\bm J$, multiplying Eq.~\eqref{eq:app_dexp} by $i$ yields $\bm\Omega=\sum_{n\ge0}\Theta^n\dot{\bm\theta}/(n+1)!$. As $\Theta^3=-|\bm\theta|^2\Theta$, all powers reduce to $\Theta$ and $\Theta^2$, with $\Theta^{2k+1}=(-|\bm\theta|^2)^k\Theta$ and $\Theta^{2k+2}=(-|\bm\theta|^2)^k\Theta^2$. Separating the $n=0$ term from the odd and even powers, the sum becomes
\begin{eqnarray}
\sum_{n\ge0}\frac{\Theta^n}{(n+1)!}
=
1
+\Theta\sum_{k\ge0}\frac{(-1)^k|\bm\theta|^{2k}}{(2k+2)!}
+\Theta^2\sum_{k\ge0}\frac{(-1)^k|\bm\theta|^{2k}}{(2k+3)!}
=
1
+\frac{1-\cos|\bm\theta|}{|\bm\theta|^2}\,\Theta
+\frac{|\bm\theta|-\sin|\bm\theta|}{|\bm\theta|^3}\,\Theta^2 ,
\label{eq:app_dexp_series}
\end{eqnarray}
where the two series are the Taylor series of $1-\cos|\bm\theta|$ and $|\bm\theta|-\sin|\bm\theta|$ divided by $|\bm\theta|^2$ and $|\bm\theta|^3$, respectively. Applying Eq.~\eqref{eq:app_dexp_series} to $\dot{\bm\theta}$, with $\Theta\dot{\bm\theta}=\bm\theta\times\dot{\bm\theta}$ and $\Theta^2\dot{\bm\theta}=\bm\theta\times(\bm\theta\times\dot{\bm\theta})$, one finds the rotation-vector kinematics,
\begin{eqnarray}
\bm\Omega
=\dot{\bm\theta}
+\frac{1-\cos|\bm\theta|}{|\bm\theta|^2}\,\bm\theta\times\dot{\bm\theta}
+\frac{|\bm\theta|-\sin|\bm\theta|}{|\bm\theta|^3}\,
\bm\theta\times(\bm\theta\times\dot{\bm\theta}) .
\label{eq:app_angular_velocity_explicit}
\end{eqnarray}
For a fixed axis one has $\bm\theta(t)=\phi(t)\hat{\bm n}$ leading to $\bm\Omega(t)=\dot\phi(t)\hat{\bm n}$. Further details on the Euler-angle parametrization of $R(t)$ used for the Krylov amplitudes are given in App.~\ref{app:wigner_d}.

Inserting $|\Psi(t)\rangle=R(t)|\chi(t)\rangle$ into
$i\partial_t|\Psi\rangle=H(t)|\Psi\rangle$ gives
$i\partial_t|\chi\rangle=R^\dagger[H-\bm\Omega\cdot\bm J]R|\chi\rangle$. The driven evolution therefore reduces to that of the static $H(0)$, as in
Eq.~\eqref{eq:Psi_exact_su2_main}, if $H(t)$ obeys the $\mathfrak{su}(2)$ form of the frame relation in Eq.~\eqref{eq:Ht_frame_general},
\begin{eqnarray}
H(t)=R(t)\,H(0)\,R^\dagger(t)+\bm\Omega(t)\cdot\bm J.
\label{eq:Ht_rotated_main}
\end{eqnarray}
Writing $H(0)=\bm h_0\!\cdot\!\bm J+\sum_i\bm u_i^{(0)}\!\cdot\!\bm s_i
+\sum_{i<j}J_{ij}\bm s_i\!\cdot\!\bm s_j$ and using
$R^\dagger\bm s_iR=\mathcal R(t)\bm s_i$ with $\mathcal R\in\mathrm{SO}(3)$,
so that $R(\bm u\cdot\bm s_i)R^\dagger=(\mathcal R\bm u)\cdot\bm s_i$, the isotropic exchange is invariant and
Eq.~\eqref{eq:Ht_rotated_main} reproduces precessing fields of fixed magnitude,
\begin{eqnarray}
\bm u_i(t)=\mathcal R(t)\,\bm u_i^{(0)},\qquad
\bm h(t)=\mathcal R(t)\,\bm h_0+\bm\Omega(t).
\end{eqnarray}
Differentiating them with $\dot{\mathcal R}\mathcal R^{-1}\bm v=\bm\Omega\times\bm v$,
following from Eq.~\eqref{eq:app_angular_velocity_def} and
$R(\bm v\cdot\bm s_i)R^\dagger=(\mathcal R\bm v)\cdot\bm s_i$, gives
$\dot{\bm u}_i=\bm\Omega\times\bm u_i$ and $\dot{\bm h}=\bm\Omega\times(\bm h-\bm\Omega)+\dot{\bm\Omega}$,
and thus the consistency conditions of Eq.~\eqref{eq:su2_consistency}.

\section{Wigner rotation matrix elements and the Krylov states}
\label{app:wigner_d}
In this Appendix, we give the Euler-angle and Wei-Norman forms of the rotation $R(t)$ together with the resulting Krylov wave functions.
The finite $\mathfrak{su}(2)$ tower $\{|j,q\rangle\}$, $q=0,\dots,2j$, of
Sec.~\ref{sec:rot_invariant} is given by
\begin{eqnarray}
J_0|j,q\rangle=(-j+q)|j,q\rangle,\qquad
J_+|j,q\rangle=\beta_{q+1,j}\,|j,q+1\rangle,\qquad
J_-|j,q\rangle=\beta_{q,j}\,|j,q-1\rangle ,
\quad\beta_{q,j}=\sqrt{q(2j-q+1)}.
\label{eq:app_su2_lanczos}
\end{eqnarray}
Inside each tower, the protocol rotation $R(t)$ acts as the Wigner matrix
\begin{eqnarray}
\mathcal D^{(j)}_{qn}(t)
=\langle j,q|R(t)|j,n\rangle
=D^{j}_{-j+q,\,-j+n}\!\big(\alpha(t),\theta(t),\phi(t)-\pi\big) ,
\label{eq:app_wigner_D}
\end{eqnarray}
where the $\mathfrak{su}(2)$ form of the propagator $\mathcal D^{(j)}_{qn}$ was introduced in Sec.~\ref{sec:rot_invariant}, with
$D^{j}_{m'm}(\alpha,\theta,\phi-\pi)=e^{-im'\alpha}d^{j}_{m'm}(\theta)e^{-im(\phi-\pi)}$
and $d^{j}$ the real Wigner $d$ matrix~\cite{Varshalovich_1988_QuantumTheoryAngularMomentum},
\begin{eqnarray}
d^{j}_{m'm}(\theta)
=
\sum_{s}
\frac{(-1)^{m'-m+s}\sqrt{(j+m')!\,(j-m')!\,(j+m)!\,(j-m)!}}
{(j+m-s)!\,s!\,(m'-m+s)!\,(j-m'-s)!}
\left(\cos\tfrac{\theta}{2}\right)^{2j-m'+m-2s}
\left(\sin\tfrac{\theta}{2}\right)^{m'-m+2s},
\label{eq:app_wigner_d_explicit}
\end{eqnarray}
with the sum running over all integers $s$ for which the factorials are nonnegative.

The Euler angles follow from the rotation vector $\bm\theta=\vartheta\,\hat{\mathbf m}$ of Eq.~\eqref{eq:app_angular_velocity_def}, with $\vartheta=|\bm\theta|$ and $\hat{\mathbf m}=(\sin\vartheta_m\cos\varphi_m,\sin\vartheta_m\sin\varphi_m,\cos\vartheta_m)$ the unit vector along the rotation axis. As the spin-$\tfrac12$ representation of $\mathrm{SU}(2)$ is faithful, it suffices to compare the two forms of $R$ there, where $e^{-i\bm\theta\cdot\bm\sigma/2}=\cos\tfrac{\vartheta}{2}-i\,(\hat{\mathbf m}\cdot\bm\sigma)\sin\tfrac{\vartheta}{2}$, while the propagator from Eq.\eqref{eq:app_wigner_d_explicit} for $j=1/2$ becomes
\begin{eqnarray}
D^{1/2}(\alpha,\theta,\phi-\pi)
=
\begin{pmatrix}
e^{-i(\alpha+\phi-\pi)/2}\cos\tfrac{\theta}{2} & -e^{-i(\alpha-\phi+\pi)/2}\sin\tfrac{\theta}{2}\\[2pt]
e^{i(\alpha-\phi+\pi)/2}\sin\tfrac{\theta}{2} & e^{i(\alpha+\phi-\pi)/2}\cos\tfrac{\theta}{2}
\end{pmatrix}.
\label{eq:app_D_half}
\end{eqnarray}
Matching the diagonal and the off-diagonal entries yields
\begin{eqnarray}
\cos\tfrac{\theta}{2}\,e^{-i(\alpha+\phi-\pi)/2}
=
\cos\tfrac{\vartheta}{2}-i\cos\vartheta_m\sin\tfrac{\vartheta}{2},
\qquad
\sin\tfrac{\theta}{2}\,e^{-i(\alpha-\phi+\pi)/2}
=
i\sin\vartheta_m\,e^{-i\varphi_m}\sin\tfrac{\vartheta}{2},
\label{eq:app_euler_from_theta}
\end{eqnarray}
which fix $\theta$, $\alpha+\phi$ and $\alpha-\phi$ including their quadrants. Taking the moduli gives the polar angle,
\begin{eqnarray}
\cos\theta=\cos\vartheta+\cos^2\vartheta_m\,(1-\cos\vartheta),
\label{eq:app_costheta_from_theta}
\end{eqnarray}
with $\cos\theta=\hat{\mathbf z}\cdot\mathcal R\hat{\mathbf z}$. Taking the phases gives $\alpha-\phi=2\varphi_m-2\pi$ and $\tfrac{\alpha+\phi-\pi}{2}=\arg\big(\cos\tfrac{\vartheta}{2}+i\cos\vartheta_m\sin\tfrac{\vartheta}{2}\big)$, so that $\phi=\tfrac{3\pi}{2}-\varphi_m+\arg\big(\cos\tfrac{\vartheta}{2}+i\cos\vartheta_m\sin\tfrac{\vartheta}{2}\big)$. Geometrically, $\theta(t)$ and $-\phi(t)$ are the polar and azimuthal angles of $\mathcal R^{-1}(t)\hat{\mathbf z}=(\sin\theta\cos\phi,\,-\sin\theta\sin\phi,\,\cos\theta)$, which is the $z$ axis in the rotating frame, and $\theta(t)$ is also the polar angle of the rotated $z$ axis, $\mathcal R(t)\hat{\mathbf z}$. For a rotation axis in the $xy$ plane, $\vartheta_m=\tfrac{\pi}{2}$, these reduce to $\theta=\vartheta$, $\alpha+\phi=\pi$, and $\alpha=\varphi_m-\tfrac{\pi}{2}$, with $\theta$ continued beyond $\pi$ at fixed $\alpha$ and $\phi$, covering the single-axis protocols of Sec.~\ref{sec:rot_invariant}.

For an initial eigenstate expanded as in Eq.~\eqref{eq:nu_expansion_general}, $|\nu\rangle=\sum_{j,n}c^{(\nu)}_{n,j}|j,n\rangle$, the exact solution $|\Psi_\nu(t)\rangle=e^{-iE_\nu t}R(t)|\nu\rangle$ of Eq.~\eqref{eq:Psi_exact_su2_main} thus yields the Krylov wave function
\begin{eqnarray}
\psi^{(\nu)}_{q,j}(t)
=
\langle j,q|\Psi_\nu(t)\rangle
=
e^{-iE_\nu t}
\sum_{n=0}^{2j}
e^{-i(-j+q)\alpha(t)}\,
d^{j}_{-j+q,\,-j+n}\!\big(\theta(t)\big)\,
e^{-i(-j+n)[\phi(t)-\pi]}\,
c^{(\nu)}_{n,j} ,
\label{eq:app_su2_psi_general}
\end{eqnarray}
with the Krylov distribution
\begin{eqnarray}
\big|\psi^{(\nu)}_{q,j}(t)\big|^2
=
\Big|
\sum_{n=0}^{2j}
d^{j}_{-j+q,\,-j+n}\!\big(\theta(t)\big)\,
(-1)^{n}\,e^{-i(-j+n)\phi(t)}\,
c^{(\nu)}_{n,j}
\Big|^2 ,
\label{eq:app_su2_prob_general}
\end{eqnarray}
where the first Euler angle $\alpha(t)$ and the energy phase drop out, as they multiply each amplitude by a phase independent of $n$. The Krylov probabilities are therefore fixed by the pair $(\theta,\phi)$ alone, and depend on the model only through the overlaps $c^{(\nu)}_{q,j}$.

Equivalently, up to the phase factor $e^{-i[\alpha(t)+\phi(t)-\pi]J_z}$,
which drops out of the Krylov probabilities, the disentangled $\mathrm{SU}(2)$ evolution is the Wei-Norman
factorization
\begin{eqnarray}
R(t)=e^{z(t)J_+}e^{\eta(t)J_0}e^{w(t)J_-},\qquad
z=\tan\tfrac{\theta}{2}\,e^{i\phi},\quad
e^{\eta}=\cos^{-2}\tfrac{\theta}{2},\quad
w=-z^*.
\label{eq:app_su2_WN}
\end{eqnarray}
In the spin-$\tfrac12$ representation, $e^{zJ_+}=\left(\begin{smallmatrix}1&z\\0&1\end{smallmatrix}\right)$, $e^{\eta J_0}=\left(\begin{smallmatrix}e^{\eta/2}&0\\0&e^{-\eta/2}\end{smallmatrix}\right)$ and $e^{wJ_-}=\left(\begin{smallmatrix}1&0\\w&1\end{smallmatrix}\right)$, the product has the entries $R_{11}=e^{\eta/2}+zw\,e^{-\eta/2}$, $R_{12}=z\,e^{-\eta/2}$, $R_{21}=w\,e^{-\eta/2}$, $R_{22}=e^{-\eta/2}$. Unitarity requires $e^{-\eta/2}=\cos\tfrac{\theta}{2}$, which fixes $e^{\eta}$ as given, and with $zw=-\tan^2\tfrac{\theta}{2}$ one obtains $R_{11}=(1-\sin^2\tfrac{\theta}{2})/\cos\tfrac{\theta}{2}=\cos\tfrac{\theta}{2}$, so that
\begin{eqnarray}
R
=
\begin{pmatrix}
\cos\tfrac{\theta}{2} & e^{i\phi}\sin\tfrac{\theta}{2}\\[2pt]
-e^{-i\phi}\sin\tfrac{\theta}{2} & \cos\tfrac{\theta}{2}
\end{pmatrix}.
\label{eq:app_WN_half}
\end{eqnarray}
Compared with Eq.~\eqref{eq:app_D_half}, the two forms coincide for
$\alpha+\phi=\pi$, where $e^{-i(\alpha-\phi+\pi)/2}=-e^{i\phi}$, and otherwise
differ by the left factor $e^{-i[\alpha+\phi-\pi]J_z}$, commuting with $J_0$. The same matrix yields $R^\dagger J_0R=\cos\theta\,J_0+\tfrac{\sin\theta}{2}(e^{i\phi}J_++e^{-i\phi}J_-)$, which is used in App.~\ref{app:su2_complexity_derivation}.

Applied to the lowest-weight state $|j,0\rangle$, the three factors act in turn as follows. The rightmost one is trivial, $e^{wJ_-}|j,0\rangle=|j,0\rangle$, since $J_-|j,0\rangle=0$. The middle one contributes the normalization, $e^{\eta J_0}|j,0\rangle=e^{-j\eta}|j,0\rangle=\cos^{2j}\tfrac{\theta}{2}\,|j,0\rangle=(1+|z|^2)^{-j}|j,0\rangle$, as $J_0|j,0\rangle=-j|j,0\rangle$. The leftmost one generates the tower, $e^{zJ_+}|j,0\rangle=\sum_{q=0}^{2j}\frac{z^q}{q!}J_+^q|j,0\rangle$, where by Eq.~\eqref{eq:app_su2_lanczos} the repeated action of the raising operator gives $J_+^q|j,0\rangle=\prod_{p=1}^{q}\beta_{p,j}\,|j,q\rangle=\sqrt{q!\,(2j)!/(2j-q)!}\,|j,q\rangle$. Altogether, the Krylov wave function is
\begin{eqnarray}
\psi_{q,j}(t)
=
\binom{2j}{q}^{1/2}
z(t)^{q}\,\big(1+|z(t)|^2\big)^{-j}
=
\binom{2j}{q}^{1/2}
\left(\sin\tfrac{\theta(t)}{2}\right)^{q}
\left(\cos\tfrac{\theta(t)}{2}\right)^{2j-q}
e^{iq\phi(t)} ,
\label{eq:app_su2_psi_binomial}
\end{eqnarray}
so that the Krylov distribution is binomial, $|\psi_{q,j}|^2=\binom{2j}{q}\,p^{q}(1-p)^{2j-q}$ with $p=\sin^2\tfrac{\theta(t)}{2}$, in agreement with Eq.~\eqref{eq:app_su2_prob_general} for $c^{(\nu)}_{n,j}=\delta_{n,0}$.

\section{Adjoint derivation of the rotation-invariant spread complexity}
\label{app:su2_complexity_derivation}

Here we obtain the closed form of Eq.~\eqref{eq:K_su2_superposition_main}
along the same lines as for the scale-invariant case in
App.~\ref{app:su11_complexity_derivation}, with the rotation $R(t)$ and its
Euler angles taken from Apps.~\ref{app:angular_velocity}
and~\ref{app:wigner_d}. For an initial eigenstate evolving as
\begin{eqnarray}
|\Psi_\nu(t)\rangle
=
e^{-iE_\nu t}R(t)|\nu\rangle,\quad H(0)|\nu\rangle=E_\nu|\nu\rangle,
\label{eq:app_su2_eigenstate_solution}
\end{eqnarray}
the spread complexity reads
\begin{eqnarray}
C_K^{(\nu)}(t)
=
\langle\nu|R^\dagger(t)N_KR(t)|\nu\rangle
=
\langle\nu|R^\dagger(t)J_0R(t)|\nu\rangle
+\langle\nu|\hat J|\nu\rangle,
\label{eq:app_su2_CK_start}
\end{eqnarray}
where the complexity operator $N_K=J_0+\hat J$ counts the level $q$ within
each tower, $N_K|j,q\rangle=q|j,q\rangle$, and $\hat J|j,q\rangle=j|j,q\rangle$
commutes with $R(t)$, as $j$ is set by the Casimir.

To obtain $R^\dagger J_0R$, it suffices to work with the spin-$\tfrac12$
matrix of Eq.~\eqref{eq:app_WN_half}. It differs from $R(t)$ only by the left
factor $e^{-i[\alpha(t)+\phi(t)-\pi]J_z}$, which commutes with $J_0$. For
$J_0=\tfrac12\sigma_z$ and $J_\pm=\sigma_\pm=\tfrac12(\sigma_x\pm i\sigma_y)$,
multiplying out the matrices gives
\begin{eqnarray}
R^\dagger\sigma_zR
=
\begin{pmatrix}
\cos\theta & e^{i\phi}\sin\theta\\[2pt]
e^{-i\phi}\sin\theta & -\cos\theta
\end{pmatrix}
=
\cos\theta\,\sigma_z+\sin\theta\left(e^{i\phi}\sigma_++e^{-i\phi}\sigma_-\right).
\label{eq:app_su2_J0_half}
\end{eqnarray}
Since the conjugation maps the generators linearly onto each other, with
coefficients set by the commutation relations alone, the same combination
appears in any tower,
\begin{eqnarray}
R^\dagger(t)J_0R(t)
=
\cos\theta(t)\,J_0
+
\frac{\sin\theta(t)}{2}
\left[
e^{i\phi(t)}J_+
+
e^{-i\phi(t)}J_-
\right],
\label{eq:app_su2_adjoint_J0}
\end{eqnarray}
or equivalently $R^\dagger J_0R=(\mathcal R^{-1}\hat{\mathbf z})\cdot\bm J$
with $\mathcal R^{-1}\hat{\mathbf z}=(\sin\theta\cos\phi,\,-\sin\theta\sin\phi,\,\cos\theta)$.

Next, the initial eigenstate is decomposed over the $\mathfrak{su}(2)$
towers, which introduces the overlaps and the moments
\begin{eqnarray}
|\nu\rangle
&=&
\sum_j\sum_{q=0}^{2j}
c_{q,j}^{(\nu)}|j,q\rangle,
\qquad
c_{q,j}^{(\nu)}=\langle j,q|\nu\rangle,\\
\bar J_\nu
&=&
\sum_{j,q}
j\,|c_{q,j}^{(\nu)}|^2,
\qquad
M_{0,\nu}
=
\sum_{j,q}
(-j+q)|c_{q,j}^{(\nu)}|^2,\qquad
M_{+,\nu}
=
\sum_j\sum_{q=0}^{2j-1}
\sqrt{(q+1)(2j-q)}
\left(c_{q+1,j}^{(\nu)}\right)^*
c_{q,j}^{(\nu)} .
\label{eq:app_su2_moments}
\end{eqnarray}
With the ladder actions of Eq.~\eqref{eq:app_su2_lanczos},
$J_+|j,q\rangle=\sqrt{(q+1)(2j-q)}\,|j,q+1\rangle$ and
$J_-|j,q\rangle=\sqrt{q(2j-q+1)}\,|j,q-1\rangle$, together with
$J_0|j,q\rangle=(-j+q)|j,q\rangle$ and $\hat J|j,q\rangle=j|j,q\rangle$, these
moments are the expectation values
\begin{eqnarray}
\langle\nu|\hat J|\nu\rangle=\bar J_\nu,
\qquad
\langle\nu|J_0|\nu\rangle=M_{0,\nu},
\qquad
\langle\nu|J_+|\nu\rangle=M_{+,\nu},
\qquad
\langle\nu|J_-|\nu\rangle=M_{+,\nu}^*.
\label{eq:app_su2_expectations}
\end{eqnarray}
Taking the expectation value of Eq.~\eqref{eq:app_su2_adjoint_J0} in
$|\nu\rangle$ therefore gives
\begin{eqnarray}
\langle\nu|R^\dagger J_0R|\nu\rangle
=\cos\theta\,M_{0,\nu}
+\frac{\sin\theta}{2}\left(e^{i\phi}M_{+,\nu}+e^{-i\phi}M_{+,\nu}^*\right)
=\cos\theta\,M_{0,\nu}+\sin\theta\,{\rm Re}\!\left[e^{i\phi}M_{+,\nu}\right],
\label{eq:app_su2_J0_expectation}
\end{eqnarray}
and, together with Eq.~\eqref{eq:app_su2_CK_start}, the complexity takes the closed form
\begin{eqnarray}
C_K^{(\nu)}(t)
=
\bar J_\nu
+
\cos\theta(t)\,M_{0,\nu}
+
\sin\theta(t)\,
{\rm Re}\!\left[
e^{i\phi(t)}M_{+,\nu}
\right].
\label{eq:app_su2_CK_final}
\end{eqnarray}
A single tower state $|\nu\rangle=|j,q\rangle$ has $\bar J_\nu=j$,
$M_{0,\nu}=-j+q$ and $M_{+,\nu}=0$, which reduces
Eq.~\eqref{eq:app_su2_CK_final} to
\begin{eqnarray}
C_K^{(j,q)}(t)
=
q+(j-q)\left[1-\cos\theta(t)\right].
\label{eq:app_su2_CK_single_tower}
\end{eqnarray}

\section{Spin-tower thermal density matrix in Liouville space}
\label{app:su2_density_CG}

In this Appendix, we give the details about the finite $\mathfrak{su}(2)$ version of
App.~\ref{app:density_CG}. Within
one spin tower, the states $|j,q\rangle\equiv|j,m=-j+q\rangle$,
$q=0,\ldots,2j$, are linked by the ladder coefficients
$\beta_{q,j}=\sqrt{q(2j-q+1)}$ of Eq.~\eqref{eq:app_su2_lanczos}, with
$J_-|j,q\rangle=\beta_{q,j}|j,q-1\rangle$,
$J_+|j,q\rangle=\beta_{q+1,j}|j,q+1\rangle$ and
$J_0|j,q\rangle=(-j+q)|j,q\rangle$. On operators, the generators act through
the adjoint maps
\begin{eqnarray}
\mathcal X_a=J_a\otimes\mathbb I-\mathbb I\otimes J_a^T,\quad a=\pm,0,
\end{eqnarray}
which close the same $\mathfrak{su}(2)$ algebra. The thermal state of Eq.~\eqref{eq:HS_thermal}, with equally spaced
levels and $Q=e^{-\beta A_0}$, $\rho_\beta^{(j)}(0)=Z_j^{-1}\sum_{q}Q^q\,\Pi_q$ with
$Z_j=(1-Q^{2j+1})/(1-Q)$, is a combination of the projectors
$\Pi_q=|j,q\rangle\langle j,q|$. These commute with $J_0$, so the Liouville
Casimir
$\mathcal C=\mathcal X_0^2+\tfrac12(\mathcal X_+\mathcal X_-+\mathcal X_-\mathcal X_+)$
acts on them only through the double commutators
$\mathcal C\,\Pi_q=\tfrac12\big([J_+,[J_-,\Pi_q]]+[J_-,[J_+,\Pi_q]]\big)$.
Each double commutator shifts a projector by at most one level, which gives
\begin{eqnarray}
\mathcal C\,\Pi_q
=
-\beta_{q+1,j}^2\Pi_{q+1}
+\left(\beta_{q+1,j}^2+\beta_{q,j}^2\right)\Pi_q
-\beta_{q,j}^2\Pi_{q-1},
\qquad
\beta_{0,j}=\beta_{2j+1,j}=0 .
\label{eq:app_su2_jacobi}
\end{eqnarray}
As the projectors are orthonormal in Liouville space,
$(\Pi_p|\Pi_q)=\mathrm{Tr}(\Pi_p\Pi_q)=\delta_{pq}$, the Casimir is
represented by a real symmetric tridiagonal matrix of dimension $2j+1$. An
eigenvector $\sum_q\psi_q(L)|\Pi_q)$ with eigenvalue $L(L+1)$ therefore has
components obeying the three-term relation of Eq.~\eqref{eq:app_su2_jacobi},
with $\mathcal C$ replaced by $L(L+1)$. Since
$-\beta_{q+1,j}^2=(q+1)(q-2j)$ and $-\beta_{q,j}^2=q(q-2j-1)$, this relation
is the difference equation in $q$ of the Hahn polynomials with vanishing
parameters on $q=0,\ldots,2j$~\cite{NIST_DLMF}. Its polynomial solution of
degree $L$ has the eigenvalue $L(L+1)$ with
$L=0,\ldots,2j$. The corresponding solutions are orthogonal with uniform weight,
\begin{eqnarray}
\mathcal Q_L(q)={}_3F_2\!\left(\!\begin{array}{c}
-L,\,L+1,\,-q\\
1,\,-2j
\end{array};1\right),\quad
\sum_{q=0}^{2j}\mathcal Q_L(q)\mathcal Q_{L'}(q)=h_L\,\delta_{LL'},\quad
h_L=\frac{\Gamma(2j-L+1)\Gamma(2j+L+2)}{(2L+1)\,\Gamma^2(2j+1)},
\label{eq:app_su2_hahn_def}
\end{eqnarray}
so that $\psi_q(L)=(-1)^L\mathcal Q_L(q)/\sqrt{h_L}$ is an orthogonal
matrix, with $\sum_q\psi_q(L)\psi_q(L')=\delta_{LL'}$ and
$\sum_L\psi_p(L)\psi_q(L)=\delta_{pq}$.

These eigenvectors are the $M=0$ members of the Liouville-space spin towers. The operator space of one tower decomposes as
$j\otimes j^*=0\oplus1\oplus\cdots\oplus2j$, and the tower states read
\begin{eqnarray}
|L,M))
&=&
\sum_{m,m'=-j}^{j}
(-1)^{j-m'}
\langle j,m;\,j,-m'|L,M\rangle
|j,m\rangle\langle j,m'| ,
\label{eq:app_su2_tensor_basis}
\end{eqnarray}
where $L=0,\ldots,2j$ and $M=-L,\ldots,L$. For $M=0$, only the diagonal
terms $m=m'$ remain, and comparing with the eigenvectors of $\mathcal C$
identifies the Clebsch-Gordan coefficients of two equal spins with the Hahn
polynomials of Eq.~\eqref{eq:app_su2_hahn_def},
\begin{eqnarray}
\psi_q(L)=(-1)^{2j-q}\langle j,-j+q;\,j,j-q|L,0\rangle
=
\frac{(-1)^L}{\sqrt{h_L}}\,
\mathcal Q_L(q),
\label{eq:app_su2_hahn}
\end{eqnarray}
where the sign $(-1)^L$ is fixed by the standard phase convention of the
Clebsch-Gordan coefficients~\cite{Edmonds_1957_AngularMomentum,Varshalovich_1988_QuantumTheoryAngularMomentum}, which makes
$\langle j,j;\,j,-j|L,0\rangle$ positive.
Inverting this relation, each projector spreads over all towers as
\begin{eqnarray}
|j,m\rangle\langle j,m|
=
\sum_{L=0}^{2j}
(-1)^{j-m}
\langle j,m;\,j,-m|L,0\rangle
|L,0)).
\label{eq:app_su2_diagonal_projector}
\end{eqnarray}

Taking the overlap of the thermal state with each $|L,0))$ yields its
decomposition over the towers,
\begin{eqnarray}
\rho_\beta^{(j)}(0)
&=&
Z_j^{-1}
\sum_{q=0}^{2j}
Q^q\,\Pi_q
=
\sum_{L=0}^{2j}
\bigg[
Z_j^{-1}\sum_{q=0}^{2j}Q^q\,\psi_q(L)
\bigg]
|L,0))
=\sum_{L=0}^{2j}
\rho_{0,L}^{(\beta)}(0)|L,0)),
\label{eq:app_su2_tower_thermal}
\end{eqnarray}
with every Liouville tower entered at its $M=0$ level and the amplitudes
\begin{eqnarray}
\rho_{0,L}^{(\beta)}(0)
=
Z_j^{-1}{\cal R}_{jL}(Q),\qquad
{\cal R}_{jL}(Q)
=
\sum_{q=0}^{2j}
Q^q\,\psi_q(L)
=
\frac{(-1)^L}{\sqrt{h_L}}\sum_{q=0}^{2j}Q^q\,\mathcal Q_L(q).
\label{eq:app_su2_R_CG}
\end{eqnarray}
The $L=0$ polynomial is constant, $\mathcal Q_0=1$ with $h_0=2j+1$, and the sum is geometric,
\begin{eqnarray}
{\cal R}_{j0}(Q)
&=&
\frac{1}{\sqrt{2j+1}}\,
\frac{1-Q^{2j+1}}{1-Q}.
\label{eq:app_su2_R0}
\end{eqnarray}
For $1\le L\le 2j$, the sum is evaluated with the discrete Rodrigues
formula of the Hahn polynomials~\cite{NIST_DLMF},
\begin{eqnarray}
\mathcal Q_L(q)=\frac{(-1)^L L!}{(-2j)_L}\,\nabla^L\!\left[\binom{q+L}{L}\binom{2j-q}{L}\right],
\qquad
\nabla^L f(q)=\sum_{k=0}^{L}(-1)^k\binom{L}{k}f(q-k),
\end{eqnarray}
where $\nabla^L$ is the $L$-th backward difference, obtained by applying
$\nabla f(q)=f(q)-f(q-1)$ $L$ times, and
$(-2j)_L=(-2j)(1-2j)\cdots(L-1-2j)$ is the Pochhammer symbol. The bracket
vanishes outside $0\le q\le 2j-L$, so the formula holds on all
$q=0,\ldots,2j$, and every lower difference $\nabla^kf$, $k<L$, vanishes at
$q=-1$ and $q=2j$. Summation by parts,
$\sum_qQ^q\nabla f(q)=(1-Q)\sum_qQ^qf(q)$, therefore has no boundary
terms, and each of the $L$ differences produces a factor $(1-Q)$. With
$\binom{q+L}{L}=(L+1)_q/q!$ and
$(-1)^LL!\binom{2j-q}{L}/(-2j)_L=(L-2j)_q/(-2j)_q$, what remains is a
terminating hypergeometric series,
\begin{eqnarray}
\sum_{q=0}^{2j}Q^q\,\mathcal Q_L(q)
&=&
\frac{(-1)^LL!}{(-2j)_L}(1-Q)^L
\sum_{q=0}^{2j-L}Q^q\binom{q+L}{L}\binom{2j-q}{L}
\nonumber\\
&=&
(1-Q)^L
\sum_{q=0}^{2j-L}\frac{(L+1)_q(L-2j)_q}{(-2j)_q\,q!}Q^q
=
(1-Q)^L\,
{}_2F_1\!\left(L-2j,\,L+1;\,-2j;\,Q\right),
\end{eqnarray}
and the amplitudes become
\begin{eqnarray}
{\cal R}_{jL}(Q)
&=&
\frac{(-1)^L}{\sqrt{h_L}}\,
(1-Q)^L
{}_2F_1
\left(
L-2j,\,
L+1;\,
-2j;\,
Q
\right),
\label{eq:app_su2_R_closed}
\end{eqnarray}
which also reproduces Eq.~\eqref{eq:app_su2_R0} for $L=0$.
Because $\psi_q(L)$ is an orthogonal matrix, the mixed products
$Q^{p+q}\psi_p(L)\psi_q(L)$ with $p\neq q$ cancel in the sum over $L$, and
only the squared thermal weights remain,
\begin{eqnarray}
\sum_{L=0}^{2j}
\left|
\rho_{0,L}^{(\beta)}(0)
\right|^2
=
{\rm Tr}\,[\rho_\beta^{(j)}]^2
=
Z_j^{-2}
\sum_{q=0}^{2j}Q^{2q}
=
Z_j^{-2}
\frac{1-Q^{4j+2}}{1-Q^2}.
\label{eq:app_su2_density_purity}
\end{eqnarray}

\section{Rotation-invariant derivation of the density-matrix Krylov variance}
\label{app:su2_density_variance}

In this section, we derive the $\mathfrak{su}(2)$ analogue of
App.~\ref{app:density_principal_complexity} for the thermal state of
App.~\ref{app:su2_density_CG}. On the vectorized density matrix, the underlying Krylov dynamics and the corresdpnind Krylov variance measuring the spread of the density matrix. On the vectorized density matrix, the
rotation acts as $R_{\cal L}=R\otimes R^*$. Since the adjoint maps satisfy the
commutation relations of the $J_a$, the conjugation of
App.~\ref{app:su2_complexity_derivation} carries over with the same Euler
angles,
\begin{eqnarray}
R_{\cal L}^\dagger\mathcal X_0R_{\cal L}
=
\cos\theta(t)\,\mathcal X_0
+\frac{\sin\theta(t)}{2}\left[e^{i\phi(t)}\mathcal X_++e^{-i\phi(t)}\mathcal X_-\right].
\label{eq:app_su2_density_adjoint}
\end{eqnarray}
Within each Liouville tower, the adjoint maps and the Casimir act as
\begin{eqnarray}
\mathcal X_0|L,M))=M|L,M)),\quad
\mathcal X_\pm|L,M))=\sqrt{(L\mp M)(L\pm M+1)}\,|L,M\pm1)),\quad
\mathcal C|L,M))=L(L+1)|L,M)).
\label{eq:app_su2_density_ladders}
\end{eqnarray}
The rotation $R_{\cal L}$ commutes with $\mathcal C$, and is thus block diagonal in the towers $L$. Within a tower, its matrix elements are those of the Wigner
matrix of App.~\ref{app:wigner_d} in the adjoint representation,
$((L,M|R_{\cal L}|L,M'))=D^L_{MM'}(\alpha,\theta,\phi-\pi)$. Starting from the
$M=0$ amplitudes of Eq.~\eqref{eq:app_su2_R_CG}, one obtains
\begin{eqnarray}
\rho^{(\beta)}_{M,L}(t)=\rho_{0,L}^{(\beta)}(0)\,D^L_{M0}(\alpha,\theta,\phi-\pi)
=\rho_{0,L}^{(\beta)}(0)\,e^{-iM\alpha(t)}\,d^L_{M0}\big(\theta(t)\big),
\label{eq:app_su2_density_amplitudes}
\end{eqnarray}
in which the third Euler angle does not appear at $M=0$ and $\alpha$
enters only through a phase. As $d^L_{-M,0}=(-1)^Md^L_{M0}$, the occupation
of each tower is symmetric about its initial level $M=0$.

To compare with the pure-state result of
App.~\ref{app:su2_complexity_derivation}, the levels of each Liouville tower
are counted from its lowest weight $M=-L$, as for $|j,q\rangle$, with
$q_L=M+L=0,\ldots,2L$. The initial state then occupies the middle level
$q_L=L$ of every tower. As the occupation over $M$ remains symmetric about
$M=0$, the mean of $q_L$ stays at $L$ in each tower, and its thermal average
is time independent in agreement with the result for quantum states, Eq.~\eqref{eq:app_su2_CK_final} applied in Liouville space. Accordingly the overlaps $c^{(\nu)}_{q,j}$ of
Eq.~\eqref{eq:app_su2_moments} take the form of the coefficients of
$|\rho_\beta^{(j)}(0))$ divided by the square root of the purity,
Eq.~\eqref{eq:app_su2_density_purity},
\begin{eqnarray}
c_{q_L,L}^{(\beta)}
=
\frac{((L,q_L|\rho_\beta^{(j)}(0))}{\sqrt{{\rm Tr}\,[\rho_\beta^{(j)}]^2}}
=
\frac{\rho_{0,L}^{(\beta)}(0)}{\sqrt{{\cal P}_\beta}}\,
\delta_{q_L,L},
\qquad
{\cal P}_\beta
\equiv{\rm Tr}\,[\rho_\beta^{(j)}]^2
=
\sum_{L=0}^{2j}
\left|\rho_{0,L}^{(\beta)}(0)\right|^2,
\label{eq:app_su2_density_coefficients}
\end{eqnarray}
where the Kronecker delta enforces the initial condition of starting from the middle of the chain $M=0$.
Here, with $j\to L$ and $q\to q_L$ in Eq.~\eqref{eq:app_su2_moments}, the three
moments become
\begin{eqnarray}
\bar J_\beta
&=&
\frac{1}{{\cal P}_\beta}
\sum_{L=0}^{2j}
L\left|\rho_{0,L}^{(\beta)}(0)\right|^2,\,
M_{0,\beta}
=
\frac{1}{{\cal P}_\beta}
\sum_{L=0}^{2j}
(-L+q_L)
\left|\rho_{0,L}^{(\beta)}(0)\right|^2
=
0,\,\\
M_{+,\beta}&=&\frac{1}{{\cal P}_\beta}\sum_{L=0}^{2j}\sum_{q_L=0}^{2L-1}\sqrt{(q_L+1)(2L-q_L)}\,\big(c^{(\beta)}_{q_L+1,L}\big)^*c^{(\beta)}_{q_L,L}=0 ,
\label{eq:app_su2_density_moments}
\end{eqnarray}
as the occupied level has $-L+q_L=M=0$ in every tower, with no coherence term. In
Eq.~\eqref{eq:app_su2_CK_final}, both angle-dependent terms then drop out,
and the first moment
$\langle q_L\rangle_\beta(t)={\cal P}_\beta^{-1}\sum_{L,q_L}q_L\,|((L,q_L|\rho_\beta^{(j)}(t))|^2$
stays at the static offset
\begin{eqnarray}
\langle q_L\rangle_\beta
&=&
\bar J_\beta
=
\frac{
\sum_{L=0}^{2j}
L\left|\rho_{0,L}^{(\beta)}(0)\right|^2
}{
\sum_{L=0}^{2j}
\left|\rho_{0,L}^{(\beta)}(0)\right|^2
}
=
\frac{1-Q^2}{1-Q^{4j+2}}
\sum_{L=1}^{2j}
L\left|{\cal R}_{jL}(Q)\right|^2 .
\label{eq:app_su2_density_CK_final}
\end{eqnarray}
As the first moment does not capture the time dependence of the complexity, the spreading within each tower is
quantified by the Krylov variance,
\begin{eqnarray}
M_{2,L}(t)
=
\sum_{M=-L}^{L}
M^2
\left|d^L_{M0}\big(\theta(t)\big)\right|^2
=
((L,0|
R_{\cal L}^\dagger\mathcal X_0^2R_{\cal L}
|L,0)).
\end{eqnarray}
In the square of Eq.~\eqref{eq:app_su2_density_adjoint}, every term
containing $\mathcal X_0$ has a vanishing diagonal element as
$\mathcal X_0|L,0))=0$, and the terms with $\mathcal X_\pm^2$ shift $M$ by two.
With $((L,0|\mathcal X_\pm\mathcal X_\mp|L,0))=L(L+1)$ from
Eq.~\eqref{eq:app_su2_density_ladders}, only the ladder products contribute,
\begin{eqnarray}
M_{2,L}(t)
=
\frac{\sin^2\theta}{4}\,
((L,0|\mathcal X_+\mathcal X_-+\mathcal X_-\mathcal X_+|L,0))
=
\frac{L(L+1)}{2}\sin^2\theta .
\label{eq:app_su2_density_M2_sector}
\end{eqnarray}

For the thermal state, Eq.~\eqref{eq:CK_density_sum} is evaluated with the
amplitudes of Eq.~\eqref{eq:app_su2_density_amplitudes} and the purity
${\rm Tr}\,\rho^2={\cal P}_\beta$, which is conserved under $R_{\cal L}$. Each
tower contributes $|\rho_{0,L}^{(\beta)}(0)|^2M_{2,L}(t)$, so the angle
dependence factorizes from the thermal average of the Casimir,
\begin{eqnarray}
M_2[\rho_\beta^{(j)}](t)
&=&
\frac{1}{{\cal P}_\beta}
\sum_{L=0}^{2j}
\left|\rho_{0,L}^{(\beta)}(0)\right|^2
M_{2,L}(t)
=
\frac{\sin^2\theta(t)}{2}\,\big\langle{\cal C}\big\rangle_\beta,
\label{eq:app_su2_density_M2_def}\\
\big\langle{\cal C}\big\rangle_\beta
&=&
\frac{(\rho_\beta^{(j)}(0)|{\cal C}|\rho_\beta^{(j)}(0))}{(\rho_\beta^{(j)}(0)|\rho_\beta^{(j)}(0))}
=
\frac{1-Q^2}{1-Q^{4j+2}}
\sum_{L=1}^{2j}
L(L+1)
\left|{\cal R}_{jL}(Q)\right|^2.
\label{eq:app_su2_density_M2_sum}
\end{eqnarray}
The first equality in Eq.~\eqref{eq:app_su2_density_M2_def} only uses the
$M=0$ occupation of every tower, and therefore holds for the general diagonal
weights of Eq.~\eqref{eq:su2_thermal_state}, with
$\langle{\cal C}\rangle_\beta=\langle L(L+1)\rangle_\beta$, which gives
Eq.~\eqref{eq:su2_density_M2_general}. For the weights $Q^q$, the Casimir
average does not require the closed-form amplitudes of
Eq.~\eqref{eq:app_su2_R_closed}, and as $Z_j$ cancels in the ratio of
Eq.~\eqref{eq:app_su2_density_M2_sum}, it can be evaluated with the
unnormalized thermal operator
$\tilde\rho=Z_j\rho_\beta^{(j)}(0)=\sum_qQ^q\,\Pi_q$. By
Eq.~\eqref{eq:app_su2_tower_thermal},
$\tilde\rho=\sum_L{\cal R}_{jL}(Q)|L,0))$, and as the $|L,0))$ are orthonormal
eigenvectors of ${\cal C}$ with eigenvalue $L(L+1)$, the sum over $L$ equals
$(\tilde\rho|{\cal C}|\tilde\rho)$. This expectation value follows equally from the recurrence of Eq.~\eqref{eq:app_su2_jacobi} in the
projector basis, whose matrix elements are
$(\Pi_q|{\cal C}|\Pi_q)=\beta_{q,j}^2+\beta_{q+1,j}^2$ and
$(\Pi_{q+1}|{\cal C}|\Pi_q)=(\Pi_q|{\cal C}|\Pi_{q+1})=-\beta_{q+1,j}^2$,
\begin{eqnarray}
(\tilde\rho|{\cal C}|\tilde\rho)
=
\sum_{p,q=0}^{2j}Q^{p+q}(\Pi_p|{\cal C}|\Pi_q)
=
\sum_{q=0}^{2j}\left(\beta_{q,j}^2+\beta_{q+1,j}^2\right)Q^{2q}
-2\sum_{q=0}^{2j-1}\beta_{q+1,j}^2\,Q^{2q+1}.
\end{eqnarray}
Shifting $q\to q+1$ in the term with $\beta_{q,j}^2$, which is allowed as
$\beta_{0,j}=0$, gives
$\sum_q\beta_{q,j}^2Q^{2q}=\sum_q\beta_{q+1,j}^2Q^{2q+2}$, while
$\beta_{2j+1,j}=0$ restricts all sums to $q\le2j-1$. Combining the diagonal
and neighboring terms in this way, the quadratic form reduces to squared
differences of neighboring weights, and with $\beta_{q+1,j}^2=(q+1)(2j-q)$,
\begin{eqnarray}
\sum_{L=1}^{2j}
L(L+1)
\left|{\cal R}_{jL}(Q)\right|^2
&=&
(\tilde\rho|{\cal C}|\tilde\rho)
=
\sum_{q=0}^{2j-1}\beta_{q+1,j}^2\left(Q^q-Q^{q+1}\right)^2
=
(1-Q)^2
\sum_{q=0}^{2j-1}
(q+1)(2j-q)Q^{2q}\nonumber\\
&=&
2j(1-Q)^2
\,{}_2F_1
\left(
2,\,
1-2j;\,
-2j;\,
Q^2
\right),
\label{eq:app_su2_density_Casimir_sum}
\end{eqnarray}
where in the last step
$(q+1)(2j-q)=2j\,(2)_q(1-2j)_q/[(-2j)_q\,q!]$, which follows from
$(2)_q/q!=q+1$ and $(1-2j)_q/(-2j)_q=(2j-q)/(2j)$, was used to write the
finite sum as a terminating hypergeometric series. Dividing by the norm
$(\tilde\rho|\tilde\rho)=\sum_qQ^{2q}=(1-Q^{4j+2})/(1-Q^2)$ gives
\begin{eqnarray}
\big\langle{\cal C}\big\rangle_\beta
=
\frac{(\tilde\rho|{\cal C}|\tilde\rho)}{(\tilde\rho|\tilde\rho)}
=
2j\,
\frac{(1-Q^2)(1-Q)^2}{1-Q^{4j+2}}
\,{}_2F_1
\left(
2,\,
1-2j;\,
-2j;\,
Q^2
\right),
\label{eq:app_su2_density_C_avg_closed}
\end{eqnarray}
which decreases from $2j$ at $Q\to0$, where only the lowest level is
occupied, to zero at $Q\to1$, where $\rho_\beta^{(j)}(0)$ is proportional to
the identity and lies entirely in the $L=0$ tower. The Krylov variance is
then
\begin{eqnarray}
M_2[\rho_\beta^{(j)}](t)
&=&
j\,
\frac{(1-Q^2)(1-Q)^2}{1-Q^{4j+2}}
\,{}_2F_1
\left(
2,\,
1-2j;\,
-2j;\,
Q^2
\right)
\sin^2\theta(t),
\label{eq:app_su2_density_M2_final}
\end{eqnarray}
where $\theta(t)$ is the polar Euler angle of App.~\ref{app:wigner_d}, in
agreement with Eq.~\eqref{eq:HS_density_M2}.

\section{Central-spin two-tower mixing}
\label{app:central_spin}

In this Appendix, we derive the two-tower mixing of the initial eigenstate of the central spin model of Sec.~\ref{sec:central_spin}.
At fixed bath spin $S$ (eigenvalue of $\mathbf S_B^2$) and fixed magnetization
$\mathcal M$ (eigenvalue of $J_z$), the central coupling
$\mathbf S_B\!\cdot\!\mathbf s_0=\tfrac12[j(j+1)-S(S+1)-\tfrac34]$ is diagonal
in the two total-spin towers $j_\pm=S\pm\tfrac12$,
\begin{eqnarray}
\mathbf S_B\!\cdot\!\mathbf s_0\big|_{j_+}=\frac{S}{2},
\qquad
\mathbf S_B\!\cdot\!\mathbf s_0\big|_{j_-}=-\frac{S+1}{2} .
\end{eqnarray}
The staggered field $\delta\,\mathbf n_0\!\cdot\!(\mathbf S_B-\mathbf s_0)$ with
$\mathbf n_0=\hat{\mathbf z}$ is $\delta A$ with the operator $A=S_{B,z}-s_{0,z}$, which at fixed
$\mathcal M$ is the $2\times2$ block
\begin{eqnarray}
\langle j_+,q_+|A|j_+,q_+\rangle=\frac{\mathcal M(2S-1)}{2S+1},\quad
\langle j_-,q_-|A|j_-,q_-\rangle=\frac{\mathcal M(2S+3)}{2S+1},\quad
\langle j_+,q_+|A|j_-,q_-\rangle=\frac{2\sqrt{(S+\tfrac12)^2-\mathcal M^2}}{2S+1} .
\end{eqnarray}
With the exchange amplitude $J$, the block Hamiltonian has diagonal entries
\begin{eqnarray}
E_+^{(0)}=\frac{JS}{2}+\delta\frac{\mathcal M(2S-1)}{2S+1},\qquad
E_-^{(0)}=-\frac{J(S+1)}{2}+\delta\frac{\mathcal M(2S+3)}{2S+1},
\end{eqnarray}
and off-diagonal coupling $V_{\mathcal M}=2\delta\sqrt{(S+\tfrac12)^2-\mathcal M^2}/(2S+1)$.
Diagonalization gives the mixing angle and energies
\begin{eqnarray}
\tan 2\alpha_{\mathcal M}=\frac{2V_{\mathcal M}}{E_+^{(0)}-E_-^{(0)}},\qquad
E_\pm(\mathcal M)=\frac{E_+^{(0)}+E_-^{(0)}}{2}
\pm\frac12\sqrt{\big(E_+^{(0)}-E_-^{(0)}\big)^2+4V_{\mathcal M}^2} ,
\label{eq:app_central_spin_mixing}
\end{eqnarray}
fixing $\alpha_{\mathcal M}$ in Eq.~\eqref{eq:central_spin_K}. The mixing vanishes at the
band edges $\mathcal M=\pm(S+\tfrac12)$, where $V_{\mathcal M}=0$ and the state lies in a single tower.

\section{$2\times 2$ representation of the doublet matrix}
\label{app:R_s}

In this section, we show the analytical details of the Schwinger boson extension for the rotation-invariant systems of Sec.~\ref{sec:su2_hw_examples}. The Schwinger doublets $\mathbf b_{r}=(b_{r,\uparrow},b_{r,\downarrow})^T$
realize the spin-$\tfrac12$ generators
$S^{(b)}_{a,r}=\tfrac12\mathbf b_r^\dagger\sigma_a\mathbf b_r$, with
$[S^{(b)}_{a,r},\mathbf b_r]=-\tfrac12\sigma_a\mathbf b_r$. For
$R=e^{-i\bm\theta\cdot\bm J}$, this gives
$R^\dagger\mathbf b_rR=e^{-i\bm\theta\cdot\bm\sigma/2}\mathbf b_r\equiv R_s\mathbf b_r$,
where $R_s(t)=D^{1/2}(\alpha,\theta,\phi-\pi)$ is the spin-$\tfrac12$ matrix of
Eq.~\eqref{eq:app_D_half}. For rotations about an axis in the $xy$ plane,
$\alpha+\phi=\pi$, and it reduces to the Wei-Norman form of
Eq.~\eqref{eq:app_WN_half},
\begin{eqnarray}
R_s(t)=\begin{pmatrix}
\cos\tfrac{\theta}{2} & e^{i\phi}\sin\tfrac{\theta}{2}\\[2pt]
-e^{-i\phi}\sin\tfrac{\theta}{2} & \cos\tfrac{\theta}{2}
\end{pmatrix}.
\label{eq:app_Rs}
\end{eqnarray}
In interaction picture, $R^\dagger(\mathbf b_r^\dagger\cdot\boldsymbol\Lambda_r)R
=\mathbf b_r^\dagger\cdot R_s^\dagger\boldsymbol\Lambda_r$, which gives
Eq.~\eqref{eq:su2_hw_interaction}. As in App.~\ref{app:HW_interaction},
the commutator of this Hamiltonian at different times is a constant, so the
interaction-picture evolution is, up to a global phase, the displacement of
each doublet by
$\boldsymbol\alpha_r(t)=-i\int_0^tdt'\,R_s^\dagger(t')\boldsymbol\Lambda_r(t')$.
As $R(t)$ conserves the total occupation $N_b$ of every doublet of
Sec.~\ref{sec:su2_hw_examples}, the vacuum of the doublet $r$ contributes
$\langle N_b\rangle=|\boldsymbol\alpha_r(t)|^2$ to the spread complexity, and
together with the spin tower this gives Eq.~\eqref{eq:su2_hw_total_K}.

\twocolumngrid
\bibliographystyle{apsrev4-2}
\bibliography{referencesMB}

@article{Kinoshita_Science_2004_TonksGas,
  author  = {Kinoshita, Toshiya and Wenger, Trevor and Weiss, David S.},
  title   = {Observation of a One-Dimensional {Tonks-Girardeau} Gas},
  journal = {Science},
  volume  = {305},
  pages   = {1125--1128},
  year    = {2004},
  doi     = {10.1126/science.1100700}
}

@article{Paredes_Nature_2004_TonksGas,
  author  = {Paredes, Bel{\'e}n and Widera, Artur and Murg, Valentin and Mandel, Olaf and F{\"o}lling, Simon and Cirac, Ignacio and Shlyapnikov, Gora V. and H{\"a}nsch, Theodor W. and Bloch, Immanuel},
  title   = {{Tonks--Girardeau} gas of ultracold atoms in an optical lattice},
  journal = {Nature},
  volume  = {429},
  pages   = {277--281},
  year    = {2004},
  doi     = {10.1038/nature02530}
}

@article{Pla_Nature_2012_SiliconSpinQubit,
  author  = {Pla, Jarryd J. and Tan, Kuan Y. and Dehollain, Juan P. and Lim, Wee H. and Morton, John J. L. and Jamieson, David N. and Dzurak, Andrew S. and Morello, Andrea},
  title   = {A single-atom electron spin qubit in silicon},
  journal = {Nature},
  volume  = {489},
  pages   = {541--545},
  year    = {2012},
  doi     = {10.1038/nature11449}
}

@article{Sias_PRL_2008_PhotonAssisted,
  author  = {Sias, C. and Lignier, H. and Singh, Y. P. and Zenesini, A. and Ciampini, D. and Morsch, O. and Arimondo, E.},
  title   = {Observation of Photon-Assisted Tunneling in Optical Lattices},
  journal = {Phys. Rev. Lett.},
  volume  = {100},
  pages   = {040404},
  year    = {2008},
  doi     = {10.1103/PhysRevLett.100.040404}
}

@article{Grass_EPJQT_2014_TunableHeisenberg,
  author  = {Gra{\ss}, Tobias and Lewenstein, Maciej},
  title   = {Trapped-ion quantum simulation of tunable-range {Heisenberg} chains},
  journal = {EPJ Quantum Technology},
  volume  = {1},
  pages   = {8},
  year    = {2014},
  doi     = {10.1140/epjqt8}
}

@misc{martinezazcona2026,
      title={On Information Propagation in Krylov Subspaces}, 
      author={Pablo Martinez-Azcona and Maximilian Meyer-Moelleringhof and Adolfo del Campo and Apollonas S. Matsoukas-Roubeas},
      year={2026},
      eprint={2610.03883},
      archivePrefix={arXiv},
      primaryClass={quant-ph},
      url={https://arxiv.org/abs/2610.03883}, 
}

@book{AbramowitzStegun_1964_Handbook,
  author    = {Abramowitz, Milton and Stegun, Irene A.},
  title     = {Handbook of Mathematical Functions with Formulas, Graphs, and Mathematical Tables},
  publisher = {Dover},
  address   = {New York},
  year      = {1964},
  note      = {Eqs.~9.1.42--9.1.45}
}

@book{Watson_1944_BesselFunctions,
  author    = {Watson, G. N.},
  title     = {A Treatise on the Theory of Bessel Functions},
  edition   = {2nd},
  publisher = {Cambridge University Press},
  address   = {Cambridge},
  year      = {1944},
  isbn      = {9780521483919}
}

@article{GurappaPanigrahi1999Equivalence,
  author = {N. Gurappa and P. K. Panigrahi},
  title = {Equivalence of the Calogero-Sutherland model to free harmonic oscillators},
  journal = {Physical Review B},
  volume = {59},
  pages = {R2490--R2493},
  year = {1999},
  doi = {10.1103/PhysRevB.59.R2490},
  eprint = {cond-mat/9710035},
  archivePrefix = {arXiv}
}

@article{Girardeau1960,
    author = {Girardeau, M.},
    title = {Relationship between Systems of Impenetrable Bosons and Fermions in One Dimension},
    journal = {Journal of Mathematical Physics},
    volume = {1},
    number = {6},
    pages = {516-523},
    year = {1960},
    month = {11},
    issn = {0022-2488},
    doi = {10.1063/1.1703687},
    url = {https://doi.org/10.1063/1.1703687}
}

@article{Girardeau2001,
  title = {Ground-state properties of a one-dimensional system of hard-core bosons in a harmonic trap},
  author = {Girardeau, M. D. and Wright, E. M. and Triscari, J. M.},
  journal = {Phys. Rev. A},
  volume = {63},
  issue = {3},
  pages = {033601},
  numpages = {6},
  year = {2001},
  month = {Feb},
  publisher = {American Physical Society},
  doi = {10.1103/PhysRevA.63.033601},
  url = {https://link.aps.org/doi/10.1103/PhysRevA.63.033601}
}

@article{WeiNorman1963,
  author  = {J. Wei and E. Norman},
  title   = {Lie Algebraic Solution of Linear Differential Equations},
  journal = {Journal of Mathematical Physics},
  volume  = {4},
  number  = {4},
  pages   = {575--581},
  year    = {1963},
  doi     = {10.1063/1.1703993}
}

@article{WeiNorman1964,
  author  = {J. Wei and E. Norman},
  title   = {On Global Representations of the Solutions of Linear Differential Equations as a Product of Exponentials},
  journal = {Proceedings of the American Mathematical Society},
  volume  = {15},
  number  = {2},
  pages   = {327--334},
  year    = {1964},
  doi     = {10.1090/S0002-9939-1964-0160922-2}
}

@article{CharzynskiKus2013WeiNorman,
  author = {Charzy{\'n}ski, Szymon and Ku{\'s}, Marek},
  title = {Wei--Norman equations for a unitary evolution},
  journal = {Journal of Physics A: Mathematical and Theoretical},
  volume = {46},
  number = {26},
  pages = {265208},
  year = {2013},
  doi = {10.1088/1751-8113/46/26/265208}
}

@article{LewisRiesenfeld_JMP_1969_Invariants,
  author = {Lewis, H. R. and Riesenfeld, W. B.},
  title = {An Exact Quantum Theory of the Time-Dependent Harmonic Oscillator and of a Charged Particle in a Time-Dependent Electromagnetic Field},
  journal = {Journal of Mathematical Physics},
  volume = {10},
  number = {8},
  pages = {1458--1473},
  year = {1969},
  doi = {10.1063/1.1664991}
}

@incollection{Torrontegui_AdvAMOP_2013_STA,
  author = {Torrontegui, Erik and Ib{\'a}{\~n}ez, Sergio and Mart{\'i}nez-Garaot, Susana and Modugno, Michele and del Campo, Adolfo and Gu{\'e}ry-Odelin, David and Ruschhaupt, Andreas and Chen, Xi and Muga, Juan Gonzalo},
  title = {Shortcuts to Adiabaticity},
  booktitle = {Advances in Atomic, Molecular, and Optical Physics},
  editor = {Arimondo, Ennio and Berman, Paul R. and Lin, Chun C.},
  volume = {62},
  pages = {117--169},
  publisher = {Academic Press},
  year = {2013},
  doi = {10.1016/B978-0-12-408090-4.00002-5}
}

@article{Perelomov_CMP_1972_CoherentStates,
  author = {Perelomov, A. M.},
  title = {Coherent States for Arbitrary Lie Group},
  journal = {Communications in Mathematical Physics},
  volume = {26},
  pages = {222--236},
  year = {1972},
  doi = {10.1007/BF01645091}
}

@book{Gilmore_1974_LieGroups,
  author    = {Gilmore, Robert},
  title     = {Lie Groups, Lie Algebras, and Some of Their Applications},
  publisher = {Wiley},
  address   = {New York},
  year      = {1974}
}

@book{Perelomov_1986_GeneralizedCoherentStates,
  author    = {Perelomov, A. M.},
  title     = {Generalized Coherent States and Their Applications},
  series    = {Texts and Monographs in Physics},
  publisher = {Springer},
  address   = {Berlin},
  year      = {1986},
  doi       = {10.1007/978-3-642-61629-7}
}

@article{Ban_JOSAB_1993_SU11SU2,
  author = {Ban, Masashi},
  title = {Decomposition Formulas for SU(1,1) and SU(2) Lie Algebras and Their Applications in Quantum Optics},
  journal = {Journal of the Optical Society of America B},
  volume = {10},
  number = {8},
  pages = {1347--1359},
  year = {1993},
  doi = {10.1364/JOSAB.10.001347}
}

@article{Bargmann_AnnMath_1947_IrreducibleUnitary,
  author  = {Bargmann, V.},
  title   = {Irreducible Unitary Representations of the {Lorentz} Group},
  journal = {Annals of Mathematics},
  volume  = {48},
  number  = {3},
  pages   = {568--640},
  year    = {1947},
  doi     = {10.2307/1969129}
}

@book{DodonovManko1989,
  author    = {V. V. Dodonov and V. I. Man'ko},
  title     = {Invariants and the Evolution of Nonstationary Quantum Systems},
  series    = {Proceedings of the Lebedev Physics Institute},
  volume    = {183},
  publisher = {Nova Science Publishers},
  address   = {New York},
  year      = {1989}
}

@article{Brink1992Explicit,
  author = {L. Brink and T. H. Hansson and M. A. Vasiliev},
  title = {Explicit Solution to the $N$-Body Calogero Problem},
  journal = {Physics Letters B},
  volume = {286},
  number = {1-2},
  pages = {109--111},
  year = {1992},
  doi = {10.1016/0370-2693(92)90166-M},
  eprint = {hep-th/9206049},
  archivePrefix = {arXiv}
}

@article{UjinoWadati1994Hidden,
  author = {H. Ujino and M. Wadati},
  title = {Hidden Symmetry of the Quantum Calogero--Moser System},
  journal = {Journal of the Physical Society of Japan},
  volume = {63},
  number = {10},
  pages = {3585--3594},
  year = {1994},
  doi = {10.1143/JPSJ.63.3585}
}

@Article{ADC_Entropy_2020,
AUTHOR = {Beau, Mathieu and del Campo, Adolfo},
TITLE = {Nonadiabatic Energy Fluctuations of Scale-Invariant Quantum Systems in a Time-Dependent Trap},
JOURNAL = {Entropy},
VOLUME = {22},
YEAR = {2020},
NUMBER = {5},
ARTICLE-NUMBER = {515},
URL = {https://www.mdpi.com/1099-4300/22/5/515},
PubMedID = {33286287},
ISSN = {1099-4300},
DOI = {10.3390/e22050515}
}

@article{ADC_UltraCold_2021,
  title = {Probing Quantum Speed Limits with Ultracold Gases},
  author = {del Campo, Adolfo},
  journal = {Phys. Rev. Lett.},
  volume = {126},
  issue = {18},
  pages = {180603},
  numpages = {6},
  year = {2021},
  month = {May},
  publisher = {American Physical Society},
  doi = {10.1103/PhysRevLett.126.180603},
  url = {https://link.aps.org/doi/10.1103/PhysRevLett.126.180603}
}

@book{McLachlan_1947_MathieuFunctions,
  author    = {McLachlan, N. W.},
  title     = {Theory and Application of Mathieu Functions},
  publisher = {Oxford University Press},
  address   = {Oxford},
  year      = {1947}
}

@book{LandauLifshitz_Mechanics,
  author    = {Landau, L. D. and Lifshitz, E. M.},
  title     = {Mechanics},
  series    = {Course of Theoretical Physics},
  volume    = {1},
  edition   = {3rd},
  publisher = {Butterworth-Heinemann},
  address   = {Oxford},
  year      = {1976}
}

@incollection{Schwinger_1965_AngularMomentum,
  author    = {Schwinger, Julian},
  title     = {On Angular Momentum},
  booktitle = {Quantum Theory of Angular Momentum},
  editor    = {Biedenharn, L. C. and van Dam, H.},
  pages     = {229--279},
  publisher = {Academic Press},
  address   = {New York},
  year      = {1965}
}

@book{Auerbach_1994_InteractingElectrons,
  author    = {Auerbach, Assa},
  title     = {Interacting Electrons and Quantum Magnetism},
  series    = {Graduate Texts in Contemporary Physics},
  publisher = {Springer-Verlag},
  address   = {New York},
  year      = {1994},
  doi       = {10.1007/978-1-4612-0869-3}
}

@article{LewisSwan2021BCSCavity,
  title = {Cavity-QED Quantum Simulator of Dynamical Phases of a BCS Superconductor},
  author = {Lewis-Swan, Robert J. and Barberena, Diego and Cline, Julia R. K. and Young, Dylan J. and Thompson, James K. and Rey, Ana Maria},
  journal = {Physical Review Letters},
  volume = {126},
  pages = {173601},
  year = {2021},
  doi = {10.1103/PhysRevLett.126.173601},
  eprint = {2011.13007},
  archivePrefix = {arXiv},
  primaryClass = {quant-ph}
}

@article{Young2024BCSCavity,
  title = {Observing Dynamical Phases of BCS Superconductors in a Cavity QED Simulator},
  author = {Young, Dylan J. and Chu, Anjun and Song, Eric Yilun and Barberena, Diego and Wellnitz, David and Niu, Zhijing and Sch{\"a}fer, Vera M. and Lewis-Swan, Robert J. and Rey, Ana Maria and Thompson, James K.},
  journal = {Nature},
  volume = {625},
  pages = {679--684},
  year = {2024},
  doi = {10.1038/s41586-023-06911-x},
  eprint = {2306.00066},
  archivePrefix = {arXiv},
  primaryClass = {quant-ph}
}

@article{Harris_NatPhoton_2017_ProgrammableNanophotonic,
  author  = {Harris, Nicholas C. and Steinbrecher, Gregory R. and Mower, Jacob and Lahini, Yoav and Prabhu, Mihika and Bunandar, Darius and Chen, Changchen and Wong, Franco N. C. and Baehr-Jones, Tom and Hochberg, Michael and Lloyd, Seth and Englund, Dirk},
  title   = {Quantum transport simulations in a programmable nanophotonic processor},
  journal = {Nature Photonics},
  volume  = {11},
  pages   = {447--452},
  year    = {2017},
  doi     = {10.1038/nphoton.2017.95},
  eprint  = {1507.03406},
  archivePrefix = {arXiv}
}

@article{Bogaerts_Nature_2020_ProgrammablePhotonicCircuits,
  author  = {Bogaerts, Wim and P{\'e}rez, Daniel and Capmany, Jos{\'e} and Miller, David A. B. and Poon, Joyce and Englund, Dirk and Morichetti, Francesco and Melloni, Andrea},
  title   = {Programmable photonic circuits},
  journal = {Nature},
  volume  = {586},
  pages   = {207--216},
  year    = {2020},
  doi     = {10.1038/s41586-020-2764-0}
}

@article{Albiez_PRL_2005_BosonicJosephson,
  author  = {Albiez, Michael and Gati, Rudolf and F{\"o}lling, Jonas and Hunsmann, Stefan and Cristiani, Matteo and Oberthaler, Markus K.},
  title   = {Direct Observation of Tunneling and Nonlinear Self-Trapping in a Single Bosonic Josephson Junction},
  journal = {Phys. Rev. Lett.},
  volume  = {95},
  issue   = {1},
  pages   = {010402},
  year    = {2005},
  doi     = {10.1103/PhysRevLett.95.010402}
}

@article{Breuer_PRB_2004_SpinStar,
  author  = {Breuer, Heinz-Peter and Burgarth, Daniel and Petruccione, Francesco},
  title   = {Non-Markovian dynamics in a spin star system: Exact solution and approximation techniques},
  journal = {Physical Review B},
  volume  = {70},
  pages   = {045323},
  year    = {2004},
  doi     = {10.1103/PhysRevB.70.045323}
}

@article{Bortz_PRB_2007_InhomogeneousCentralSpin,
  author  = {Bortz, Michael and Stolze, Joachim},
  title   = {Exact dynamics in the inhomogeneous central-spin model},
  journal = {Physical Review B},
  volume  = {76},
  pages   = {014304},
  year    = {2007},
  doi     = {10.1103/PhysRevB.76.014304}
}

@misc{Bhattacharjee2023,
      title={A Lanczos approach to the Adiabatic Gauge Potential}, 
      author={Budhaditya Bhattacharjee},
      year={2023},
      eprint={2302.07228},
      archivePrefix={arXiv},
      primaryClass={quant-ph},
      url={https://arxiv.org/abs/2302.07228}, 
}

@book{Viswanath_Muller_1994_RecursionMethod,
  author = {Viswanath, V. S. and M{\"u}ller, G.},
  title = {The Recursion Method: Application to Many-Body Dynamics},
  publisher = {Springer-Verlag},
  address = {Berlin},
  year = {1994}
}

@article{Parker_PRX_2019_UniversalOperatorGrowth,
  author = {Parker, Daniel E. and Cao, Xiangyu and Avdoshkin, Alexander and Scaffidi, Thomas and Altman, Ehud},
  title = {A Universal Operator Growth Hypothesis},
  journal = {Physical Review X},
  volume = {9},
  pages = {041017},
  year = {2019},
  doi = {10.1103/PhysRevX.9.041017}
}

@article{Nandy_PhysRep_2025_KrylovReview,
  author = {Nandy, Pratik and Matsoukas-Roubeas, Apollonas S. and Mart{\'i}nez-Azcona, Pablo and Dymarsky, Anatoly and del Campo, Adolfo},
  title = {Quantum Dynamics in Krylov Space: Methods and Applications},
  journal = {Physics Reports},
  volume = {1125\textendash 1128},
  pages = {1\textendash 82},
  year = {2025},
  doi = {10.1016/j.physrep.2025.05.001},
  eprint = {2405.09628},
  archivePrefix = {arXiv},
  primaryClass = {quant-ph}
}

@article{Baiguera_PhysRep_2026_QuantumComplexityReview,
  author = {Baiguera, Stefano and Balasubramanian, Vijay and Caputa, Pawel and Chapman, Shira and Haferkamp, Jonas and Heller, Michal P. and Halpern, Nicole Yunger},
  title = {Quantum Complexity in Gravity, Quantum Field Theory, and Quantum Information Science},
  journal = {Physics Reports},
  volume = {1159},
  pages = {1\textendash 77},
  year = {2026},
  doi = {10.1016/j.physrep.2025.12.001},
  eprint = {2503.10753},
  archivePrefix = {arXiv},
  primaryClass = {hep-th}
}

@misc{Rabinovici_arXiv_2025_KrylovComplexity,
  author = {Rabinovici, Eliezer and S{\'a}nchez-Garrido, Aitor and Shir, Ruth and Sonner, Julian},
  title = {Krylov Complexity},
  year = {2025},
  eprint = {2507.06286},
  archivePrefix = {arXiv},
  primaryClass = {hep-th}
}

@article{Barbon_JHEP_2019_OperatorComplexity,
  author = {Barb{\'o}n, J. L. F. and Rabinovici, Eliezer and Shir, Ruth and Sinha, Robert},
  title = {On the Evolution of Operator Complexity Beyond Scrambling},
  journal = {Journal of High Energy Physics},
  volume = {2019},
  pages = {264},
  year = {2019},
  doi = {10.1007/JHEP10(2019)264},
  eprint = {1907.05393},
  archivePrefix = {arXiv},
  primaryClass = {hep-th}
}

@article{Rabinovici_JHEP_2021_EdgeKrylovSpace,
  author = {Rabinovici, Eliezer and S{\'a}nchez-Garrido, Aitor and Shir, Ruth and Sonner, Julian},
  title = {Operator Complexity: A Journey to the Edge of Krylov Space},
  journal = {Journal of High Energy Physics},
  volume = {2021},
  pages = {62},
  year = {2021},
  doi = {10.1007/JHEP06(2021)062},
  eprint = {2009.01862},
  archivePrefix = {arXiv},
  primaryClass = {hep-th}
}

@article{Caputa_PRR_2022_GeometryKrylov,
  author = {Caputa, Pawel and Mag{\'a}n, Javier M. and Patramanis, Dimitrios},
  title = {Geometry of Krylov Complexity},
  journal = {Physical Review Research},
  volume = {4},
  pages = {013041},
  year = {2022},
  doi = {10.1103/PhysRevResearch.4.013041},
  eprint = {2109.03824},
  archivePrefix = {arXiv},
  primaryClass = {hep-th}
}

@article{Balasubramanian_PRD_2022_ComplexitySpread,
  author = {Balasubramanian, Vijay and Caputa, Pawel and Mag{\'a}n, Javier M. and Wu, Qingyue},
  title = {Quantum Chaos and the Complexity of Spread of States},
  journal = {Physical Review D},
  volume = {106},
  pages = {046007},
  year = {2022},
  doi = {10.1103/PhysRevD.106.046007},
  eprint = {2202.06957},
  archivePrefix = {arXiv},
  primaryClass = {hep-th}
}

@article{Dymarsky_Gorsky_PRB_2020_Delocalization,
  author = {Dymarsky, Anatoly and Gorsky, Alexander},
  title = {Quantum Chaos as Delocalization in Krylov Space},
  journal = {Physical Review B},
  volume = {102},
  pages = {085137},
  year = {2020},
  doi = {10.1103/PhysRevB.102.085137},
  eprint = {1912.12227},
  archivePrefix = {arXiv},
  primaryClass = {cond-mat.stat-mech}
}

@article{Dymarsky_Smolkin_PRD_2021_CFT,
  author = {Dymarsky, Anatoly and Smolkin, Michael},
  title = {Krylov Complexity in Conformal Field Theory},
  journal = {Physical Review D},
  volume = {104},
  pages = {L081702},
  year = {2021},
  doi = {10.1103/PhysRevD.104.L081702},
  eprint = {2104.09514},
  archivePrefix = {arXiv},
  primaryClass = {hep-th}
}

@article{Rabinovici_JHEP_2022_IntegrabilityChaos,
  author = {Rabinovici, Eliezer and S{\'a}nchez-Garrido, Aitor and Shir, Ruth and Sonner, Julian},
  title = {Krylov Complexity from Integrability to Chaos},
  journal = {Journal of High Energy Physics},
  volume = {2022},
  pages = {151},
  year = {2022},
  doi = {10.1007/JHEP07(2022)151},
  eprint = {2207.07701},
  archivePrefix = {arXiv},
  primaryClass = {hep-th}
}

@article{Avdoshkin_JHEP_2024_QFTBeyond,
  author = {Avdoshkin, Alexander and Dymarsky, Anatoly and Smolkin, Michael},
  title = {Krylov Complexity in Quantum Field Theory, and Beyond},
  journal = {Journal of High Energy Physics},
  volume = {2024},
  pages = {066},
  year = {2024},
  doi = {10.1007/JHEP06(2024)066},
  eprint = {2212.14429},
  archivePrefix = {arXiv},
  primaryClass = {hep-th}
}

@article{Hornedal_CommPhys_2022_UltimateSpeedLimits,
  author = {H{\"o}rnedal, Niklas and Carabba, Niccol{\`o} and Matsoukas-Roubeas, Apollonas S. and del Campo, Adolfo},
  title = {Ultimate Speed Limits to the Growth of Operator Complexity},
  journal = {Communications Physics},
  volume = {5},
  pages = {207},
  year = {2022},
  doi = {10.1038/s42005-022-00985-1},
  eprint = {2202.05006},
  archivePrefix = {arXiv},
  primaryClass = {quant-ph}
}

@article{Takahashi_PRX_2024_STA_Krylov,
  author = {Takahashi, Kazutaka and del Campo, Adolfo},
  title = {Shortcuts to Adiabaticity in Krylov Space},
  journal = {Physical Review X},
  volume = {14},
  pages = {011032},
  year = {2024},
  doi = {10.1103/PhysRevX.14.011032},
  eprint = {2302.05460},
  archivePrefix = {arXiv},
  primaryClass = {quant-ph}
}

@article{Visuri_npjQI_2026_DigitizedCD,
  author = {Visuri, A.-M. and Gomez Cadavid, A. and Bhargava, B. A. and Romero, S. V. and Grabarits, A. and Chandarana, P. and Solano, E. and del Campo, A. and Hegade, N. N.},
  title = {Digitized Counterdiabatic Quantum Critical Dynamics},
  journal = {npj Quantum Information},
  volume = {12},
  pages = {47},
  year = {2026},
  doi = {10.1038/s41534-026-00994-0}
}

@article{Caputa_JHEP_2023_KitaevTransitions,
  author = {Caputa, Pawel and Gupta, Nitin and Haque, S. Shajidul and Liu, Sinong and Murugan, Jeff and Van Zyl, Hendrik J. R.},
  title = {Spread Complexity and Topological Transitions in the Kitaev Chain},
  journal = {Journal of High Energy Physics},
  volume = {2023},
  pages = {120},
  year = {2023},
  doi = {10.1007/JHEP01(2023)120},
  eprint = {2208.06311},
  archivePrefix = {arXiv},
  primaryClass = {hep-th}
}

@article{Bento_PRB_2024_LMGDynamicalTransition,
  author = {Bento, P. H. S. and del Campo, A. and C{\'e}leri, L. C.},
  title = {Krylov Complexity and Dynamical Phase Transition in the Quenched Lipkin-Meshkov-Glick Model},
  journal = {Physical Review B},
  volume = {109},
  pages = {224304},
  year = {2024},
  doi = {10.1103/PhysRevB.109.224304}
}

@misc{Grabarits_arXiv_2025_UniversalGrowthQPT,
  author = {Grabarits, Andr{\'a}s and del Campo, Adolfo},
  title = {Universal Growth of Krylov Complexity Across A Quantum Phase Transition},
  year = {2025},
  eprint = {2510.13947},
  archivePrefix = {arXiv},
  primaryClass = {quant-ph},
  doi = {10.48550/arXiv.2510.13947}
}

@article{Takahashi_PRB_2026_TodaFlows,
  author = {Takahashi, Kazutaka and Nandy, Pratik and del Campo, Adolfo},
  title = {Krylov Complexity under Hamiltonian Deformations and Toda Flows},
  journal = {Physical Review B},
  volume = {113},
  pages = {144312},
  year = {2026},
  doi = {10.1103/PhysRevB.113.144312}
}

@article{Nizami_PRE_2023_DrivenKrylov,
  author = {Nizami, A. A. and Shrestha, A. W.},
  title = {Krylov Construction and Complexity for Driven Quantum Systems},
  journal = {Physical Review E},
  volume = {108},
  pages = {054222},
  year = {2023},
  doi = {10.1103/PhysRevE.108.054222}
}

@article{Nizami_PRE_2024_PeriodicallyDriven,
  author = {Nizami, A. A. and Shrestha, A. W.},
  title = {Spread Complexity and Quantum Chaos for Periodically Driven Spin Chains},
  journal = {Physical Review E},
  volume = {110},
  pages = {034201},
  year = {2024},
  doi = {10.1103/PhysRevE.110.034201}
}

@article{Suchsland_PRB_2025_TrotterTransitions,
  author = {Suchsland, Philippe and Moessner, Roderich and Claeys, Pieter W.},
  title = {Krylov Complexity and Trotter Transitions in Unitary Circuit Dynamics},
  journal = {Physical Review B},
  volume = {111},
  pages = {014309},
  year = {2025},
  doi = {10.1103/PhysRevB.111.014309}
}

@article{Takahashi_PRL_2025_TimeDependentGenerators,
  author = {Takahashi, Kazutaka and del Campo, Adolfo},
  title = {Krylov Subspace Methods for Quantum Dynamics with Time-Dependent Generators},
  journal = {Physical Review Letters},
  volume = {134},
  pages = {030401},
  year = {2025},
  doi = {10.1103/PhysRevLett.134.030401}
}

@misc{Grabarits_arXiv_2026_TDKrylovLie,
  author = {Grabarits, Andr{\'a}s and Medina-Guerra, E. and del Campo, Adolfo},
  title = {Krylov Dynamics and Operator Growth in Time-Dependent Systems via Lie Algebras},
  year = {2026},
  eprint = {2605.05290},
  archivePrefix = {arXiv},
  primaryClass = {quant-ph},
  doi = {10.48550/arXiv.2605.05290}
}

@article{Chen_PRL_2010_FastCooling,
  author = {Chen, X. and Ruschhaupt, A. and Schmidt, S. and del Campo, A. and Gu{\'e}ry-Odelin, D. and Muga, J. G.},
  title = {Fast Optimal Frictionless Atom Cooling in Harmonic Traps},
  journal = {Phys. Rev. Lett.},
  volume = {104},
  pages = {063002},
  year = {2010},
  doi = {10.1103/PhysRevLett.104.063002}
}

@article{delCampo_PRA_2011_FrictionlessQuenches,
  author = {del Campo, A.},
  title = {Frictionless Quantum Quenches in Ultracold Gases: A Quantum-Dynamical Microscope},
  journal = {Phys. Rev. A},
  volume = {84},
  pages = {031606(R)},
  year = {2011},
  doi = {10.1103/PhysRevA.84.031606}
}

@article{delCampo_Boshier_SciRep_2012_Box,
  author = {del Campo, A. and Boshier, M. G.},
  title = {Shortcuts to Adiabaticity in a Time-Dependent Box},
  journal = {Sci. Rep.},
  volume = {2},
  pages = {648},
  year = {2012},
  doi = {10.1038/srep00648}
}

@article{delCampo_PRL_2013_CounterdiabaticDriving,
  author = {del Campo, A.},
  title = {Shortcuts to Adiabaticity by Counter-Diabatic Driving},
  journal = {Phys. Rev. Lett.},
  volume = {111},
  pages = {100502},
  year = {2013},
  doi = {10.1103/PhysRevLett.111.100502},
  eprint = {1306.0410},
  archivePrefix = {arXiv}
}

@article{Deffner_PRX_2014_ScaleInvariantSTA,
  author = {Deffner, S. and Jarzynski, C. and del Campo, A.},
  title = {Classical and Quantum Shortcuts to Adiabaticity for Scale-Invariant Driving},
  journal = {Phys. Rev. X},
  volume = {4},
  pages = {021013},
  year = {2014},
  doi = {10.1103/PhysRevX.4.021013},
  eprint = {1401.1184},
  archivePrefix = {arXiv}
}

@article{CastinDum_PRL_1996_BECScaling,
  author = {Castin, Y. and Dum, R.},
  title = {Bose-Einstein Condensates in Time Dependent Traps},
  journal = {Phys. Rev. Lett.},
  volume = {77},
  pages = {5315},
  year = {1996},
  doi = {10.1103/PhysRevLett.77.5315}
}

@article{Kagan_PRA_1996_BECExpansion,
  author = {Kagan, Yu. and Surkov, E. L. and Shlyapnikov, G. V.},
  title = {Evolution of a Bose-Condensed Gas under Variations of the Confining Potential},
  journal = {Phys. Rev. A},
  volume = {54},
  pages = {R1753},
  year = {1996},
  doi = {10.1103/PhysRevA.54.R1753}
}

@article{Pitaevskii_Rosch_PRA_1997_Breathing,
  author = {Pitaevskii, L. P. and Rosch, A.},
  title = {Breathing Modes and Hidden Symmetry of Trapped Atoms in Two Dimensions},
  journal = {Phys. Rev. A},
  volume = {55},
  pages = {R853},
  year = {1997},
  doi = {10.1103/PhysRevA.55.R853}
}

@article{Girardeau_OptComm_2004_TonksScaling,
  author = {Girardeau, M. D. and Nguyen, Hieu and Olshanii, M.},
  title = {Effective interactions, {Fermi--Bose} duality, and ground states of ultracold atomic vapors in tight de {Broglie} waveguides},
  journal = {Opt. Commun.},
  volume = {243},
  pages = {3--22},
  year = {2004},
  doi = {10.1016/j.optcom.2004.09.079}
}

@article{Sutherland_PRL_1998_CalogeroScaling,
  author = {Sutherland, B.},
  title = {Exact Coherent States of a One-Dimensional Quantum Fluid in a Time-Dependent Trapping Potential},
  journal = {Phys. Rev. Lett.},
  volume = {80},
  pages = {3678},
  year = {1998},
  doi = {10.1103/PhysRevLett.80.3678}
}

@article{Yang2025,
  title = {Time-dependent Jastrow ansatz: Exact quantum dynamics, shortcuts to adiabaticity, and quantum quenches in strongly correlated many-body systems},
  author = {Yang, Jing and del Campo, Adolfo},
  journal = {Phys. Rev. A},
  volume = {111},
  issue = {5},
  pages = {053315},
  numpages = {23},
  year = {2025},
  month = {May},
  publisher = {American Physical Society},
  doi = {10.1103/PhysRevA.111.053315},
  url = {https://link.aps.org/doi/10.1103/PhysRevA.111.053315}
}

@article{delcampo2016,
doi = {10.1088/1367-2630/18/1/015014},
url = {https://doi.org/10.1088/1367-2630/18/1/015014},
year = {2016},
month = {jan},
publisher = {IOP Publishing},
volume = {18},
number = {1},
pages = {015014},
author = {Campo, Adolfo del},
title = {Exact quantum decay of an interacting many-particle system: the Calogero–Sutherland model},
journal = {New Journal of Physics}
}

@article{Haldane_PRL_1988_HaldaneShastry,
  author = {Haldane, F. D. M.},
  title = {Exact Jastrow-Gutzwiller Resonating-Valence-Bond Ground State of the Spin-1/2 Antiferromagnetic Heisenberg Chain with 1/r2 Exchange},
  journal = {Physical Review Letters},
  volume = {60},
  number = {7},
  pages = {635-638},
  year = {1988},
  doi = {10.1103/PhysRevLett.60.635}
}

@article{Shastry_PRL_1988_HaldaneShastry,
  author = {Shastry, B. Sriram},
  title = {Exact Solution of an S=1/2 Heisenberg Antiferromagnetic Chain with Long-Ranged Interactions},
  journal = {Physical Review Letters},
  volume = {60},
  number = {7},
  pages = {639-642},
  year = {1988},
  doi = {10.1103/PhysRevLett.60.639}
}

@article{Gaudin_JPhys_1976_SpinHamiltonians,
  author = {Gaudin, M.},
  title = {Diagonalisation d'une classe d'hamiltoniens de spin},
  journal = {Journal de Physique},
  volume = {37},
  number = {10},
  pages = {1087-1098},
  year = {1976},
  doi = {10.1051/jphys:0197600370100108700}
}

@article{Lieb_Liniger_PR_1963_ExactAnalysis,
  author = {Lieb, Elliott H. and Liniger, Werner},
  title = {Exact Analysis of an Interacting Bose Gas. I. The General Solution and the Ground State},
  journal = {Physical Review},
  volume = {130},
  number = {4},
  pages = {1605--1616},
  year = {1963},
  doi = {10.1103/PhysRev.130.1605}
}

@article{Olshanii_PRL_1998_AtomicScattering,
  author = {Olshanii, Maxim},
  title = {Atomic Scattering in the Presence of an External Confinement and a Gas of Impenetrable Bosons},
  journal = {Physical Review Letters},
  volume = {81},
  number = {5},
  pages = {938--941},
  year = {1998},
  doi = {10.1103/PhysRevLett.81.938},
  eprint = {cond-mat/9804130},
  archivePrefix = {arXiv}
}

@article{Moshinsky_JMP_1968_HarmonicInteraction,
  author = {Moshinsky, Marcos},
  title = {How Good Is the Hartree-Fock Approximation?},
  journal = {American Journal of Physics},
  volume = {36},
  number = {1},
  pages = {52--56},
  year = {1968},
  doi = {10.1119/1.1974410}
}

@article{Caputa_JHEP_2024_DensityMatrixKrylov,
  author        = {Caputa, Pawel and Jeong, Hyun-Sik and Liu, Sinong and Pedraza, Juan F. and Qu, Le-Chen},
  title         = {{Krylov complexity of density matrix operators}},
  journal       = {Journal of High Energy Physics},
  volume        = {2024},
  number        = {5},
  pages         = {337},
  year          = {2024},
  doi           = {10.1007/JHEP05(2024)337},
  eprint        = {2402.09522},
  archivePrefix = {arXiv},
  primaryClass  = {hep-th}
}

@article{Tan_PRE_2025_FiniteTemperatureKrylov,
  author        = {Tan, Chengming and Wei, Zhiyang and Zhang, Ren},
  title         = {{Scaling relations of spectral form factor and Krylov complexity at finite temperature}},
  journal       = {Physical Review E},
  volume        = {111},
  number        = {1},
  pages         = {014135},
  year          = {2025},
  doi           = {10.1103/PhysRevE.111.014135},
  eprint        = {2401.10499},
  archivePrefix = {arXiv},
  primaryClass  = {cond-mat.stat-mech}
}

@article{Holten_PRL_2018_ScaleAnomaly,
  author        = {Holten, Marvin and Bayha, Luca and Klein, Antonia C. and Murthy, Puneet A. and Preiss, Philipp M. and Jochim, Selim},
  title         = {{Anomalous Breaking of Scale Invariance in a Two-Dimensional Fermi Gas}},
  journal       = {Physical Review Letters},
  volume        = {121},
  number        = {12},
  pages         = {120401},
  year          = {2018},
  doi           = {10.1103/PhysRevLett.121.120401},
  eprint        = {1803.08879},
  archivePrefix = {arXiv},
  primaryClass  = {cond-mat.quant-gas}
}

@article{SaintJalm_PRX_2019_DynamicalSymmetry,
  author        = {Saint-Jalm, Rapha{\"e}l and Castilho, Patricia C. M. and Le Cerf, {\'E}douard and Bakkali-Hassani, Brice and Ville, Jean-Loup and Nascimbene, Sylvain and Beugnon, J{\'e}r{\^o}me and Dalibard, Jean},
  title         = {{Dynamical Symmetry and Breathers in a Two-Dimensional Bose Gas}},
  journal       = {Physical Review X},
  volume        = {9},
  number        = {2},
  pages         = {021035},
  year          = {2019},
  doi           = {10.1103/PhysRevX.9.021035},
  eprint        = {1903.04528},
  archivePrefix = {arXiv},
  primaryClass  = {cond-mat.quant-gas}
}

@article{Monroe_RMP_2021_IonSpinSim,
  author        = {Monroe, C. and Campbell, W. C. and Duan, L.-M. and Gong, Z.-X. and Gorshkov, A. V. and Hess, P. W. and Islam, R. and Kim, K. and Linke, N. M. and Pagano, G. and Richerme, P. and Senko, C. and Yao, N. Y.},
  title         = {{Programmable quantum simulations of spin systems with trapped ions}},
  journal       = {Reviews of Modern Physics},
  volume        = {93},
  number        = {2},
  pages         = {025001},
  year          = {2021},
  doi           = {10.1103/RevModPhys.93.025001},
  eprint        = {1912.07845},
  archivePrefix = {arXiv},
  primaryClass  = {quant-ph}
}

@article{Schaff2010,
  title = {Fast optimal transition between two equilibrium states},
  author = {Schaff, Jean-Fran\ifmmode \mbox{\c{c}}\else \c{c}\fi{}ois and Song, Xiao-Li and Vignolo, Patrizia and Labeyrie, Guillaume},
  journal = {Phys. Rev. A},
  volume = {82},
  issue = {3},
  pages = {033430},
  numpages = {5},
  year = {2010},
  month = {Sep},
  publisher = {American Physical Society},
  doi = {10.1103/PhysRevA.82.033430},
  url = {https://link.aps.org/doi/10.1103/PhysRevA.82.033430}
}

@article{Jaramillo2016,
  author  = {Jaramillo, Juan and Beau, Mathieu and del Campo, Adolfo},
  title   = {Quantum Supremacy of Many-Particle Thermal Machines},
  journal = {New Journal of Physics},
  volume  = {18},
  number  = {7},
  pages   = {075019},
  year    = {2016},
  doi     = {10.1088/1367-2630/18/7/075019},
  eprint  = {1510.04633},
  archivePrefix = {arXiv},
  primaryClass  = {quant-ph}
}

@article{Beau2016,
  author  = {Beau, Mathieu and Jaramillo, Juan and del Campo, Adolfo},
  title   = {Scaling-Up Quantum Heat Engines Efficiently via Shortcuts to Adiabaticity},
  journal = {Entropy},
  volume  = {18},
  number  = {5},
  pages   = {168},
  year    = {2016},
  doi     = {10.3390/e18050168},
  eprint  = {1603.06019},
  archivePrefix = {arXiv},
  primaryClass  = {quant-ph}
}

@article{Dupays21,
  title = {Delta-kick cooling, time-optimal control of scale-invariant dynamics, and shortcuts to adiabaticity assisted by kicks},
  author = {Dupays, L\'eonce and Spierings, David C. and Steinberg, Aephraim M. and del Campo, Adolfo},
  journal = {Phys. Rev. Res.},
  volume = {3},
  issue = {3},
  pages = {033261},
  numpages = {14},
  year = {2021},
  month = {Sep},
  publisher = {American Physical Society},
  doi = {10.1103/PhysRevResearch.3.033261},
  url = {https://link.aps.org/doi/10.1103/PhysRevResearch.3.033261}
}

@incollection{Ketterle1999,
  author    = {Ketterle, W. and Durfee, D. S. and Stamper-Kurn, D. M.},
  title     = {Making, probing and understanding {Bose-Einstein} condensates},
  booktitle = {Bose-Einstein Condensation in Atomic Gases},
  series    = {Proceedings of the International School of Physics ``Enrico Fermi''},
  volume    = {140},
  note      = {Course CXL},
  editor    = {Inguscio, M. and Stringari, S. and Wieman, C. E.},
  publisher = {IOS Press},
  address   = {Amsterdam},
  year      = {1999},
  pages     = {67--176},
  eprint    = {cond-mat/9904034},
  archivePrefix = {arXiv}
}

@article{Schaff2011,
doi = {10.1209/0295-5075/93/23001},
url = {https://doi.org/10.1209/0295-5075/93/23001},
year = {2011},
month = {jan},
publisher = {},
volume = {93},
number = {2},
pages = {23001},
author = {Schaff, J.-F. and Song, X.-L. and Capuzzi, P. and Vignolo, P. and Labeyrie, G.},
title = {Shortcut to adiabaticity for an interacting Bose-Einstein condensate},
journal = {Europhysics Letters}
}

@article{rohringer2015nonequilibrium,
  title   = {Non-equilibrium scale invariance and shortcuts to adiabaticity in a one-dimensional Bose gas},
  author  = {Rohringer, W. and Fischer, D. and Steiner, F. and Mazets, I. E. and Schmiedmayer, J. and Trupke, M.},
  journal = {Scientific Reports},
  volume  = {5},
  pages   = {9820},
  year    = {2015},
  doi     = {10.1038/srep09820}
}

@article{elliott2014conformal,
  title   = {Observation of Conformal Symmetry Breaking and Scale Invariance in Expanding Fermi Gases},
  author  = {Elliott, E. and Joseph, J. A. and Thomas, J. E.},
  journal = {Physical Review Letters},
  volume  = {112},
  pages   = {040405},
  year    = {2014},
  doi     = {10.1103/PhysRevLett.112.040405}
}

@article{deng2016efimovian,
  title   = {Observation of the Efimovian expansion in scale-invariant Fermi gases},
  author  = {Deng, Shujin and Shi, Zhe-Yu and Diao, Pengpeng and Yu, Qianli and Zhai, Hui and Qi, Ran and Wu, Haibin},
  journal = {Science},
  volume  = {353},
  number  = {6297},
  pages   = {371--374},
  year    = {2016},
  doi     = {10.1126/science.aaf0666}
}

@article{Deng2018Efimov,
  title = {Observation of Dynamical Super-Efimovian Expansion in a Unitary Fermi Gas},
  author = {Deng, Shujin and Diao, Pengpeng and Li, Fang and Yu, Qianli and Yu, Shi and Wu, Haibin},
  journal = {Phys. Rev. Lett.},
  volume = {120},
  issue = {12},
  pages = {125301},
  numpages = {5},
  year = {2018},
  month = {Mar},
  publisher = {American Physical Society},
  doi = {10.1103/PhysRevLett.120.125301},
  url = {https://link.aps.org/doi/10.1103/PhysRevLett.120.125301}
}

@article{Diao2018,
doi = {10.1088/1367-2630/aae45e},
url = {https://doi.org/10.1088/1367-2630/aae45e},
year = {2018},
month = {oct},
publisher = {IOP Publishing},
volume = {20},
number = {10},
pages = {105004},
author = {Diao, Pengpeng and Deng, Shujin and Li, Fang and Yu, Shi and Chenu, Aurélia and del Campo, Adolfo and Wu, Haibin},
title = {Shortcuts to adiabaticity in Fermi gases},
journal = {New Journal of Physics}
}

@article{Deng2018,
  title = {Shortcuts to adiabaticity in the strongly coupled regime: Nonadiabatic control of a unitary Fermi gas},
  author = {Deng, Shujin and Diao, Pengpeng and Yu, Qianli and del Campo, Adolfo and Wu, Haibin},
  journal = {Phys. Rev. A},
  volume = {97},
  issue = {1},
  pages = {013628},
  numpages = {7},
  year = {2018},
  month = {Jan},
  publisher = {American Physical Society},
  doi = {10.1103/PhysRevA.97.013628},
  url = {https://link.aps.org/doi/10.1103/PhysRevA.97.013628}
}

@article{delcampo2008,
  title = {Fermionization and bosonization of expanding one-dimensional anyonic fluids},
  author = {del Campo, A.},
  journal = {Phys. Rev. A},
  volume = {78},
  issue = {4},
  pages = {045602},
  numpages = {4},
  year = {2008},
  month = {Oct},
  publisher = {American Physical Society},
  doi = {10.1103/PhysRevA.78.045602},
  url = {https://link.aps.org/doi/10.1103/PhysRevA.78.045602}
}

@article{Garttner_NatPhys_2017_OTOC,
  author        = {G{\"a}rttner, Martin and Bohnet, Justin G. and Safavi-Naini, Arghavan and Wall, Michael L. and Bollinger, John J. and Rey, Ana Maria},
  title         = {{Measuring out-of-time-order correlations and multiple quantum spectra in a trapped-ion quantum magnet}},
  journal       = {Nature Physics},
  volume        = {13},
  pages         = {781--786},
  year          = {2017},
  doi           = {10.1038/nphys4119},
  eprint        = {1608.08938},
  archivePrefix = {arXiv},
  primaryClass  = {quant-ph}
}

@article{Minguzzi_PRL_2005_TonksScaling,
  title = {Exact Coherent States of a Harmonically Confined {Tonks-Girardeau} Gas},
  author = {Minguzzi, A. and Gangardt, D. M.},
  journal = {Phys. Rev. Lett.},
  volume = {94},
  pages = {240404},
  year = {2005},
  doi = {10.1103/PhysRevLett.94.240404}
}

@article{Fang_PRL_2014_1DBreathing,
  title = {Quench-Induced Breathing Mode of One-Dimensional {Bose} Gases},
  author = {Fang, Bess and Carleo, Giuseppe and Johnson, Aisling and Bouchoule, Isabelle},
  journal = {Phys. Rev. Lett.},
  volume = {113},
  pages = {035301},
  year = {2014},
  doi = {10.1103/PhysRevLett.113.035301}
}

@article{Pollack_PRA_2010_ScatteringLengthModulation,
  title = {Collective excitation of a Bose-Einstein condensate by modulation of the atomic scattering length},
  author = {Pollack, S. E. and Dries, D. and Hulet, R. G. and Magalh{\~a}es, K. M. F. and Henn, E. A. L. and Ramos, E. R. F. and Caracanhas, M. A. and Bagnato, V. S.},
  journal = {Phys. Rev. A},
  volume = {81},
  pages = {053627},
  year = {2010},
  doi = {10.1103/PhysRevA.81.053627}
}

@article{Friebel_PRA_1998_CO2Lattice,
  title = {{CO$_2$}-laser optical lattice with cold rubidium atoms},
  author = {Friebel, S. and D'Andrea, C. and Walz, J. and Weitz, M. and H{\"a}nsch, T. W.},
  journal = {Phys. Rev. A},
  volume = {57},
  pages = {R20--R23},
  year = {1998},
  doi = {10.1103/PhysRevA.57.R20}
}

@article{Balik_PRA_2009_ImagingParametricResonance,
  title = {Imaging-based parametric resonance in an optical dipole trap},
  author = {Balik, S. and Win, A. L. and Havey, M. D.},
  journal = {Phys. Rev. A},
  volume = {80},
  pages = {023404},
  year = {2009},
  doi = {10.1103/PhysRevA.80.023404}
}

@article{Shu_PRL_2018_FloquetRaman,
  title   = {Observation of Floquet Raman Transition in a Driven Solid-State Spin System},
  author  = {Shu, Zijun and Liu, Yu and Cao, Qingyun and Yang, Pengcheng and Zhang, Shaoliang and Plenio, Martin B. and Jelezko, Fedor and Cai, Jianming},
  journal = {Phys. Rev. Lett.},
  volume  = {121},
  issue   = {21},
  pages   = {210501},
  year    = {2018},
  doi     = {10.1103/PhysRevLett.121.210501}
}

@article{Grossmann_PRL_1991_CoherentDestruction,
  title   = {Coherent Destruction of Tunneling},
  author  = {Grossmann, F. and Dittrich, T. and Jung, P. and H\"anggi, P.},
  journal = {Phys. Rev. Lett.},
  volume  = {67},
  issue   = {4},
  pages   = {516--519},
  year    = {1991},
  doi     = {10.1103/PhysRevLett.67.516}
}

@article{Hung_Nature_2011_ScaleInvariance2D,
  title = {Observation of scale invariance and universality in two-dimensional Bose gases},
  author = {Hung, Chen-Lung and Zhang, Xibo and Gemelke, Nathan and Chin, Cheng},
  journal = {Nature},
  volume = {470},
  pages = {236--239},
  year = {2011},
  doi = {10.1038/nature09722}
}

@article{Gemelke_Nature_2009_InSituImaging,
  title = {In situ observation of incompressible Mott-insulating domains in ultracold atomic gases},
  author = {Gemelke, Nathan and Zhang, Xibo and Hung, Chen-Lung and Chin, Cheng},
  journal = {Nature},
  volume = {460},
  pages = {995--998},
  year = {2009},
  doi = {10.1038/nature08244}
}

@article{Vogt_PRL_2012_ScaleInvariance2DFermi,
  title = {Scale Invariance and Viscosity of a Two-Dimensional Fermi Gas},
  author = {Vogt, Enrico and Feld, Marco and Fr\"ohlich, Bernd and Pertot, Daniel and Koschorreck, Marco and K\"ohl, Michael},
  journal = {Phys. Rev. Lett.},
  volume = {108},
  pages = {070404},
  year = {2012},
  doi = {10.1103/PhysRevLett.108.070404}
}

@article{Kinoshita_Nature_2006_NewtonsCradle,
  title = {A quantum Newton's cradle},
  author = {Kinoshita, Toshiya and Wenger, Trevor and Weiss, David S.},
  journal = {Nature},
  volume = {440},
  pages = {900--903},
  year = {2006},
  doi = {10.1038/nature04693}
}

@article{Chevy_PRL_2002_TransverseMode,
  title = {Transverse Breathing Mode of an Elongated Bose-Einstein Condensate},
  author = {Chevy, F. and Bretin, V. and Rosenbusch, P. and Madison, K. W. and Dalibard, J.},
  journal = {Phys. Rev. Lett.},
  volume = {88},
  pages = {250402},
  year = {2002},
  doi = {10.1103/PhysRevLett.88.250402}
}

@article{Moritz_PRL_2003_CollectiveOscillations1D,
  title = {Exciting Collective Oscillations in a Trapped 1D Gas},
  author = {Moritz, H. and St\"oferle, T. and K\"ohl, M. and Esslinger, T.},
  journal = {Phys. Rev. Lett.},
  volume = {91},
  pages = {250402},
  year = {2003},
  doi = {10.1103/PhysRevLett.91.250402}
}

@article{Haller_Science_2009_ExcitedGasPhase,
  title = {Realization of an Excited, Strongly Correlated Quantum Gas Phase},
  author = {Haller, Elmar and Gustavsson, Mattias and Mark, Manfred J. and Danzl, Johann G. and Hart, Russell and Pupillo, Guido and N\"agerl, Hanns-Christoph},
  journal = {Science},
  volume = {325},
  pages = {1224--1227},
  year = {2009},
  doi = {10.1126/science.1175850}
}

@article{Lignier_PRL_2007_DynamicalControlTunneling,
  title = {Dynamical Control of Matter-Wave Tunneling in Periodic Potentials},
  author = {Lignier, H. and Sias, C. and Ciampini, D. and Singh, Y. and Zenesini, A. and Morsch, O. and Arimondo, E.},
  journal = {Phys. Rev. Lett.},
  volume = {99},
  pages = {220403},
  year = {2007},
  doi = {10.1103/PhysRevLett.99.220403}
}

@article{Kierig_PRL_2008_DrivenDoubleWell,
  title = {Single-Particle Tunneling in Strongly Driven Double-Well Potentials},
  author = {Kierig, E. and Schnorrberger, U. and Schietinger, A. and Tomkovic, J. and Oberthaler, M. K.},
  journal = {Phys. Rev. Lett.},
  volume = {100},
  pages = {190405},
  year = {2008},
  doi = {10.1103/PhysRevLett.100.190405}
}

@article{Eckardt_RMP_2017_PeriodicallyDriven,
  title = {Colloquium: Atomic quantum gases in periodically driven optical lattices},
  author = {Eckardt, Andr\'e},
  journal = {Rev. Mod. Phys.},
  volume = {89},
  pages = {011004},
  year = {2017},
  doi = {10.1103/RevModPhys.89.011004}
}

@article{Britton_Nature_2012_EngineeredIsing,
  title = {Engineered two-dimensional Ising interactions in a trapped-ion quantum simulator with hundreds of spins},
  author = {Britton, Joseph W. and Sawyer, Brian C. and Keith, Adam C. and Wang, C.-C. Joseph and Freericks, James K. and Uys, Hermann and Biercuk, Michael J. and Bollinger, John J.},
  journal = {Nature},
  volume = {484},
  pages = {489--492},
  year = {2012},
  doi = {10.1038/nature10981}
}

@article{Bohnet_Science_2016_QuantumSpinDynamics,
  title = {Quantum spin dynamics and entanglement generation with hundreds of trapped ions},
  author = {Bohnet, Justin G. and Sawyer, Brian C. and Britton, Joseph W. and Wall, Michael L. and Rey, Ana Maria and Foss-Feig, Michael and Bollinger, John J.},
  journal = {Science},
  volume = {352},
  pages = {1297--1301},
  year = {2016},
  doi = {10.1126/science.aad9958}
}

@article{Hosten_Nature_2016_MeasurementNoise,
  title = {Measurement noise 100 times lower than the quantum-projection limit using entangled atoms},
  author = {Hosten, Onur and Engelsen, Nils J. and Krishnakumar, Rajiv and Kasevich, Mark A.},
  journal = {Nature},
  volume = {529},
  pages = {505--508},
  year = {2016},
  doi = {10.1038/nature16176}
}

@article{Childress_Science_2006_CoherentDynamics,
  title = {Coherent Dynamics of Coupled Electron and Nuclear Spin Qubits in Diamond},
  author = {Childress, L. and Gurudev Dutt, M. V. and Taylor, J. M. and Zibrov, A. S. and Jelezko, F. and Wrachtrup, J. and Hemmer, P. R. and Lukin, M. D.},
  journal = {Science},
  volume = {314},
  pages = {281--285},
  year = {2006},
  doi = {10.1126/science.1131871}
}

@book{Edmonds_1957_AngularMomentum,
  author    = {Edmonds, A. R.},
  title     = {Angular Momentum in Quantum Mechanics},
  series    = {Investigations in Physics},
  number    = {4},
  publisher = {Princeton University Press},
  address   = {Princeton, NJ},
  year      = {1957},
  doi       = {10.1515/9781400884186}
}

@book{Varshalovich_1988_QuantumTheoryAngularMomentum,
  author    = {Varshalovich, D. A. and Moskalev, A. N. and Khersonskii, V. K.},
  title     = {Quantum Theory of Angular Momentum},
  publisher = {World Scientific},
  address   = {Singapore},
  year      = {1988},
  doi       = {10.1142/0270}
}

@article{Schliemann_JPCM_2003_ElectronSpinDynamics,
  author  = {Schliemann, John and Khaetskii, Alexander and Loss, Daniel},
  title   = {Electron spin dynamics in quantum dots and related nanostructures due to hyperfine interaction with nuclei},
  journal = {J. Phys.: Condens. Matter},
  volume  = {15},
  number  = {50},
  pages   = {R1809},
  year    = {2003},
  doi     = {10.1088/0953-8984/15/50/R01}
}

@article{Coish_PRB_2004_HyperfineQuantumDot,
  author  = {Coish, W. A. and Loss, Daniel},
  title   = {Hyperfine interaction in a quantum dot: Non-Markovian electron spin dynamics},
  journal = {Phys. Rev. B},
  volume  = {70},
  pages   = {195340},
  year    = {2004},
  doi     = {10.1103/PhysRevB.70.195340}
}

@article{Merkulov_PRB_2002_ElectronSpinRelaxation,
  author  = {Merkulov, I. A. and Efros, Al. L. and Rosen, M.},
  title   = {Electron spin relaxation by nuclei in semiconductor quantum dots},
  journal = {Phys. Rev. B},
  volume  = {65},
  pages   = {205309},
  year    = {2002},
  doi     = {10.1103/PhysRevB.65.205309}
}

@misc{NIST_DLMF,
  title = {{NIST Digital Library of Mathematical Functions}},
  howpublished = {\url{https://dlmf.nist.gov/}},
  note = {Release 1.2.3 of 2024-12-15, F.~W.~J. Olver, A.~B. Olde Daalhuis,
          D.~W. Lozier, B.~I. Schneider, R.~F. Boisvert, C.~W. Clark,
          B.~R. Miller, B.~V. Saunders, H.~S. Cohl, and M.~A. McClain, eds.}
}

@article{Bluhm_NatPhys_2011_DephasingTime,
  title = {Dephasing time of GaAs electron-spin qubits coupled to a nuclear bath exceeding 200 $\mu$s},
  author = {Bluhm, Hendrik and Foletti, Sandra and Neder, Izhar and Rudner, Mark and Mahalu, Diana and Umansky, Vladimir and Yacoby, Amir},
  journal = {Nature Physics},
  volume = {7},
  pages = {109--113},
  year = {2011},
  doi = {10.1038/nphys1856}
}
\end{document}